\documentclass[twocolumn,pra,superscriptaddress,floatfix,preprintnumbers,nofootinbib]{revtex4-2}
\usepackage{lmodern}
\usepackage{float}

\usepackage[T1]{fontenc}
\usepackage[utf8]{inputenc}
\usepackage{geometry}
\usepackage{color}
\usepackage{comment}
\usepackage[english]{babel}
\usepackage{amsmath}
\usepackage{amssymb}
\usepackage{dsfont}
\usepackage{graphicx}
\usepackage{stackengine}
\usepackage{esint}
\usepackage{comment}
\usepackage{physics}
\usepackage[unicode=true,pdfusetitle,
 bookmarks=true,bookmarksnumbered=false,bookmarksopen=false,
 breaklinks=false,pdfborder={0 0 1},backref=false,colorlinks=true]
 {hyperref}
\hypersetup{citecolor=black,linkcolor=black,urlcolor=black}
\usepackage[hang,flushmargin]{footmisc} 
\allowdisplaybreaks
\makeatletter

\newcommand{\Tvac}{\includegraphics[height=1em]{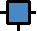}}
\newcommand{\Tfvac}{\includegraphics[height=1em]{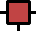}}
\newcommand{\Tkink}{\includegraphics[height=1em]{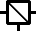}}
\newcommand{\Takink}{\includegraphics[height=1em]{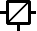}}
\newcommand{\Tmes}{\includegraphics[height=1em]{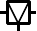}}

\DeclareMathOperator{\arccosh}{arccosh}

\begin{document}

\pagestyle{plain}

\preprint{CALT-TH-2020-010}

\title{Collisions of false-vacuum bubble walls in a quantum spin chain}

\author{Ashley Milsted}
\email{ashmilsted@gmail.com}
\affiliation{Walter Burke Institute for Theoretical Physics\\ California Institute of Technology, Pasadena, CA 91125, USA}
\affiliation{Institute for Quantum Information and Matter\\ California Institute of Technology, Pasadena, CA 91125, USA}
\affiliation{Perimeter Institute for Theoretical Physics, 31 Caroline St. N., Waterloo, Ontario, Canada, N2L 2Y5}
\affiliation{AWS Center for Quantum Computing, Pasadena, CA 91125, USA}
\thanks{AM completed this work prior to joining Amazon.}
\author{Junyu Liu}
\email{jliu2@caltech.edu}
\affiliation{Walter Burke Institute for Theoretical Physics\\ California Institute of Technology, Pasadena, CA 91125, USA}
\affiliation{Institute for Quantum Information and Matter\\ California Institute of Technology, Pasadena, CA 91125, USA}
\author{John Preskill}
\affiliation{Walter Burke Institute for Theoretical Physics\\ California Institute of Technology, Pasadena, CA 91125, USA}
\affiliation{Institute for Quantum Information and Matter\\ California Institute of Technology, Pasadena, CA 91125, USA}
\affiliation{AWS Center for Quantum Computing, Pasadena, CA 91125, USA}
\author{Guifre Vidal}
\affiliation{Perimeter Institute for Theoretical Physics, 31 Caroline St. N., Waterloo, Ontario, Canada, N2L 2Y5}
\affiliation{Sandbox@Alphabet, Mountain View, CA 94043, USA}


\begin{abstract}
We simulate, using non-perturbative methods, the real-time dynamics of small bubbles of ``false vacuum'' in a quantum spin chain near criticality, where the low-energy physics is described by a relativistic (1+1)-dimensional quantum field theory. We consider bubbles whose walls are kink and antikink quasiparticle excitations, so that wall collisions are kink-antikink scattering events. To construct these bubbles in the presence of strong correlations, we extend a recently proposed Matrix Product State (MPS) ansatz for quasiparticle wavepackets. We simulate dynamics within a window of $\sim1000$ spins embedded in an infinite chain at energies of up to $\sim5$ times the mass gap.
By choosing the wavepacket width and the bubble size appropriately, we avoid strong lattice effects and observe relativistic kink-antikink collisions. We use the MPS quasiparticle ansatz to detect scattering outcomes: (i) In the Ising model, with transverse and longitudinal fields, we do not observe particle production despite nonintegrability (supporting recent observations of nonthermalizing states in this model).
(ii) Switching on an additional interaction, we see production of confined and unconfined particle pairs.
We characterize the amount of entanglement generated as a function of energy and time and conclude that our classical simulation methods will ultimately fail as these increase. We anticipate that kink-antikink scattering in 1+1 dimensions will be an instructive benchmark problem for future quantum computers and analog quantum simulators.
\end{abstract}
\maketitle

It is possible that the known universe is built on top of a metastable, or ``false'' vacuum state. In this scenario, there is a tiny but nonzero probability of a small bubble of ``true'' vacuum forming via tunneling. The bubble interior has a lower energy density than its surroundings and so it expands, its walls accelerating, bulldozing everything in their path. If multiple bubbles of true vacuum form far apart, their walls will rush toward each other and eventually collide, producing showers of entangled particles. It is also possible that such events have already occurred, and are thus relevant for the evolution of the early universe \cite{Coleman:1977py,Callan:1977pt,Turner:1982xj,Coleman:1980aw, markkanen_cosmological_2018}.

Vacuum bubble walls are topological excitations and can be modelled as domain walls or kinks. Particle production is known to be suppressed in some weakly-coupled models of relativistic kink-antikink collision (see e.g.~\cite{giblin_2010, amin_2013}), but much less is known about scattering of topological excitations under strong interactions, where a lot of entanglement is typically generated and semi-classical methods break down. Whether or not these phenomena are important for simulations of the early universe is an open question. Simulations that can handle such non-classical dynamics could provide an important window into these high-energy, strongly-coupled dynamical processes.

However, one does not easily simulate the dynamics of a strongly interacting quantum field theory (QFT), at least using classical computers. Quantum Monte Carlo -- the workhorse for simulations of equilibrium phenomena in lattice systems (such as lattice QCD \cite{durr_2008}) -- is hard to apply efficiently to real-time dynamics, although recently some progress has been made (e.g.~\cite{kanwar_2021}). Tensor-network methods also show promise, and have been used to simulate nontrivial dynamical phenomena in (1+1)D systems, such as string breaking in lattice gauge theory \cite{milsted_2013_phi4, kuhn_nonabelian_2015, pichler_realtime_2016, buyens_realtime_2017, chanda_confinement_2020, magnifico_real_2020}. Nevertheless, although the computational cost is typically linear in spatial volume, it increases exponentially with time in the general case (due to linear scaling of entanglement entropy), the exponent increasing with the number of spatial dimensions. 

In principle, quantum computers (both analog and digital) can simulate dynamics of quantum field theories at long timescales with polynomial costs \cite{jordan_quantum_2012, jordan_quantum_2014}, a topic which has recently attracted great interest, with many digital \cite{mezzacapo_nonabelian_2015, martinez_realtime_2016, zohar_digital_2017, muschik_u1_2017, bender_digital_2018, preskill_simulating_2018, lamm_general_2019, kreshchuk_quantum_2020, Liu:2020eoa, du_quantum_2020, farrelly_discretizing_2020, shaw_quantum_2020, klco_su2_2020, vovrosh2020confinement, Li:2020kbv} and analog \cite{gonzalez-cuadra_quantum_2017, surace_lattice_2020, surace_scattering_2020, notarnicola_realtimedynamics_2020, davoudi_analog_2020, celi_emerging_2020} proposals (see \cite{banuls_simulating_2020} for a review). However, existing or near-term digital quantum devices are noisy, such that only shallow quantum circuits can avoid being overwhelmed by errors. Analog quantum simulators are currently more capable, supporting longer coherence times, but also have no general mechanism for correcting errors. As such, for the time being, the physical systems that can practically be simulated are limited to simple models and we expect classical simulations to perform better, with quantum devices catching up as the hardware improves, ultimately beating classical computers by an exponential margin with the arrival of large-scale, fault-tolerant digital quantum computers, permitting more complex simulations. Thus dynamical simulations of phenomena like false-vacuum collapse are physically-motivated applications for quantum computers and analog quantum simulators, and simple models can be used as benchmark problems for present and near-term quantum devices. To understand which problems make the most suitable quantum benchmarks, it is important to explore the limits of classical methods.

We address this question by developing a framework for classically simulating the full quantum dynamics of relativistic false-vacuum bubble-wall collisions in (1+1) dimensions on the lattice, using Matrix Product States \cite{fannes_1992, rommer_1997} to represent the evolving state. This is computationally feasible as long as the state does not become too entangled. We demonstrate our methods on a simple lattice model with emergent relativistic, strongly-coupled false-vacuum physics, showing that they can work away from perturbative approximations. By highlighting the limitations of such classical methods, our work clarifies where quantum advantage might potentially arise in relatively near-term quantum simulators. In the following paragraphs, we outline the structure of this paper.

To improve both the interpretability and the computational efficiency of our simulations, we choose our initial states to be false-vacuum bubbles whose walls are single topological particles: a kink and an antikink. We motivate this choice in Sec.~\ref{sec:bubble_structure}.

In Sec.~\ref{sec_selectingchain} we explain our selection of lattice model: Rather than a lattice-regularized QFT Hamiltonian (see e.g.~\cite{sugihara_2004, milsted_2013_phi4, jordan_quantum_2014, burak_upcoming, kuhn_nonabelian_2015, pichler_realtime_2016, buyens_realtime_2017, chanda_confinement_2020, magnifico_real_2020} for MPS simulations of these), which in the case of bosonic fields requires a truncation of an otherwise infinite Hilbert space for each lattice site, we consider a \emph{quantum spin chain}, chosen and tuned so that its low-energy physics is governed by an \emph{emergent} relativistic QFT. This is known to occur in the vicinity of many continuous phase transitions, where the emergent QFT is often a (by definition relativistic) conformal field theory (CFT) (see e.g.~\cite{friedan_conformal_1984}). To avoid strong lattice effects, we ensure that the kink and antikink lattice velocities remain below their maximum values.

The initial state preparation for our dynamical simulations is subtle, because the kink-antikink pair is not an energy eigenstate. As discussed in Sec.~\ref{sec_methods}, we take care to prepare initial kink and antikink wavepackets (efficiently represented as MPS) which are broad compared to the lattice spacing, and are not contaminated by additional unwanted excitations.

We successfully simulate inelastic kink-antikink collisions in our spin-chain model, including particle production, over $\sim 1000$ lattice sites at energies of up to $\sim 5 m_\mu$, where $m_\mu$ is the mass of the lightest quasiparticle, as we detail in Sec.~\ref{sec_result}. We develop tools for analysis of the outgoing particles produced in inelastic kink-antikink collisions and use them to show that particle production is strongly suppressed in the Ising model with intrinsic $\mathbb{Z}_2$ symmetry breaking, even though the model is nonintegrable in that case. We further show that copious particle production occurs once a $\mathbb{Z}_2$-symmetric three-site local interaction turns on. We also quantitatively track the growth of entanglement entropy during repeated kink-antikink collisions, thus inferring how large a bond dimension is needed to provide an accurate approximation to the evolving quantum state.

Sec.~\ref{sec_discussion} contains concluding remarks, and further details of our methods and results are provided in the appendices.

Before proceeding, we note that although our work is partially motivated by a potential connection with early universe cosmology, we would need to reach energies several orders of magnitude higher than those of our present simulations, as well as increase the number of spatial dimensions, in order to study models of direct relevance to the early universe. Both of these are challenging and our work can represent, at most, a small step toward performing such simulations at strong coupling. On the other hand, we feel that our studies of inelastic scattering events in a strongly coupled relativistic quantum field theory, and of the entropy generated in such events, are of intrinsic interest apart from any cosmological motivation. 

\section{Bubble structure}
\label{sec:bubble_structure}

\begin{figure}[t]
    \includegraphics[width=\linewidth]{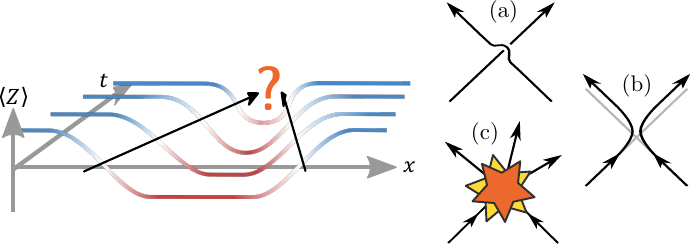}
    \caption{\label{fig:cartoon} Cartoon illustrating the relaxation of a false-vacuum bubble in a spin chain. The magnetization $\langle Z \rangle$ is positive in the true vacuum, but negative in the false vacuum. A bubble-wall collision is a scattering process, which may be (a) free (no interaction), (b) elastic (no particle production), or (c) inelastic (particle production). Note that in free and elastic scattering of topological particles the left (right) particle always remains a kink (antikink).}
\end{figure}

In (1+1) dimensions, a bubble of false vacuum separates two regions of true vacuum. Such a bubble may also be viewed as a confined pair of topological excitations -- a kink and an antikink -- with the false vacuum playing the role of a low-energy string that provides the confining force. The metastability of the false vacuum corresponds to suppression of string breaking \cite{lerose_quasilocalized_2020}.
We simulate the relaxation of bubbles in which the kink and antikink are initially single, localized, spatially separated topological particles of low mass. We henceforth use ``kink'' and ``antikink'' to refer exclusively to such topological particles, and ``bubble'' to refer to a false-vacuum bubbles so composed. Although a general false-vacuum bubble could have a more complicated wall structure (whose dynamics one could also simulate using MPS) we restrict our initial states in this way in order to make the simulation outcome easier to interpret: If the bubble walls are topological quasiparticles, then no particles will be produced \emph{until the walls collide} and the collision can be thought of as a kink-antikink scattering event with a single input channel. Such scattering events may be free (no interaction), elastic (interaction, but no particle production), or inelastic (particle production), as illustrated in Fig.~\ref{fig:cartoon}. Particles produced may include non-topological particles, which we call \emph{mesons} (following \cite{fonseca_ising_2001}) because they can be thought of as bound kink-antikink pairs in our model. These are particularly easy to observe, since outgoing mesons are not subject to a confining force and will propagate ballistically, quickly separating from any confined particles. 

\begin{figure}[t]
    \includegraphics[width=\linewidth]{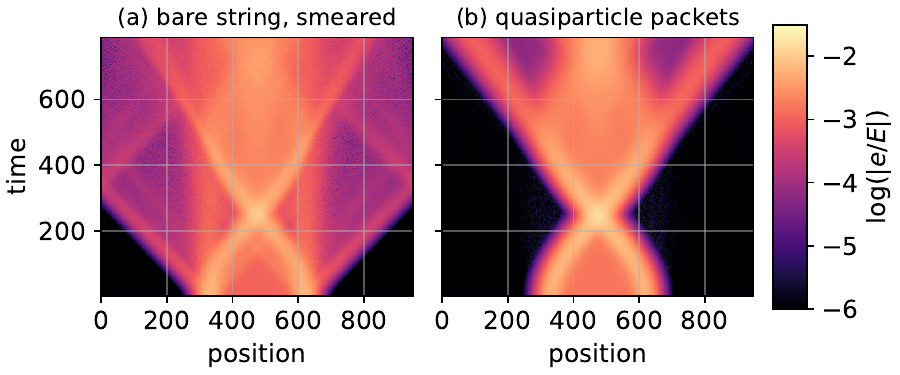}
    \caption{\label{fig:quench_main} Evolution of the excess energy density $e$ (relative to vacuum), as a fraction of total excess energy $E$, in a spin chain for two initial states: (a) created by applying a spatially smeared string operator to the vacuum and (b) constructed from MPS tensors to contain kink and antikink quasiparticle wavepackets. In (a) meson pairs are produced immediately at the string edges, whereas in (b) there is no particle production until the initial kink and antikink collide. The dynamics are restricted to a window of $\sim 1000$ sites, leading to boundary effects in (a). For more details, see App.~\ref{app:quenches}.}
\end{figure}

Previous work on related phenomena includes MPS simulations of string breaking in the Schwinger model \cite{kuhn_nonabelian_2015, pichler_realtime_2016, buyens_realtime_2017, Nagele_2019, chanda_confinement_2020, magnifico_real_2020}, where the initial state is typically prepared by applying a bare string operator to the vacuum. This generally creates excitations involving multiple particles of different energies, highly localized at the string edges, leading to relatively complex dynamics. Although such dynamics can still be simulated using standard MPS techniques, the rapid resulting entanglement growth can make it difficult to reach long times and to treat large systems. Such strings can be smeared out into wavepackets (as considered, for example in some recent work on confined excitations in spin chains \cite{surace_scattering_2020,Karpov:2020pqe}), which focuses the wavepacket momenta, significantly reducing the energy and entanglement growth. Nevertheless, the smeared string will still generally create multiple species of topological excitation, as shown in Fig.~\ref{fig:quench_main}.

Recently, techniques have been developed \cite{vandamme_realtime_2019a} to construct wavepacket states with selective particle content in generic (1+1)-dimensional systems using MPS\footnote{Also see \cite{vlijm_quasisoliton_2015} for a Bethe-ansatz approach.}. A main result of our paper is that we can extend those techniques to build initial bubbles whose walls are kink and antikink quasiparticle wavepackets. Apart from improving the interpretability of results, as discussed above, this further reduces the energy and the rate of entanglement growth compared to smeared string excitations, which enables us to treat larger systems and simulate for longer times.

\section{Selecting a spin chain}\label{sec_selectingchain}

\begin{figure}
    \includegraphics[width=0.95\linewidth]{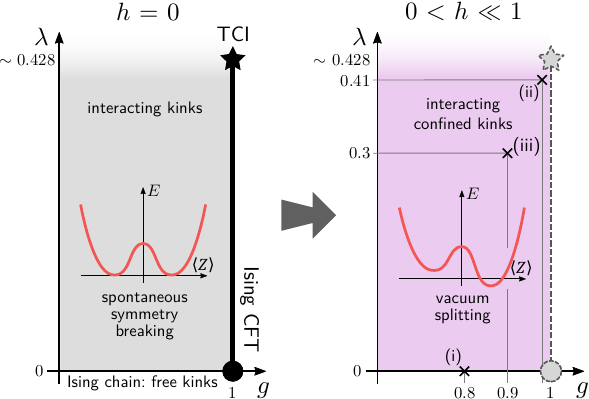}
    \caption{Partial phase diagrams of the extended Ising chain \cite{obrien_lattice_2018, sannomiya_supersymmetry_2019}, both without and with a small longitudinal field $h$ that breaks the $\mathbb{Z}_2$ symmetry. At $h=0$, $g=1$ there is a continuous (symmetry-breaking) phase transition described by the Ising CFT for $\lambda \alt 0.428$, and by the Tri-Critical Ising (TCI) CFT at $\lambda \approx 0.428$. By studying the spin chain near to these transitions ($g\rightarrow 1$, $h\rightarrow 0$), we can access emergent, relativistic quantum field theories with confined kinks \cite{fonseca_ising_2006, mussardo_effective_2009}. Points (i), (ii), and (iii) correspond to the data shown in the figures below.
    \label{fig:phasediag}}
\end{figure}

We seek a spin chain whose IR physics is described by a relativistic emergent field theory supporting confined kinks. In principle, there are many suitable models: An emergent field theory of confined kink-antikink pairs can be engineered by starting with a spontaneously-broken discrete symmetry, which provides multiple vacua and topological excitations. We then tune close to a symmetry-breaking phase transition, typically described by a CFT, and finally add a weak symmetry-breaking field to lift the vacuum degeneracy and confine the kinks. We must take care, however, since emergent field theories of such spin chains are sometimes \emph{integrable} \cite{zamolodchikov_integrable_1989, zamolodchikov_integrals_1989a, delfino_nonintegrable_1996, mussardo_integrability_2011}, in which case scattering, including kink collision, is always \emph{elastic}. If we want to observe particle production in the emergent field theory, we must avoid integrability.

We choose an extension \cite{obrien_lattice_2018, sannomiya_supersymmetry_2019} of the transverse-field Ising chain 
\begin{multline} \label{eq:H_OF}
  H = \sum_{j=1}^N \big[ -Z_j Z_{j+1} - g X_j - h Z_j +\\
    \lambda \left(X_j Z_{j+1} Z_{j+2} + Z_j Z_{j+1} X_{j+2} \right)\big],
\end{multline}
where $X$, $Z$ are Pauli matrices. At $h=0$, this model has a $\mathbb{Z}_2$ symmetry ($Z_j \rightarrow -Z_j$, $X_j \rightarrow X_j$) that is spontaneously-broken when $g < 1$ for a large range of $\lambda$: see Fig.~\ref{fig:phasediag} for a phase diagram. 
For $\lambda=0$, we have the transverse-field Ising chain, which already supports confined kinks for $g < 1$ and $0 < |h| \ll 1$ \cite{mccoy_twodimensional_1978, fonseca_ising_2006}. However, for small $|h|$ it is very close to being integrable \cite{zamolodchikov_integrals_1989a, fonseca_ising_2006} (both the emergent field theory and the spin chain itself are noninteracting for $h=0$). Previous work has shown that kink-antikink pair states in this model can have extremely long lifetimes \cite{Rakovszky_2016, kormos_real_2017, lin_quasiparticle_2017, hodsagi_quench_2018, castro-alvaredo_entanglement_2020}, with recent numerical studies suggesting that these long-lived states can have energies well above the threshold for inelastic scattering \cite{james_nonthermal_2019, wurtz_emergent_2020, lerose_quasilocalized_2020}. This is despite the lack of any exact conservation law protecting these excited meson states from decay.

Turning on $\lambda$ allows us to go beyond this ``almost-integrable'' regime, since both the spin chain itself and the emergent field theory are nonintegrable for $\lambda > 0$, $g < 1$, even at $h=0$ \cite{lassig_scaling_1991, lepori_particle_2008, mussardo_effective_2009, mussardo_integrability_2011}. 
We present simulations at a point along the Ising line $\lambda=0$, labelled (i) in Fig.~\ref{fig:phasediag}, as well as at two points, labelled (ii) and (iii), closer to the Tri-Critical Ising (TCI) point at $\lambda \rightarrow 0.428$, $g = 1$, $h = 0$.

In the vicinity of the TCI point, \eqref{eq:H_OF} exhibits the same universal properties \cite{lassig_scaling_1991, lepori_particle_2008} as the Landau-Ginzburg theory with action
\begin{equation}
    \mathcal{A}_{LG} = \int \mathrm{d}^2 x \; \left(
      \frac{1}{2}(\partial_\mu \phi)^2 + a_2 \phi^2 + a_4 \phi^4 + \phi^6
    \right),
\end{equation}
with the parameters corresponding as $g \sim a_2$, $\lambda \sim a_4$, and the $\mathbb{Z}_2$-breaking field controlled by $h$ mapping to odd powers of $\phi$.

\section{Methods}\label{sec_methods}

We first describe how to construct the states relevant to our simulations. This includes the bubble states, consisting of a localized kink and antikink pair, which we use to initialize our simulations. We choose the kink and the antikink to be topological quasiparticles of our spin chain model.

For simplicity, we begin with the construction of states in the \emph{bare} setting $\lambda = 0$, $g = 0$, where quantum fluctuations vanish, before moving onto the \emph{dressed} setting, where we use MPS to capture, non-perturbatively, the fluctuations that appear. In both cases we first describe the true and false vacua, then the kink and antikink states, before explaining how to combine them into a bubble.
We work in the $Z$-basis throughout: ${Z|\mkern-4.0mu\uparrow\rangle = |\mkern-4.0mu\uparrow\rangle}$, ${Z|\mkern-4.0mu\downarrow\rangle = -|\mkern-4.0mu\downarrow\rangle}$.

\subsection{Bare states}\label{sec_methods_bare_states}

Let us first consider $\lambda = 0$, $g = 0$, for which the terms in \eqref{eq:H_OF} commute and there are no quantum fluctuations (eigenstates of $H$ can always be chosen to have definite spin orientations in the $Z$ basis). In this bare case, the true and false vacua are simply $|\Omega^{\textrm{bare}}\rangle := |\dots \uparrow \uparrow \uparrow \dots\rangle$ and $|\overline{\Omega}{}^{\textrm{bare}}\rangle := |\dots \downarrow \downarrow \downarrow \dots\rangle$, respectively. A kink is a domain wall $|\kappa_j^{\textrm{bare}}\rangle := |\dots \uparrow \downarrow_j \dots \rangle$, here located at position $j$, and an antikink is $|\overline{\kappa}{}_k^{\textrm{bare}}\rangle := |\dots  \downarrow_k \uparrow \dots\rangle$. These highly-localized excitations have maximal momentum uncertainty. By smearing them out into \emph{wavepackets}, e.g.\ $\sum_j f_j |\kappa_j^{\textrm{bare}}\rangle$, we can make them quasilocal in both position and momentum space. We consider Gaussian packets
\begin{equation} \label{eq:gaussian}
    f_j(x, p) := e^{\mathrm{i}pj} e^{\frac{-(j - x)^2}{\sigma^2}},
\end{equation}
centered at position $x$ and momentum $p$, with spatial width $\sigma$. In the maximally delocalized limit $\sigma \rightarrow \infty$ we obtain a momentum eigenstate with momentum $p$. By combining kink and antikink wavepackets we can construct a false-vacuum bubble with quasilocalized walls at positions $x_L$ and $x_R$
\begin{equation} \label{eq:dbl_pkt_bare}
    |\Psi^{\textrm{bare}}\rangle = \sum_{j<k} f_j(x_L, p_L) f_k(x_R, p_R) |\kappa \overline{\kappa}{}^{\textrm{bare}}_{jk}\rangle,
\end{equation}
where $x_R - x_L$ determines the size of the bubble, $p_L$ and $p_R$ specify the expected momenta of the bubble walls, and we define the localized kink-antikink pair states
\begin{equation} \label{eq:bubble_bare}
|\kappa \overline{\kappa}{}^{\textrm{bare}}_{jk}\rangle := |\dots  \uparrow  \downarrow_j \dots \downarrow_k  \uparrow \dots\rangle.
\end{equation}
Note that the restriction $j < k$ (the kink must be to the left of the antikink) means that the Gaussian packets \eqref{eq:gaussian} are truncated. In practice, one can ensure that this truncation is negligible by choosing $x_L$, $x_R$, and $\sigma$ so that the coefficients are very small when $j \sim k$.

In addition to kinks and kink-antikink pairs, the bare ``meson'' states $|\mu^{\textrm{bare}}_{j}\rangle := |\dots \uparrow \downarrow_j \uparrow \dots \rangle$ and pairs thereof $|\mu\mu^{\textrm{bare}}_{jk}\rangle := |\dots \uparrow \downarrow_j \uparrow \dots \uparrow \downarrow_k \uparrow \dots\rangle$ represent classes of topologically \emph{trivial} excitations. The dressed counterparts of these ``scalar'' particles, are among the possible outcomes of kink-antikink scattering events.

\subsection{Dressed states as Matrix Product States}

While the bare states introduced above illustrate many relevant properties of the states we wish to construct for ${H(g > 0, \lambda \ge 0)}$, they are all \emph{eigenstates} of the bare Hamiltonian ${H(g=0,\lambda=0)}$, implying that kinks and antikinks do not propagate\footnote{$H(g=0, \lambda=0)$ is an RG fixed-point and one can think of the kinks and antikinks as having infinite mass.}. They are also \emph{product states}, devoid of entanglement. To obtain interesting dynamics, we need $g>0$, for which all the bare states have counterparts, non-perturbatively dressed by fluctuations, possessing exponentially decaying correlations and entanglement between lattice sites. In addition to this entanglement, for $g>0$ evolution by $H$ can \emph{generate} new entanglement, in contrast to the bare case.

The dressed states and their dynamics under $H$ can be represented, non-perturbatively, using Matrix Product States (MPS)\cite{fannes_1992, rommer_1997, vidal_2003, vidal_2004, verstraete_2006}, a variational class of states with the form
\begin{equation} \label{eq:MPS_finite}
    |\psi\rangle = \sum_{\{s\}} A_1^{(s_1)} A_2^{(s_2)} \dots A_N^{(s_N)} |s_1 s_2 \dots s_N\rangle, 
\end{equation}
where $N$ is the number of spins (lattice sites), $s_j = \; \uparrow,\downarrow$ for our model and each $A_j^{(s)}$ is a $D_{j-1} \times D_j$ matrix, making $A_j$ a rank-3 tensor. At the ends of the chain we have $D_0 = D_N = 1$. MPS can also be illustrated using tensor network diagrams, which we will use in the following for convenience. For example, we can rewrite~\eqref{eq:MPS_finite} as
\begin{equation}
    |\psi\rangle = \vcenter{\hbox{\includegraphics[height=1.6em]{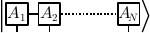}}},
\end{equation}
where $\vcenter{\hbox{\includegraphics[height=1em]{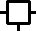}}}$ represents a rank-3 tensor.

Computations with MPS typically scale as $\mathcal{O}(D^3)$, with linear scaling in the number of spins involved in the computation. The dimensions $D_j$, called \emph{bond dimensions}, limit the amount of entanglement that can be represented. We define the \emph{cut entropy} at location~$j$ 
\begin{equation}
    S(\rho_{>j}) = -\tr(\rho_{>j} \log_2(\rho_{>j}))
\end{equation}
to be the von Neumann entropy of the subsystem consisting of all sites $>j$. In an MPS, the cut-entropy at $j$ is upper-bounded as $S(\rho_{>j}) \leq \log_2 D_j$.

With appropriate choice of bond dimensions $D_j$, any state of a quantum spin chain with $N$ spins can be represented exactly as an MPS \cite{vidal_2003}. Furthermore, low-energy eigenstates of locally-interacting one-dimensional lattice systems with a spectral gap, such as our model \eqref{eq:H_OF} at off-critical parameters, can be represented exponentially accurately in the bond dimensions $D_j$, independently of the system size $N$ \cite{hastings2007}. We therefore expect an MPS to accurately represent both the initial and evolved states of our simulations, provided that we use sufficiently large bond dimensions $D_j$. If the bond dimensions are too small, this will introduce errors.

\begin{figure*}
    \includegraphics[width=0.65\linewidth]{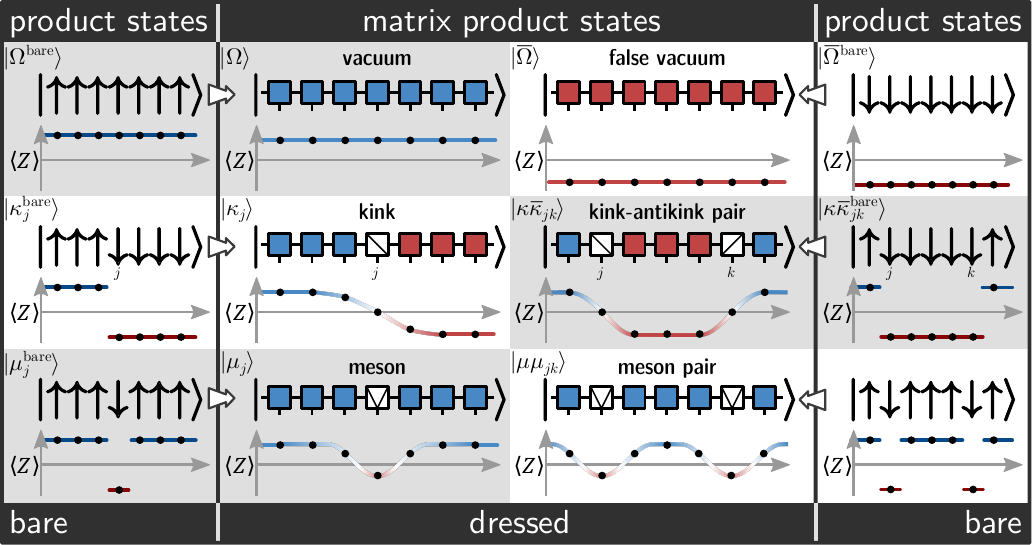}
    \caption{Diagram illustrating the various types of states, and their spin profiles, relevant for simulations. For example, our initial states are \emph{wavepackets} constructed from kink-antikink pairs. The product states listed are eigenstates of $H$ when $g=0$, $\lambda=0$. Away from this regime, we use MPS to accurately capture fluctuations in the vacua, as well as the \emph{quasilocal} nature of the excited quasiparticle states.}
    \label{fig:states}
\end{figure*}

In our constructions, to avoid boundary effects, we work directly in the thermodynamic limit $N\rightarrow \infty$, using infinite MPS (see App.~\ref{app:imps}), which we illustrate as
\begin{equation}
|\psi\rangle = \vcenter{\hbox{\includegraphics[height=1.6em]{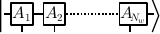}}},
\end{equation}
using trailing legs on the left and right to indicate that the chain of tensors extends from $-\infty$ to $\infty$. The numbered sites $1 \dots N_w$ indicate a finite window inside the bulk of the infinite chain. In Fig.~\ref{fig:states}, we use similar diagrams to indicate how the our dressed states are constructed as infinite MPS.

\emph{Vacuum and false vacuum.---} The MPS $|\Omega\rangle$ and $|\overline{\Omega}\rangle$ are approximations to the dressed true and false vacua. They are uniform, infinite MPS built from tensors $\Tvac$ and $\Tfvac$, respectively. We optimize $\Tvac$ using variational methods \cite{haegeman_2011, haegeman_2012_excite, haegeman_2013_el_ex, milsted_2013_phi4, milsted_2013_sand} to minimize the energy of $|\Omega\rangle$. To find the metastable false vacuum $|\overline{\Omega}\rangle$, we first apply a global spin-flip to $|\Omega\rangle$, resulting in another uniform MPS, whose energy we then minimize. For sufficiently small $g \ll 1$, we observe that the energy-minimization procedure does not find a path to the true ground state, resulting in a \emph{metastable} false vacuum state $|\overline{\Omega}\rangle$, with MPS tensor~$\Tfvac$, that behaves as an energy eigenstate for all practical purposes (see App.~\ref{app:falsevac}).

\emph{Kinks and antikinks.---} The MPS $|\kappa_j\rangle$ approximates a dressed, localized kink state. It is constructed by introducing a new tensor $\Tkink$ that sits at position $j$, between two semi-infinite chains, one consisting of $\Tvac$ on the left and one of $\Tfvac$ on the right. The tensor $\Tkink$ parameterizes the spatial transition between the true and false vacuum regions. Unlike in the bare case, the transition region may encompass many lattice sites, as illustrated by the $\langle Z\rangle$ plots in Fig.~\ref{fig:states}. The antikink $|\overline{\kappa}{}_k\rangle$ is similarly constructed by introducing a tensor $\Takink$ between a chain of $\Tfvac$ on the left and $\Tvac$ on the right. We select $\Tkink$ and $\Takink$ using an MPS Bloch-state approach \cite{rommer_1997, haegeman_2012_excite, haegeman_2013_el_ex, vanderstraeten_2015} so that $|\kappa_j\rangle$ and $|\overline{\kappa}{}_k\rangle$ states can be thought of as ``position bases'' for the kink and antikink \emph{quasiparticles} of lowest energy. We may use these states to construct topological quasiparticle wavepackets $\sum_j f_j |\kappa_j\rangle$ and $\sum_k f_k |\overline{\kappa}{}_k\rangle$ \cite{vandamme_realtime_2019a}. If $D$ is the bond dimension of the vacuum MPS, such wavepackets have MPS representations with bond dimension $2D$ (see App.~\ref{app:ex_basis} for details).

The Bloch-state approach for finding $\Tkink$ and $\Takink$ is conceptually simpler when $h = 0$, so that there is no confining force acting on the kinks and antikinks. We consider the $h \neq 0$ case further below. For $h = 0$, we solve an effective Hamiltonian for $\Tkink$ and $\Takink$ such that the momentum eigenstates, $\sum_j e^{\mathrm{i}pj} |\kappa_j\rangle$ and $\sum_k e^{\mathrm{i}pk} |\overline{\kappa}{}_k\rangle$, approximate the lowest-energy topological eigenstates of $H$ with momentum $p$ \cite{haegeman_2013_el_ex}. The resulting tensors $\Tkink$, $\Takink$ (and hence $|\kappa_j\rangle$, $|\overline{\kappa}{}_k\rangle$) generally depend on the momentum $p$, but we ignore this dependence when building wavepackets, aside from choosing the $p$ used to solve for $\Tkink$ and $\Takink$ to match the expectation value of momentum in the wavepacket state. This is justified for Gaussian wavepackets with large $\sigma$, and hence small momentum variance, if the tensors vary sufficiently slowly with $p$.

There is also an important physical reason for choosing $\sigma$ to be large: In the presence of fluctuations, localized packets can no longer be truly static, since they are not eigenstates of $H$ even for $h=0$. Instead, they will spread out as time passes, at a rate dependent on $\sigma$. Wavepackets can be made to spread slowly relative to other processes, such as the collapse of a false-vacuum bubble, by choosing $\sigma \gg \xi$, where $\xi$ is the correlation length in lattice units. It is desirable for the kink and antikink wavepackets comprising a bubble to spread only minimally prior to collision, since then the wavepackets of \emph{outgoing} quasiparticles also tend to be well localized, which makes them easier to characterize.

We now explain how to find $\Tkink$ and $\Takink$ in case $|h| > 0$, where the confining force on kinks and antikinks means that $\sum_j e^{\mathrm{i}pj} |\kappa_j\rangle$ and $\sum_k e^{\mathrm{i}pk} |\overline{\kappa}{}_k\rangle$ can no longer be eigenstates of $H$. In this case, we find $\Tkink$ and $\Takink$ by optimizing \emph{modified} energy functions that subtract away the false-vacuum contributions, which in $|\kappa_j\rangle$ and $|\overline{\kappa}{}_k\rangle$ depend on the positions $j$ and $k$ (thus providing an accelerating force). We explain this for the case of $|\kappa_j\rangle$ and $\Tkink$, since the procedure is completely analogous for $|\overline{\kappa}{}_k\rangle$ and $\Takink$. We first note that it is possible to choose $\Tkink$, by exploiting a redundancy in the representation of momentum eigenstates, to achieve ${\langle \kappa_j | \kappa_{k}\rangle = \delta_{jk}}$ (see App.~\ref{app:ex_basis}). After making this choice, we minimize
\begin{equation} \label{eq:Etilde}
    \tilde E = \sum_{jk} e^{\mathrm{i}p(k-j)} \langle \kappa_j | (H - \Delta E_j \; \mathds{1}) | \kappa_{k}\rangle,
\end{equation}
where $\Delta E_j := \sum_{-\infty}^{j-1} e_{\textrm{true}} + \sum_{j}^{\infty} e_{\textrm{false}}$ captures the infinite bulk contributions to the energy present in $|\kappa_{j}\rangle$, coming from the true and false vacua ($e_{\textrm{true}}$ and $e_{\textrm{false}}$ are the energy densities of the true and false vacua). Subtracting them in this position-dependent way makes the contribution of each $|\kappa_j\rangle$ term to $\tilde E$ finite and independent of $j$. Note that the $\Delta E_j$ correction does not affect off-diagonal terms $j\neq k$ in the sum of \eqref{eq:Etilde}. The energy minimization procedure is easily carried out by slightly adapting the methods of \cite{haegeman_2012_excite} (see App.~\ref{app:ex_basis} for details). 

\emph{False vacuum bubbles.---} To build dressed bubble states $|\Psi\rangle$ as MPS, we proceed analogously to the bare case by combining a kink and an antikink wavepacket
\begin{equation} \label{eq:dbl_pkt}
    |\Psi\rangle = \sum_{j<k} f_j(x_L, p_L) f_k(x_R, p_R) |\kappa \overline{\kappa}{}_{jk}\rangle
\end{equation}
where, in our simulations, we choose the momenta $p_L=p_R=0$ and set $x_R - x_L \gg \sigma$ so that $f_j(x_L) f_k(x_R)$ is small for small $k-j$. The kink-antikink pair states $|\kappa \overline{\kappa}{}_{jk}\rangle$ are constructed by combining the tensors $\Tvac$, $\Tkink$, $\Tfvac$ and $\Takink$ (already optimized to represent the vacua, kinks, and antikinks) without further modification, as illustrated in Fig.~\ref{fig:states} (see App.~\ref{app:ex_basis} for further details). With this scheme, $|\kappa \overline{\kappa}{}_{jk}\rangle$ accurately describes a kink-antikink pair at asymptotically large separations $k-j$. However, at small separations, corrections would generally be needed due to interaction effects\footnote{A priori, the MPS tensors may also need modifying at short distances even in the absence of interactions.}. We again rely on $x_R - x_L \gg \sigma$ here, which ensures that terms with small separation are strongly suppressed, so that the error incurred by ignoring interactions is small.

\emph{Mesons and meson pairs.---} In addition to the topological excitations $|\kappa_j\rangle$ and $|\overline{\kappa}{}_j\rangle$, we can construct MPS representations $|\mu_j\rangle$ of topologically trivial ``meson'' states. The meson states are built from the vacuum tensor $\Tvac$ and the meson tensor $\Tmes$ as illustrated in Fig.~\ref{fig:states}, where we optimize $\Tmes$ to represent the topologically trivial particle of lowest energy using the same Bloch state approach that we use to find the kink and antikink tensors (except that no special accommodation is needed for $h\neq 0$ as mesons remain unconfined). As in the kink-antikink case, we can construct meson-pair states $|\mu\mu_{jk}\rangle$ by combining two instances of $\Tmes$, separated by vacuum tensors $\Tvac$ (see Fig.~\ref{fig:states}). As discussed below, we can use these meson-pair states to detect the presence of outgoing mesons (even before they are visible as particle tracks distinct from an outgoing false-vacuum bubble).

\subsection{Time evolution}\label{sec_methods_time}
To classically evolve an initial MPS $|\Psi(t=0)\rangle$ in time, we apply the time-dependent variational principle (TDVP) \cite{haegeman_2011} within a finite window of the infinite chain surrounding the initial bubble \cite{milsted_2013_sand}. The TDVP provides effective equations of motion for the MPS tensors so that the evolution of the MPS approximates evolution by the Hamiltonian $H$. Assuming these equations are integrated accurately, any systematic errors come from restricting the bond dimension $D$, and hence the entanglement, in the MPS. If $D$ were allowed to grow arbitrarily large, the evolution of the state could be computed exactly. In our simulations, we allow $D$ to grow \cite{haegeman_2013_post, haegeman_unifying_2016} up to a predetermined maximum value in each simulation, running the evolution several times with different limits on $D$ in order to detect any systematic errors due to the resulting entanglement restriction. Note that we do not restrict the MPS tensors within the window surrounding the initial bubble in any other way: \emph{during the evolution} the state is not constrained to the forms illustrated in Fig.~\ref{fig:states}, but is allowed to be a general MPS.

For the numerical integration of the TDVP equations of motion, we primarily use the Runge-Kutta 4/5 algorithm, which we find provides a good balance of speed and accuracy except at very early times, where we use the better-conditioned, but more computationally intensive, ``split-step'' integrator of \cite{haegeman_unifying_2016}. These methods are implemented in the \emph{evoMPS} python package \cite{milsted_evomps}.

As the state evolves, we monitor its spin and energy expectation values as well as its entanglement properties. This allows us to draw conclusions about collision (scattering) outcomes. For instance, elastic and inelastic scattering are easily distinguished from the trivial case, as interaction generically results in \emph{entanglement} between any outgoing kinks or particles\footnote{This is true for wavepackets of finite width. In the limit of infinitely-broad spatial wavepackets (momentum eigenstates), elastic scattering does not produce entanglement in (1+1)D. see App.~\ref{app:elastic_ent}.}, whereas trivial scattering never does. We can also easily distinguish elastic and inelastic scattering in many cases. For example, if a collision produces a pair of mesons, their wavepackets will spread \emph{ballistically} since two mesons are not subject to a confining force. Indeed, any sustained ballistic spread of energy implies particle production. Importantly, the converse does not always hold, since confined topological particles different from those of the initial state may also be produced.

\subsection{Particle detection}\label{sec_methods_particledetect}
Aside from constructing the initial state $|\Psi\rangle$, we can also use MPS such as $|\kappa \overline{\kappa}{}_{jk}\rangle$, representing kink-antikink pairs, as a kind of \emph{particle detector}, the inner product $\langle\kappa \overline{\kappa}{}_{jk}|\Psi(t)\rangle$ corresponding approximately to the amplitude of a kink-antikink pair with position $j,k$ at time~$t$. Conveniently, these states can be made to fulfill $\langle\kappa \overline{\kappa}{}_{jk}|\kappa \overline{\kappa}{}_{lm}\rangle = \delta_{jl}\delta_{kl}$ (see App.~\ref{app:ex_basis}). We can treat the subspace spanned by these basis states as an approximate kink-antikink pair ``sector'', which we denote $\kappa\overline{\kappa}$. One reason for its approximate nature should be familiar from the discussion above: The basis captures a kink-antikink pair most accurately if the kink and antikink are smeared out into wavepackets that are sufficiently broad, so that the wavepacket momenta are focused around the momentum $p$ used to compute $\Tkink$ and $\Takink$. Additionally, the kink and antikink must be sufficiently separated so that interaction effects are insignificant. Fortunately, these two properties can be checked \emph{after} projecting the wavefunction into the $|\kappa \overline{\kappa}{}_{jk}\rangle$ subspace. The simplest way to deal with terms in which the kink and antikink are too close together is to simply exclude them from the projection subspace. Inaccuracies due to the momentum-dependence of $\Tkink$ and $\Takink$ can be mitigated in a few ways: Assuming the momentum dependence is not too strong, the simplest strategy is to tune $\Tkink$ and $\Takink$ to match the expected momentum of the projected wavefunction. A more precise result can be had via a Fourier analysis, in which the detection subspace is further restricted to a range of momenta that match $\Tkink$ and $\Takink$. See App.~\ref{app:detection} for a more detailed, technical discussion.

The $\kappa\overline{\kappa}$ subspace is already sufficient to detect inelastic scattering: if the portion of the wavefunction within the subspace drops significantly during evolution, particle production has likely occurred. Going further, we can construct quasiparticle position bases for other quasiparticle types, both topological and nontopological. To this end, we define the MPS $|\kappa^{(a)}_j\rangle$, $|\overline{\kappa}{}^{(a)}_j\rangle$, and $|\mu^{(a)}_j\rangle$, with corresponding tensors $\Tkink^{(a)}$, $\Takink^{(a)}$, and $\Tmes^{(a)}$, to be approximate position bases for the $a^{th}$ kink, antikink, and meson quasiparticles, with $a=0,1,\dots$ in ascending order of energy. We sometimes suppress the superscript $a$ when considering the lowest-energy quasiparticles of each type $a=0$. We compute $\Tkink^{(a)}$, $\Takink^{(a)}$, and $\Tmes^{(a)}$ by simply solving for multiple energies in the Bloch-state approach used above to generate the lowest-energy kink and antikink tensors $\Tkink$, $\Takink$ \cite{haegeman_2012_excite, vanderstraeten_2015}\footnote{These methods also deliver approximate quasiparticle dispersion relations.}. This procedure can deliver accurate quasiparticle states for quasiparticles with energy $E_a$ below the two-particle threshold \cite{haegeman_2013_el_ex}. Above that threshold, these tensors may correspond to unstable excitations. The procedure also guarantees that $\langle \kappa^{(a)}_j|\kappa^{(b)}_k\rangle = \langle \overline{\kappa}{}^{(a)}_j|\overline{\kappa}{}^{(b)}_k\rangle = \langle\mu^{(a)}_j|\mu^{(b)}_k\rangle = \delta_{jk}\delta_{ab}$. The kink, antikink and meson single-particle bases are mutually orthogonal by construction, due to the orthogonality of the true and false vacua in the thermodynamic limit.

We can construct pair states $|\kappa \overline{\kappa}{}_{jk}^{(a,b)}\rangle$ and $|\mu \mu_{jk}^{(a,b)}\rangle$ from this extended set of single-quasiparticle states following Fig.~\ref{fig:states}. These extended bases are \emph{not} orthonormal at small separations $k-j$ due to interaction effects. Nevertheless, we can compute a minimum separation $d$ for each set of Hamiltonian parameters $g,\Delta,h$ so that the bases are \emph{approximately} orthonormal when $k-j \ge d$ (see App.~\ref{app:ex_basis}). These restricted bases give us access to extended kink-antikink $\kappa \overline{\kappa}{}^{(a,b)}$ and meson-pair $\mu \mu^{(a,b)}$ ``sectors'', allowing a much finer analysis of particle content.

Note that, while it is possible to construct basis states containing $k$ particles, including $k \ge 3$, the cost of computing inner products of these basis states with the evolved state $|\Psi(t)\rangle$ scales as $\mathcal{O}(N_w^k)$, where $N_w$ is the size of the lattice window where the state is allowed to differ from the vacuum. For $N_w \sim 1000$ this makes accessing sectors with $k > 2$ more challenging. That said, the presence of outgoing scattering channels can still be inferred indirectly by looking for a probability deficit after accounting for all the kinematically allowed $k=2$ sectors. We further note that this limitation does not affect our dynamical simulations, which represent the state as a general MPS rather than using the quasiparticle basis states, and as such do not affect the suitability of such simulations as benchmark problems for quantum dynamical simulations.

\subsection{Sources of error}
\label{sec_methods_error}

\emph{Vacuum and false vacuum.---} The quality of our MPS approximations, $|\Omega\rangle$ and $|\overline{\Omega}\rangle$, to the true and false vacua, is dependent on the MPS bond dimension $D$ and on the success of the optimization procedure used to find the tensors $\Tvac$ and $\Tfvac$. We choose $D$ high enough to ensure that after optimization, the smallest Schmidt coefficient under a cut is $\mathcal{O}(10^{-6})$ or smaller (in amplitude -- this corresponds to probabilities $\mathcal{O}(10^{-12})$). We optimize the vacua until the norm of the energy gradient vector is $< 10^{-11}$, indicating that we have, to very good approximation, an energy eigenstate. It is worth noting that, since the vacua are forced to be translation invariant, in simulations any inaccuracy leads to spatially uniform (global quench) dynamics that are easily distinguished from localized quasiparticle wavepackets.

\emph{Kinks, antikinks, and mesons.---} The accuracy of our kink, antikink, and meson Fourier modes is limited by the vacuum bond dimension. The quality of the localized wavepackets constructed from the tensors $\Tkink$, $\Takink$, which we use to initialize our simulations, can be evaluated by projecting the wavepacket states onto the basis of Fourier modes. We do this for individual kinks in App.~\ref{app:ex_basis}, finding errors of $\mathcal{O}(10^{-6})$.
Although we do not directly check the accuracy of the Fourier modes themselves, the observation that our kinks and antikinks propagate, to very good accuracy, in a stable fashion (see the results section below) until they collide demonstrates that we are successfully capturing the targeted topological quasiparticles.

\emph{Bubble states.---} Beyond any errors in the individual kink and antikink wavepackets, we ensure that, in our bubble states, the kink and antikink are sufficiently well separated such that the initial kink and antikink do not interact significantly. What can be considered a sufficiently large separation is evaluated in App.~\ref{app:detection}. 

\emph{Time evolution.---} We simulate the full quantum dynamics of our initial states under our spin chain Hamiltonian. The only sources of error, aside from the storage and manipulation of complex numbers using a floating point representation (we use 128-bit, ``double'' precision for complex numbers), are (i) the numerical integration method used to step through time and (ii) the restriction of the allowed MPS bond dimensions $D_j$ to a chosen maximum value (for computational efficiency). In our simulations, (ii) is by far the dominant source of error (see App.~\ref{app:tdvp}). For this reason, we carry out our simulations using several different values for the maximum allowed bond dimension, in order to probe the sensitivity of each result to this parameter.

\emph{Particle detection.---} As for the bubble states, the particle-pair states we use to detect outgoing particles are dependent, for accuracy, on the quality of the kink, antikink, and meson tensors, as well as the validity of neglecting interaction effects. Again, a minimum separation between excitation tensors in the pair states is needed to avoid the latter source of error: see App.~\ref{app:detection}. In some cases, components of the evolving state may not neatly fall into particle sectors due to lack of separation. For example, confinement may prevent all terms in an outgoing bubble state from being sufficiently well separated. We note that, even in such cases, our main result -- that we can accurately simulate relativistic kink-antikink collisions -- is not affected by these limitations of the particle detection scheme.

\section{Results}\label{sec_result}

In the following text and figures, positions are given in lattice sites (relative to the leftmost point in the simulation window) and times are scaled so that the maximum kink velocity, determined from dynamical simulation, is 1 lattice site per unit time. Note that we \emph{only} scale times in this way: energies are given, where not specified as ratios, in unscaled lattice units of \eqref{eq:H_OF}.  Momenta $p$ are given in lattice units $-\pi < p \leq \pi$.

\subsection{Kink dynamics}\label{sec_result_kink}

\begin{figure}
    \centering
    \includegraphics[width=0.98\linewidth]{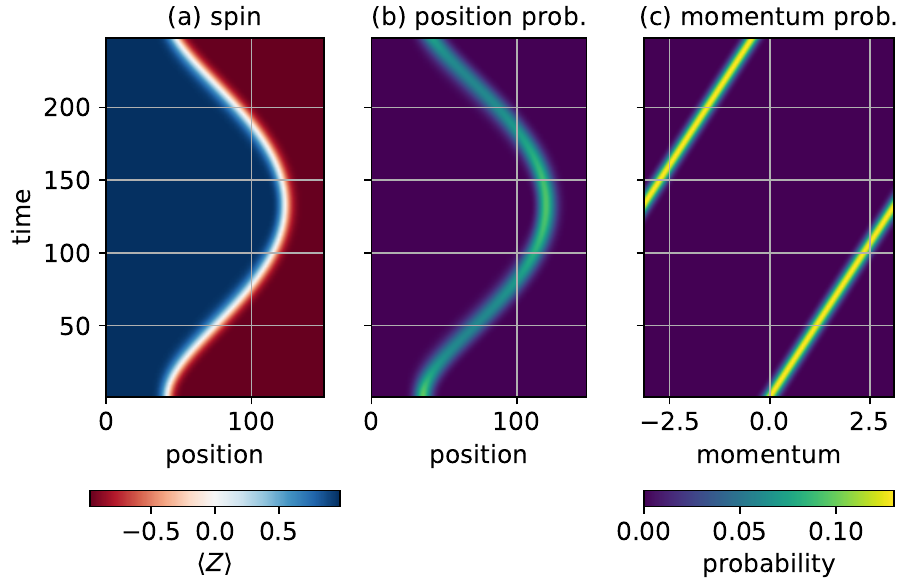}
    \caption{Evolution of a single kink in the $\mathbb{Z}_2$-broken Ising model (parameters $\lambda=0$, $g=0.4$, $h=0.01$). The vacuum bond dimension $D=6$, was allowed to evolve to a maximum $D \leq 32$. The meson mass is $m_\mu \approx 2.36$ and $v_{\max} \approx 0.83$, both in unscaled units of \eqref{eq:H_OF}. Spin expectation values $\langle Z \rangle$ are shown (a), as is the position-basis probability (b) $|\langle\kappa_j|\psi(t)\rangle|^2$. The momentum (c) is obtained from the Fourier transform of the position-basis projection.}
    \label{fig:1kink_pos_mom_main}
\end{figure}

\begin{figure}
    \centering
    \includegraphics[width=0.9\linewidth]{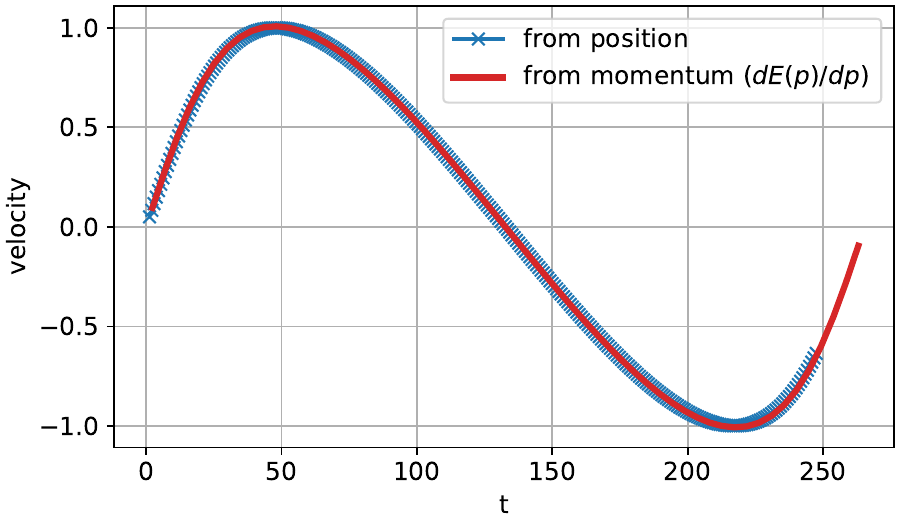}
    \caption{Evolution of the kink velocity, relative to its maximum value, for the single-kink simulation of Fig.~\ref{fig:1kink_pos_mom_main}, computed from the projection of the wavefunction into the $|\kappa_j\rangle$ position basis in two different ways: from finite-differences of the position expectation value and from the the momentum expectation value, via the numerical dispersion relation $E(p)$. That there is a good match shows that the kink-quasiparticle ansatz accurately captures the confined quasiparticles present in the $\mathbb{Z}_2$-broken Ising model.}
    \label{fig:1kink_vel_main}
\end{figure}

In the following we consider kinks, but the discussion applies equally to antikinks. The evolution of a kink-quasiparticle wavepacket will generically involve propagation and spreading (delocalization). We wish to construct wavepackets that are sufficiently broad so that they spread slowly, relative to propagation. Broader spatial wavepackets lead to slower spread because they have narrower momentum support; furthermore, spreading is reduced for wavepackets with higher momentum, because the relevant part of the kink-quasiparticle dispersion relation $E_\kappa(p)$ looks increasingly linear.

For $h=0$, we observe that our kink wavepackets indeed spread slowly as they propagate at their initial set momentum (see App.~\ref{app:zero_field}). In the presence of a confining force from a symmetry-breaking field $h > 0$, kinks undergo acceleration, as expected. A stationary kink is initially accelerated in the direction of the false vacuum, as the energy of the false vacuum is converted into kinetic energy of the kink, as would also be expected in a relativistic QFT, but the long term behavior is strongly influenced by the lattice. The lattice momentum $p$ is bounded $-\pi < p \le \pi$, and the momentum expectation value of the kink wavepacket precesses around the unit circle with $\dot p = \textit{constant}$. To understand how the \emph{position} of the kink evolves as this happens, we must consider the wavepacket group velocity $v(p) := \partial E_\kappa(p) / \partial p$. With an emergent relativistic QFT governing the IR physics, the dispersion relation is approximately relativistic ($E_\kappa(p) \sim \sqrt{p^2 + m_\kappa^2}$ for a kink of mass $m_\kappa$) for small $|p|$, becoming almost linear as $p$ increases. However, due to the bounded nature of $p$ on the lattice, $E_\kappa(p)$ must deviate from relativistic behavior as $|p|$ continues to increase. Indeed, assuming $E_\kappa(p)$ is smooth, including at the boundary value $E_\kappa(\pi) = E_\kappa(-\pi)$, it is \emph{also} bounded from above and below. As such, a wavepacket will typically reach a maximum group velocity for some $p(v_{\max})$, after which it will begin to \emph{slow down}. Assuming $E_\kappa(p) = E_\kappa(-p)$ it will ultimately \emph{reverse} and retrace its path back to its original position and momentum (with some wavepacket spread), performing \emph{Bloch oscillations}. These effects are demonstrated in the single-kink simulation of Figs.~\ref{fig:1kink_pos_mom_main} and~\ref{fig:1kink_vel_main}.

\subsection{Bubble dynamics}\label{sec_result_bubble}
Instead of Bloch oscillations of individual kinks, we wish to study the emergent relativistic dynamics of false-vacuum \emph{bubbles} comprised of a kink wavepacket and an antikink wavepacket. In particular, we want to simulate kink-antikink collisions at large kinetic energies (to increase particle-production amplitudes). Since the kink and antikink accelerate toward each other under the confining force, we can increase the kinetic energy at the time of collision by increasing the initial bubble size $x_R-x_L$ (and hence the amount of energy stored in the false vacuum). However, if we allow the kink and antikink to evolve for too long prior to collision, their momenta will exceed $|p(v_{\max})|$ and they will begin to undergo Bloch oscillations, deviating from their relativistic behavior. We can ensure that this does not occur by limiting the initial bubble size, with the maximum size depending on the Hamiltonian parameters $g,\lambda,h$. In general, a smaller mass gap (since $h \neq 0$, this is the meson mass $m_\mu$), measured in lattice units, increases the maximum bubble size, measured in physical units (multiples of the lattice correlation length $\xi$). Moving closer to criticality thus allows us to reach higher collision energies relative to the mass gap while keeping $|p|$ below $|p(v_{\max})|$.

We simulated bubble dynamics for the Ising model ($\lambda=0$) as well as near to the Tri-Critical Ising point of the extended model ($\lambda > 0$) for a range of parameters. We first focus on the two points marked (i) and (ii) in Fig.~\ref{fig:phasediag}.

\begin{figure}
    \includegraphics[width=\linewidth]{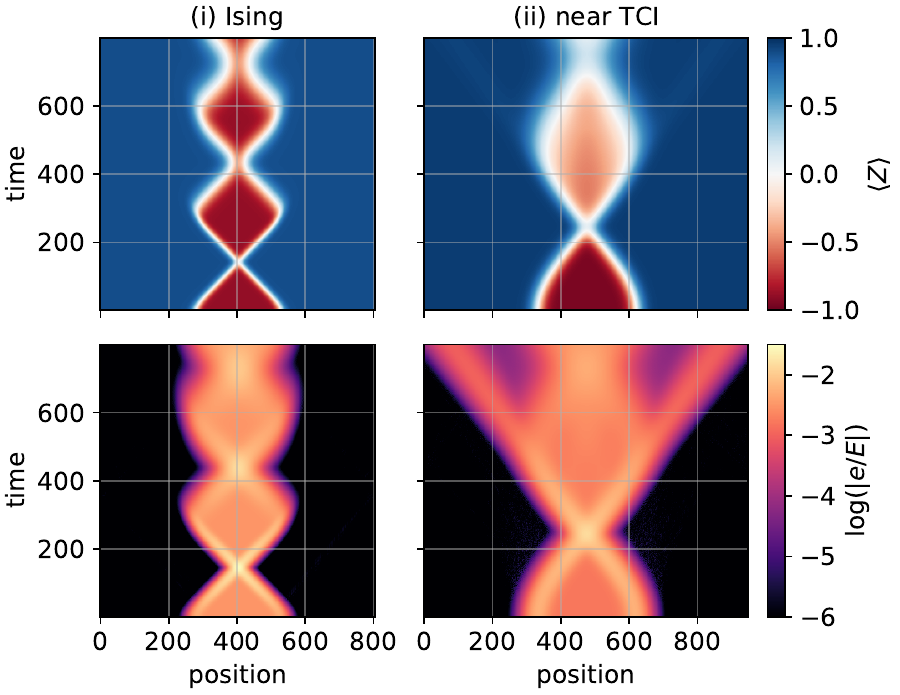}
    \caption{Spin expectation values and relative energy density $e/E$ for (i) the Ising model ($\lambda=0$, $g=0.8$, $h=0.007$, $m_\mu \approx 1.04$, $v_{\max}=1.6$, $\xi \approx 1.64$) and (ii) the generalized Ising model nearer to the Tricritical Ising CFT fixed point (TCI) ($\lambda=0.41$, $g=0.98$, $h=0.001$, $m_\mu \approx 0.43$, $v_{\max}\approx 1.58$, $\xi \approx 3.63$). For (i), the initial wavepackets have $\sigma=25$ and are $248.5$ sites apart ($E/m_\mu = 3.72$).  For (ii), $\sigma=40$ with separation $287.4$ ($E/m_\mu = 2.62$). In the plots, the position is given in lattice sites and the time units are rescaled such that the maximum kink velocity is 1 lattice site per unit time (as opposed to the unscaled $v_{\max}$ given above). The MPS bond dimensions are $D=10$ and $D=18$ for the vacua of (i) and (ii), respectively. During the simulation the dimensions are restricted to $D\le 128$ and the integration step size is $\delta t \approx 0.08$ ($0.05$ in lattice units of the unscaled Hamiltonian).
    }
    \label{fig:implots}
\end{figure}

\begin{figure}
    \includegraphics[width=0.95\linewidth]{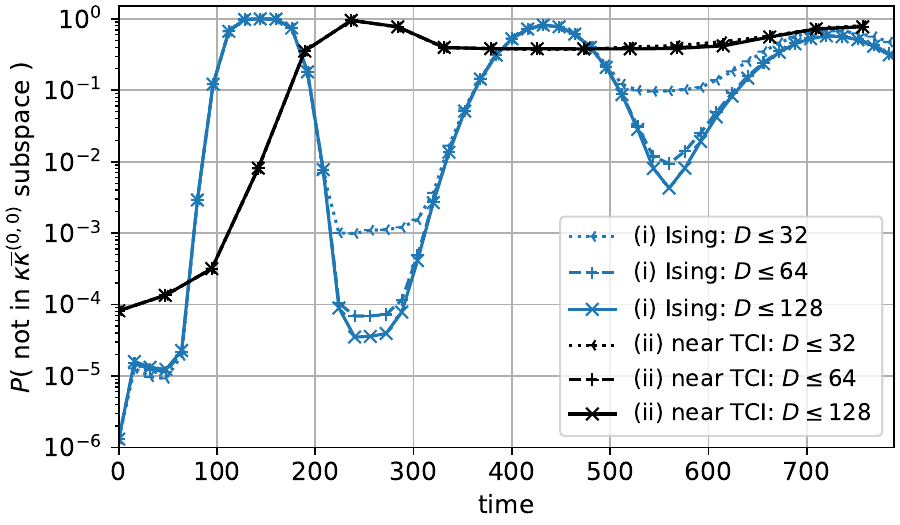}
    \caption{Portion of state (by probability) outside of the MPS kink-antikink subspace $\kappa \overline{\kappa}^{(0,0)}$ for simulations (i) and (ii) of Fig.~\ref{fig:implots}. Here we fully account for momentum dependence of the basis states $|\kappa \overline{\kappa}_{j,k}\rangle$ via a Fourier analysis and count only contributions with $k-j \ge 60$ (see App.~\ref{app:detection}). For Ising (i), the small probability after the first collision of $t \approx 140$ indicates elastic scattering of kinks, in stark contrast with the TCI case (ii), where the probability remains high after the first collision at $t\approx 250$. In (i), the growth of the post-collision probability with subsequent collisions is consistent with delocalization of the wavepackets, since contributions from kink-antikink pairs with $k-j < 60$ are not counted. Hence the larger values for (i) after the second collision at $t\approx 430$ should not be taken as evidence of inelastic scattering.}
    \label{fig:ising_error}
\end{figure}

\subsection{The Ising model}\label{sec_result_ising}
In the Ising case ($\lambda=0$) with $h=0$, known to be a theory of free kinks, our simulations reproduce the expected trivial scattering: kinks given an initial nonzero momentum collide without generating any additional entanglement. With explicit symmetry-breaking $0 < h \ll 1$ we find nontrivial scattering, as evidenced by entanglement between the post-collision kink wavepackets. However, even when the energy is significantly above the meson pair-production threshold $E > 2m_{\mu}$ there is no obvious ballistic spread to indicate production of unconfined particles. In Fig.~\ref{fig:implots}, we show results for $g=0.8$, $h=0.007$, where we observe the model to have a mass gap (meson mass) of $m_\mu \approx 1.04$ and a maximum kink velocity of $v_{\max} = 1.6$, both in unscaled units of the lattice Hamiltonian \eqref{eq:H_OF}. The vacuum correlation length (the length scale of the exponentially-decaying vacuum correlations) is $\xi \approx 1.64$ lattice sites.

To confirm that no particle production (even of confined particles) is occurring, we compute overlaps $\langle \kappa\overline{\kappa}_{jk} | \Psi(t)\rangle$. We compute the probability of being in the $\kappa\overline{\kappa}^{(0,0)}$ ``sector'' by approximating the integral
\begin{equation}
    P_{\kappa\overline{\kappa}} = \int \mathrm{d} p \; \mathrm{d} p'\; |\langle \kappa\overline{\kappa}(p,p') | \Psi(t)\rangle|^2,
\end{equation}
where 
\begin{equation} \label{eq:mom_overlap}
|\kappa\overline{\kappa}(p,p')\rangle := \sum_{k-j \ge d} e^{\mathrm{i}(pj + p'k)} |\kappa\overline{\kappa}_{jk}\rangle
\end{equation}
and the minimum separation $d$ is chosen to avoid interaction effects between the kink and antikink. See App.~\ref{app:detection} for details of the approximation. In this case $d=60$ lattice sites.

We plot the probability $1-P_{\kappa\overline{\kappa}}$ of \emph{not} being in the $\kappa\overline{\kappa}^{(0,0)}$ sector as a function of time for $g=0.8$ in Fig.~\ref{fig:ising_error}. The results are consistent with purely elastic scattering of kinks:
The probability is estimated to be around $\mathcal{O}(10^{-5})$ before the first collision, which occurs at $t\approx 140$. This is the same order of magnitude as our numerical estimate of the accuracy of the kink-antikink quasiparticle basis states, as detailed in App.~\ref{app:ex_basis}. After the first collision, $1-P_{\kappa\overline{\kappa}}$ again drops to $\mathcal{O}(10^{-5})$, at a value slightly higher than the pre-collision value. The difference closes as the maximum allowed MPS bond dimension is increased, leaving little room for any inelastic scattering process.

\emph{During} the first collision, components of the state leave the space of well-separated localized quasiparticles as the kink and antikink approach each other and begin to interact. We \emph{explicitly exclude} these interacting states from the $\kappa\overline{\kappa}$ subspace via the minimum separation $d = 60$ in \eqref{eq:mom_overlap}, hence $1-P_{\kappa\overline{\kappa}}$ increases significantly until the kink and and antikink wavefronts separate again.

After the second collision, at $t\approx 430$, $1-P_{\kappa\overline{\kappa}}$ does not drop as low as $\mathcal{O}(10^{-5})$. However, this should not be interpreted as evidence of inelastic scattering. As is apparent in both Fig.~\ref{fig:implots} and Fig.~\ref{fig:ising_error}, the kink and antikink wavepackets broaden significantly with time and successive collisions. This broadening leads to more terms in the wavefunction in which the separation of the kink and antikink remains less than $d$ even between collisions. We cannot unambiguously count these terms toward the $\kappa\overline{\kappa}$ sector due to interaction effects.

While we cannot entirely rule out inelastic scattering using our data, the very small value of $1-P_{\kappa\overline{\kappa}}$ after the first collision tells us that any inelastic process would have to be extremely unlikely to be consistent with these results. This observation is surprising given that the spin chain and its emergent field theory are not integrable, but consistent with recent observations of nonthermalizing states in the Ising model \cite{james_nonthermal_2019, wurtz_emergent_2020, lerose_quasilocalized_2020}. We further find that elastic scattering persists even if we allow the kink lattice momentum to exceed $p(v_{\max})$, as it does in simulation (i) of Figs.~\ref{fig:implots} and~\ref{fig:ising_error} (see App.~\ref{app:velocity}), so that the emergent relativistic field theory is no longer a good description of the physics. This is strong evidence that, in the Ising chain with a weak longitudinal field, bubbles are stable up to \emph{arbitrarily} high energies: When a bubble is large enough, its walls will not meet due to Bloch oscillations, so no scattering can occur while it remains localized. When bubbles are small enough for the kinks to collide, our evidence suggests they do so elastically with extremely high probability.

\begin{figure}
    \includegraphics[width=0.95\linewidth]{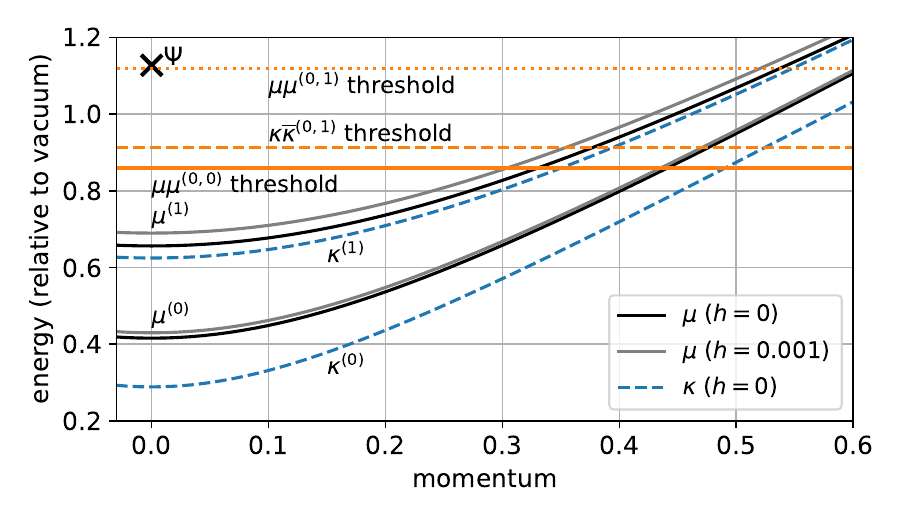}
    \caption{Dispersion relations (numerical, using MPS) of kinks~$\kappa$ and mesons~$\mu$ for $\lambda=0.41$, $g=0.98$ in lattice units
    for the Hamiltonian \eqref{eq:H_OF}. Momentum ranges from  $-\pi$ to $\pi$. For mesons, energies are shown with and without a weak longitudinal field. Individual kinks do not have a finite energy for $h>0$. Threshold energies for pair production are shown (computed assuming $h=0$ for kinks and $h=0.001$ for mesons), as is the energy (labelled $\Psi$) of the simulation shown in Fig.~\ref{fig:implots} for parameter-set~(ii). The dispersion relations reach their maximum gradient at momentum $p(v_{\max}) \approx 0.6$.}
    \label{fig:ens_tci}
\end{figure}

\begin{figure}
    \includegraphics[width=\linewidth]{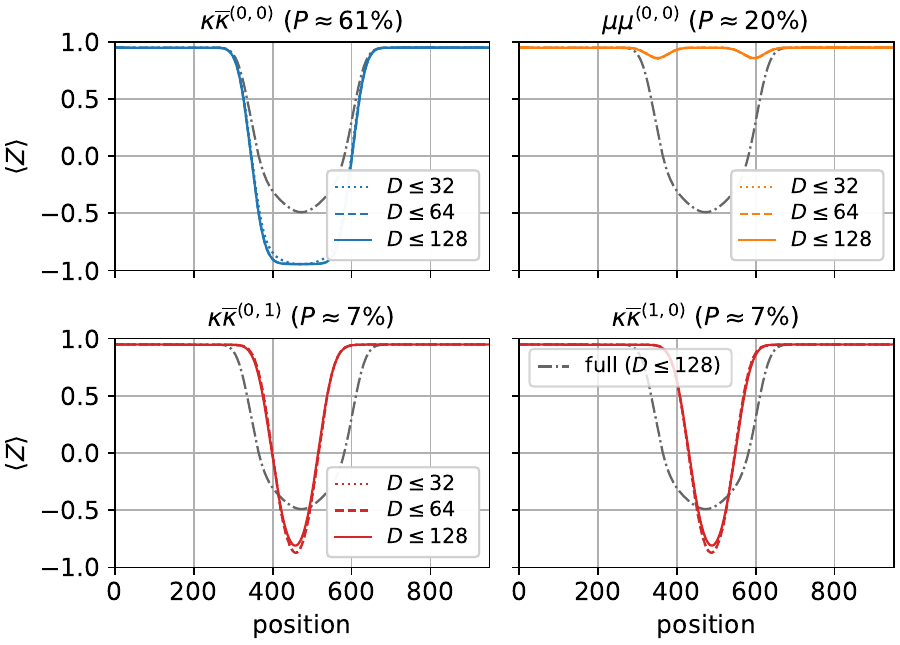}
    \caption{Spin expectation values, after projecting into selected quasiparticle subspaces and normalizing, for simulation (ii) of Fig.~\ref{fig:implots}, bond dimension $D\le 128$, after the first collision ($t\approx 426$). The amount of wavefunction captured by each (approximately orthogonal) subspace is given as a probability $P$ (see App.~\ref{app:detection}).
    Included subspaces are
    $\mu\mu^{(0,0)}$, a pair of mesons of lowest energy, and $\kappa \overline{\kappa}{}^{(a,b)}$, a bubble made of a kink of type $a$ and an antikink of type $b$ (where $0$ is the lowest-energy kink quasiparticle, and $1$ is the next highest --  see Fig.~\ref{fig:ens_tci}).}
    \label{fig:spin_comps}
\end{figure}

\subsection{Near the Tri-Critical Ising point}\label{sec_result_triising}
Going to nonzero $\lambda = 0.41$, with $g=0.98$, $h=0.001$, we find a mass gap of $m_\mu \approx 0.43$ and a maximum kink velocity $v_{\max} \approx 1.58$, both in lattice units of \eqref{eq:H_OF}. The vacuum correlation length is $\xi \approx 3.6$ lattice sites. We choose the initial bubble size so that the energy, shown in Fig.~\ref{fig:ens_tci}, is well above the pair-production threshold, but still low enough to keep the kink velocity $\ll v_{\max}$ at all times. Here we find clear evidence that unconfined particles are produced. Most apparently, Fig.~\ref{fig:implots}~(ii) clearly shows \emph{ballistic} spread of wavepackets emanating from the first collision event. To further resolve the scattering outcomes, we project onto meson-pair and kink-antikink-pair quasiparticle bases, finding four dominant ``sectors'', illustrated in Fig.~\ref{fig:spin_comps}, where we tune the quasiparticle basis MPS to match the momentum expectation value of the outgoing quasiparticle wavepackets (as estimated from the Fourier transform of the projected wavefunctions) and compute the spin expectation values of the projected wavefunction for each sector. We also compute the scattering outcome probabilities (the norms of the projected wavefunctions)\footnote{For this we use a more sophisticated Fourier analysis that more fully accounts for the momentum dependence of the quasiparticle basis MPS. see App.~\ref{app:detection}.}. We find the most likely outgoing configurations to be: a bubble made of type-$0$ kinks $\kappa\overline{\kappa}{}^{(0,0)}$ (elastic channel) with probability $P = 62\%$, then a type-$0$ meson pair $\mu\mu^{(0,0)}$ with $P = 19\%$, and finally a bubble made either of a type-$0$ kink paired with a type-$1$ antikink (higher energy) $\kappa\overline{\kappa}{}^{(0,1)}$, or a type-$1$ kink paired with a type-$0$ antikink $\kappa\overline{\kappa}{}^{(1,0)}$, each with $P \approx 7\%$ (reflection symmetry). These outcomes are all kinematically allowed, according to the energetic thresholds shown in Fig.~\ref{fig:ens_tci}.

We note that the (rounded) projection probabilities in Fig.~\ref{fig:spin_comps} only add to $95\%$. This may indicate the presence of other sectors we have not accounted for, such as a $\mu^{(0)}$ paired with a small $\kappa \overline{\kappa}^{(0,0)}$ bubble, or a $\kappa \overline{\kappa}^{(0,0)}$ bubble containing one or more quasiparticle-excitations of the false vacuum. Unfortunately, since these ``sectors'' each involve at least three quasiparticles, the corresponding position bases have many more terms ($\mathcal{O}(N_w^3)$ versus $\mathcal{O}(N_w^2)$ for pairs), making it difficult to compute these projections\footnote{Not only computationally, but also because basis orthogonality is harder to achieve.}. We emphasize, however, that this limitation only hinders this particular form of analysis of the evolved state. The dynamical simulation itself is not affected as it is not restricted to these multi-particle basis states.

It is also possible that various sources of error have affected results: (i) when excitation tensors in a 2-quasiparticle MPS are close together, so that interactions are relevant, the state may not accurately represent quasiparticles, (ii) the MPS representations of the quasiparticle position states are variational approximations subject to some error (which also affects the initial state of the simulation), and (iii) although we allow the MPS bond dimension to increase up to some maximum during simulations ($D \le 128$ in this case), errors can still accumulate if that maximum is insufficient to capture all entanglement, as well as due to errors in the numerical integration steps. We did not explicitly characterize the effects of (ii), but we have indirect evidence that they are small: see Sec.~\ref{sec_methods_error}. By varying the minimum quasiparticle separation used in the projection, as well as the maximum bond dimension of the simulation, we were able to characterize effects (i) and (iii), finding them to amount to changes in the outcome probabilities of $\ll 0.01$, except in the case of $\kappa \overline{\kappa}^{(0,1)}$ and $\kappa \overline{\kappa}^{(1,0)}$, in which one of the quasiparticles is heavier than the other, leading to a smaller separation between the kink and antikink. In this case, our analysis suggests the error here amounts to a change of around $\pm 0.01$ in the outcome probability, possibly more (see App.~\ref{app:detection}). This outcome might be better resolved at higher energies, at which the kink-antikink separations would increase.

In case of the $\mu\mu^{(0,0)}$ outcome, we cross-check the computed outcome probability by comparing it with the excess energy (relative to the vacuum) $E_{\textrm{pkts}}$ of the regions containing the ballistic wavepackets, visible in Fig.~\ref{fig:implots} (ii). If these wavepackets belong to a two-meson ``branch'' of the wavefunction, that branch (the portion of the wavefunction in the $\mu\mu^{(0,0)}$ subspace) must contribute $EP$ to the energy, where $E$ is the total energy and $P$ is the probability of the $\mu\mu^{(0,0)}$ scattering outcome. We can therefore estimate $P$ as $E_{\textrm{pkts}} / E$. This gives us a $P$ within the range $19\%$ to $20\%$ at $t\approx 757$ (after separation), depending on the precise extent of the region we sum over (e.g.\ from site 0 to site 250 for the left packet), compatible with the projected $\mu\mu^{(0,0)}$ wavefunction.

\begin{figure}
    \includegraphics[width=\linewidth]{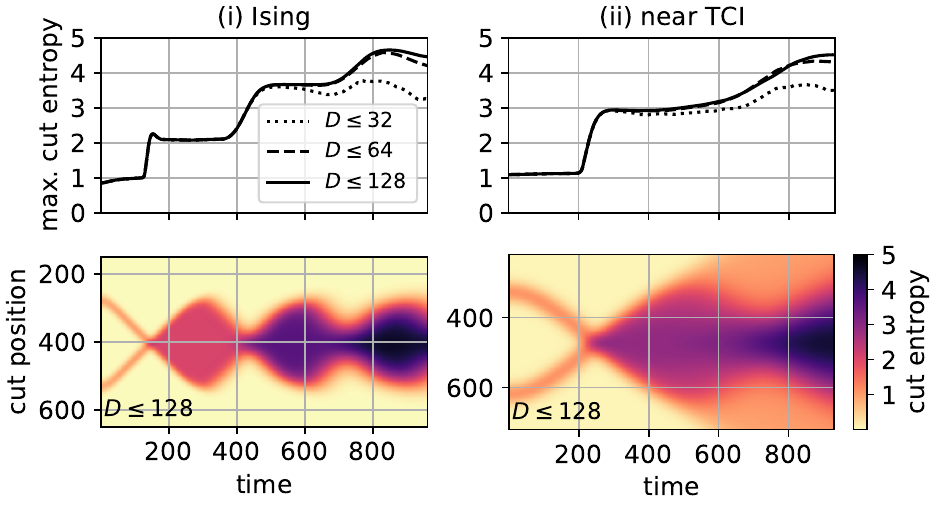}
    \caption{Entanglement entropy (base 2) for cuts (left-right bipartitions) of the spin chain as a function of time for the simulations (i) and (ii) of Fig.~\ref{fig:implots}. Convergence with the bond dimension $D$ slows as time goes on. For example, in (i) the max.\ cut entropy at $D\leq 128$ is very likely not converged after $t\approx 700$.}
    \label{fig:entropy_time}
\end{figure}

\begin{figure}
    \includegraphics[width=0.95\linewidth]{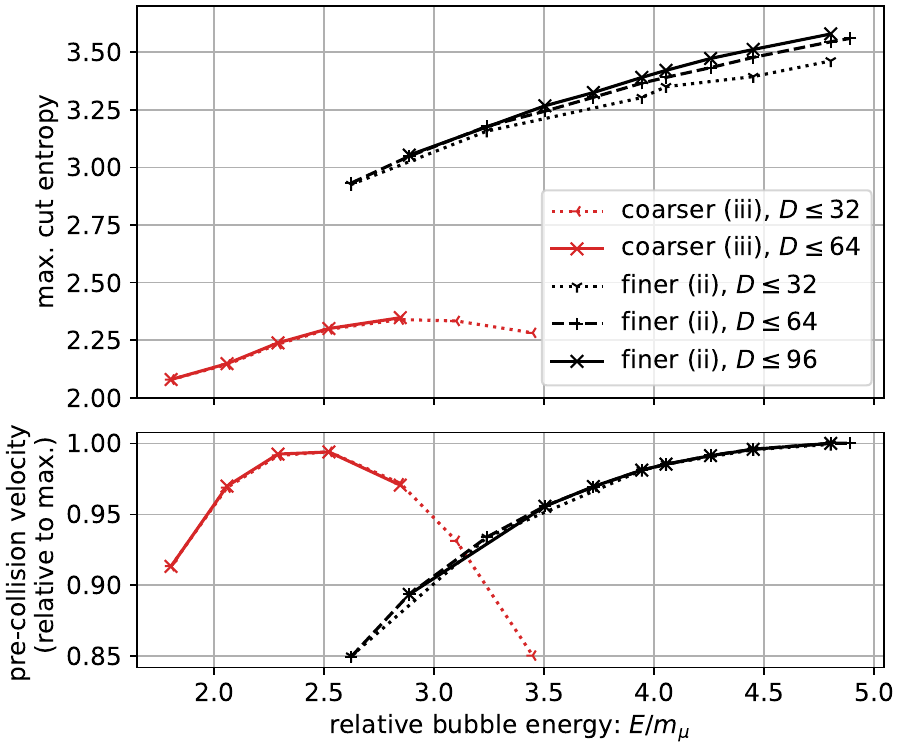}
    \caption{Peak maximum cut entropy during the first collision as a function of energy (ii) close to ($\lambda=0.41$ $g=0.98$, $h=0.001$) and (iii) further from the TCI point ($\lambda=0.3$, $g=0.9$, $h=0.0069$, $m_{\mu} \approx 0.97$, $v_{\max} \approx 1.43$), with the energy controlled by the initial kink-antikink separation. The vacuum correlation length $\xi$ is 3.6 sites for (ii) and 1.8 sites for (iii), indicating that (ii) is closer to criticality. The initial wavepacket width is $\sigma=40$ for (ii) and $\sigma=19$ for (iii). The kink velocity at the start of the first entropy jump (see Fig.~\ref{fig:entropy_time}), normalized so that the maximum is 1, is also shown. Decreasing velocity with energy indicates the onset of Bloch oscillations. Simulation (ii) of Fig.~\ref{fig:implots} corresponds to the leftmost point of the black curves.}
    \label{fig:entropy_energy}
\end{figure}

\subsection{Entropy and computational cost}\label{sec_result_cost}
As evidenced by Fig.~\ref{fig:ising_error}, the bond dimension of the MPS representing the evolving state must continue to grow as time goes on, in order to maintain accuracy. The cut entropy at location $j$ is a proxy for the required bond dimension $D_j$. Fig.~\ref{fig:entropy_time} shows the evolution of the maximum cut entropy for the simulations of Fig.~\ref{fig:implots}. At early times, we observe that the maximum cut entropy jumps dramatically during scattering events, whether elastic or inelastic, remaining almost constant in between. This is consistent with a model of interacting quasiparticle wavepackets: Separated wavepackets undergo stable propagation until they collide, at which point interactions generate entanglement corresponding to the different possible scattering outcomes. At late times, we observe a temporal broadening of the jumps, consistent with spatial broadening of the wavepackets involved. Fig.~\ref{fig:entropy_time} also shows that, although there is cut entropy associated with the wavepackets themselves, this is quickly surpassed by the cut entropy in the center of the chain, associated with entanglement between the left and right outgoing packets. It is this entanglement between outgoing quasiparticles that is responsible for the post-collision plateaus visible in the maximum cut entropy.

The entropy jumps clearly make MPS simulations of long-time dynamics demanding. However, for the purposes of studying the quasiparticle content of scattering outcomes, with the incoming quasiparticles chosen via the initial state, it is enough to accurately simulate a single collision and then wait until the outgoing wavepackets have separated sufficiently so that interactions between outgoing quasiparticles may be neglected (we assume the simulation parameters are chosen such that wavepackets remain localized for sufficiently long times)\footnote{Simulating for only relatively short times also increases the likelihood that these tasks can be carried out on NISQ-era quantum devices.}.

We expect the entropy generated in a collision of localized quasiparticle wavepackets to depend on the collision energy relative to the masses of quasiparticles: the more scattering outcomes there are, and the greater the probability of those outcomes, the larger the post-collision entanglement entropy can be. In Fig.~\ref{fig:entropy_energy}, we explore the maximum cut entropy occurring during the first collision as a function of energy, controlled via the initial kink separation, for two sets of Hamiltonian parameters, one (ii) closer and one (iii) further from criticality. As each datapoint on this Figure requires a full dynamical simulation, we limited the bond dimension to $D \le 96$ (versus $D\le 128$ of Fig.~\ref{fig:implots}) to save some computational time. We find that the trend with increasing bond dimension is nevertheless clearly visible. We find that the entropy indeed grows with energy, smoothly increasing even as thresholds are crossed, e.g.\ the $2m_{\mu^{(1)}}$, $3m_{\mu^{(0)}}$ and $4m_{\mu^{(0)}}$ thresholds in case of (ii) (see also Fig.~\ref{fig:ens_tci}).

The entropy continues to increase at least until the energy is sufficient for the kinks to approach the maximum possible kink velocity prior to collision, at which point we expect deviations from the emergent relativistic dynamics to become apparent as Bloch oscillations emerge. In the case of (iii), Fig.~\ref{fig:entropy_energy} shows that the post-collision entropy eventually decreases as lattice effects kick in, coincident with deceleration of the kinks prior to collision. Note that we are able to reach much higher relative energies with parameters (ii) before encountering obvious lattice effects. This illustrates the general principle that more of the emergent relativistic QFT is revealed as one approaches criticality: the relative energies accessible by quasiparticles, while avoiding Bloch oscillations (momenta $|p| < p(v_{\max})$), grows as the lattice meson mass drops.

We also observe that much more entropy is generated in the first collision for parameters (ii) than for parameters (iii), even when the relative energy is similar\footnote{The difference in post-collision entropy is not attributable to different vacuum entropies, since these are $< 0.1$ in both cases.}. A significant part of this difference likely comes from a much higher probability of meson pair production, as well as the availability of the $\kappa\overline{\kappa}^{(0,1)}$ outcomes, in case (ii): the probability of particle production is $< 10\%$ in case (iii) at energy $E / m_\mu \approx 2.52$, in contrast with $\sim 38 \%$ at energy $E / m_\mu \approx 2.62$ in case (ii), according to $\kappa\overline{\kappa}^{(0,0)}$ basis overlaps. This is possible, since these two parameter sets were not chosen to be part of an RG trajectory, so that their emergent QFTs need not be the same.

\section{Discussion}\label{sec_discussion}
Building on recent innovations in the classical simulation of quasiparticle dynamics using Matrix Product States \cite{vandamme_realtime_2019a}, we proposed a framework for simulating and characterizing the full (nonperturbative) quantum dynamics of false-vacuum bubbles in relativistic QFTs that govern the IR physics of one-dimensional lattice systems. While we chose to simulate a quantum spin chain, the methods we use are general and could also be applied directly to, for instance, a spatially-discretized QFT such as the Schwinger model or $\lambda\phi^4$ theory \cite{sugihara_2004, milsted_2013_phi4, jordan_quantum_2014, burak_upcoming, kuhn_nonabelian_2015, pichler_realtime_2016, buyens_realtime_2017, chanda_confinement_2020, magnifico_real_2020}.
We also demonstrated that the MPS quasiparticle ansatz, with which we initialized our simulations, can be used to \emph{detect} quasiparticles that are produced as time evolves. This allowed us to verify quasiparticle pair-production in the modified Ising model we studied, including production of different species of confined kink that were not obvious from examining energy density and spin expectation values alone. We used the same kind of analysis to confirm a \emph{lack of} particle-production in the unmodified Ising model (with transverse field and small longitudinal field), supporting other recent studies that suggest particle production is very strongly suppressed \cite{james_nonthermal_2019, wurtz_emergent_2020, lerose_quasilocalized_2020}.

We were able to significantly improve the efficiency and interpretability of our simulations by carefully choosing our initial states in two different ways: Firstly, constructing spatially broad wavepackets allowed us to access the dynamics of the emergent IR QFT without the spoiling effects of UV, high-momentum components that are strongly influenced by the lattice. Broad wavepackets also lead to localization of quasiparticles over long times, making it easier to characterize scattering outcomes, and improve the numerical conditioning of the dynamical simulation (see App.~\ref{app:quenches}). Secondly, by precisely tuning the quasiparticle content of the initial wavepackets \cite{vandamme_realtime_2019a}, we were able to study individual scattering events in isolation, while further reducing the computational demands of the simulation by lowering entanglement. 

Entanglement growth is the most significant barrier to dynamical simulations with MPS, as the computational cost of each time step scales exponentially with the cut entropy. By choosing broad quasiparticle wavepackets, we reduce entanglement growth at the expense of growing the number of lattice sites involved in the simulation. This is a good tradeoff for MPS simulations, as the computational cost scales only linearly in the number of lattice sites in our simulation window. Even with this tradeoff, we found that the large jumps in cut entanglement with each collision (of confined quasiparticles in the system) preclude simulating more than a handful of successive collisions. Furthermore, we found clear evidence of entanglement growth with the collision energy, although the onset of Bloch oscillations prevented us from drawing strong conclusions about how this growth continues in the emergent IR QFT. Nevertheless, in the absence of lattice effects that obscure the IR QFT, it seems reasonable to expect the entropy to continue to grow with energy, which would eventually preclude accurate simulation using MPS.

We note that our current simulations are already well beyond the reach of present digital NISQ devices, as they would require at least $\sim 1000$ qubits and circuit depths that are larger still.

A natural next step would be to perform simulations along RG trajectories in the Hamiltonian parameter space, so that results can be extrapolated to the continuum. This is equivalent to finding paths toward criticality of the lattice model, along which the low-energy spectrum remains consistent with a particular emergent (IR) QFT. Moving closer to the continuum would also allow us to reach higher (relative) collision energies while avoiding lattice effects, such as Bloch oscillations. In turn, this would permit a more thorough exploration of the energy-dependence of the entanglement generated in collisions.

As we approach criticality, the bond dimension of the MPS vacua must grow to maintain accuracy, as must the size of the simulation window, since the wavepacket width in lattice units would have to increase with the lattice correlation length in order to maintain localization. Getting closer to criticality seems feasible: The simulations featured in the main text, with maximum MPS bond dimension $128$, took between one and two weeks to complete on 8 cores each and this time could likely be reduced significantly with further work to optimize the code\footnote{GPU acceleration may also be useful \cite{li_tddmrg_2020}.} and the use of a numerical integrator with an adaptive time-step size.

Increasing the number of spatial dimensions presents a significantly greater challenge for classical algorithms: while the computational cost of MPS simulations scales with the bond dimension $D$ as $\mathcal{O}(D^3)$, the scaling for tensor-networks capable of handling large (2+1)-dimensional systems, such as PEPS \cite{verstraete_peps_2006}, is much worse (albeit still polynomial)\footnote{Some methodological changes would also be needed: false-vacuum bubbles in $>1$ spatial dimensions, having extensive boundaries, no longer look like pairs of confined particles. As such, localized quasiparticle techniques (such as those developed for PEPS \cite{vanderstraeten_2019}) are not directly applicable. Instead, suitable initial states could likely be prepared via energy minimization in the presence of a nonuniform symmetry-breaking field.}. As an intermediate step, one could consider systems with a small, compactified second dimension of space, which are often within reach of MPS methods. By performing a Fourier transform of the Hamiltonian in the compactified direction only \cite{motruk_2016}, one could study scattering of quasiparticles that are spatially localized in one direction, while being momentum eigenstates of the other. Compared to the purely (1+1)-dimensional case, the additional ``Kaluza-Klein'' excitations associated with the Fourier modes of the compactified dimension would already open up a much greater range of scattering outcomes.

Compared to the simulations we performed, increasing the variety of scattering outcomes, whether by raising the relative energy in a given model, choosing a lattice model with a richer set of low-energy excitations (e.g.\ near a phase transition described by a CFT with larger central charge), or adding spatial dimensions, seems necessary in order to find problems that exhaust tensor-network methods due to the additional entanglement generated. Such problems appear more amenable to simulation on quantum hardware, which is not \emph{a priori} limited in the amount of entanglement it can deal with. However, raising the energy may be problematic for near-term quantum devices, which are limited both in their size and coherence times, since avoiding lattice effects (such as Bloch oscillations) at higher relative energies requires moving closer to criticality while increasing system sizes and evolution times. Instead, increasing the richness of low-energy excitations by changing the model or adding (compactified) dimensions, while avoiding coming too close to criticality, may be a more promising route toward quantum advantage using near-term devices.

\textit{Acknowledgments}:
We thank Alex Buser, Marcela Carena, Cliff Cheung, Natalie Klco, Ying-Ying Li, Spiros Michalakis, Nicola Pancotti, Burak \c{S}ahino\u{g}lu, Martin Savage and Eugene Tang for useful discussions and comments. 
This material is based upon work supported in part by the U.S. Department of Energy Office of Science, Office of Advanced Scientific Computing Research, (DE-NA0003525, DE-SC0020290), and Office of High Energy Physics (DE-ACO2-07CH11359, DE-SC0018407).
AM, JL and JP also acknowledge funding provided by the Institute for Quantum Information and Matter, an NSF Physics Frontiers Center (NSF Grant PHY-1733907), the Simons Foundation It from Qubit Collaboration, and the Air Force Office of Scientific Research (FA9550-19-1-0360). GV is a CIFAR fellow in the Quantum Information Science Program. 
Sandbox is a team within the Alphabet family of companies, which includes Google, Verily, Waymo, X, and others.
Research at Perimeter Institute is supported in part by the Government of Canada through the Department of Innovation, Science and Economic Development Canada and by the Province of Ontario through the Ministry of Colleges and Universities. Some of the computations presented here were conducted on the Caltech High Performance Cluster, partially supported by a grant from the Gordon and Betty Moore Foundation.

\appendix

\section{Infinite MPS}
\label{app:imps}

In the main text, we define a Matrix Product State on a finite number of sites with open boundary conditions. To represent the true and false vacua of a spin chain, we use infinite MPS (iMPS), in which the number of sites~$\rightarrow \infty$
\begin{equation} \label{eq:iMPS}
    |\psi\rangle = \sum_{\{s\}} v_L^\dagger \dots A_{-1}^{(s_{-1})} A_0^{(s_0)} A_1^{(s_1)} \dots v_R \; | \dots s_{-1} s_0 s_1 \dots \rangle, 
\end{equation}
where $A_j^{(s)}$ is a $D_{j-1} \times D_j$ matrix assigned to site $j$ in basis-state $s$ and $v_L$ and $v_R$ are appropriately-sized boundary vectors. In a \emph{uniform} (translation invariant) iMPS, we use the same tensor $A$ everywhere: $A_j^{(s)} = A^{(s)}$ $\forall j \in \mathbb{Z}$. Such a state has a well defined norm for generic choices of $v_L$ and $v_R$ if the $D^2 \times D^2$ ``transfer matrix''
\begin{equation} \label{eq:mps_tm}
    \vcenter{\hbox{\includegraphics[height=3.5em]{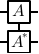}}} = 
    \sum_s A^{(s)} \otimes A^{(s)}{}^*,
\end{equation}
where $^*$ indicates the complex conjugate, has a nondegenerate eigenvalue of largest magnitude, with $A$ normalized so that this eigenvalue is equal to 1 \cite{fannes_1992}. This condition implies exponential decay of correlations with distance. By additionally normalizing $v_L$ and $v_R$ appropriately, we can achieve $\langle \psi | \psi\rangle = 1$. The precise choice of boundary vectors does not affect bulk expectation values due to the aforementioned exponential decay of correlators.

\subsection{Nonuniform windows}

To build the bubble states of the main text, and to simulate their evolution in time, we allow the tensors of an otherwise uniform iMPS to vary within a ``window'', consisting of $N_w$ contiguous sites. These states have the form
\begin{multline}
    \label{eq:iMPS_window}
    |\psi\rangle = \sum_{\{s\}} | \dots s_1 \dots s_N \dots \rangle \times \\
    v_L^\dagger \left( \prod_{i=-\infty}^0 A_L^{(s_i)} \right) A_1^{(s_1)} \dots A_{N_w}^{(s_{N_w})}
    \left( \prod_{j=N_w+1}^\infty A_R^{(s_j)} \right) v_R,
\end{multline}
where $A_L$ and $A_R$ parameterize the semi-infinite left and right bulk parts of the chain and $A_1 \dots A_{N_w}$ parameterize the nonuniform window. The above transfer-matrix conditions for a well-defined uniform iMPS must be satisfied for both the $A_L$ and $A_R$ tensors. The norm of the state is then determined by the content of the window tensors $A_1 \dots A_{N_w}$. For the bubble states, we let $A_L = A_R = \Tvac$, where $\Tvac$ is the tensor optimized for the uniform ground state (true vacuum) of the spin chain. We then choose $A_1 \dots A_{N_w}$ to represent a false-vacuum bubble, as described below in App.~\ref{app:ex_basis}. For example,  a fully localized bubble state (the kink-antikink state of Fig.~\ref{fig:states}) has $A_1 = \Tkink$ (representing a kink), $A_2 \dots A_{N_w-1} = \Tfvac$ (representing the false vacuum), and $A_{N_w} = \Takink$ (representing the antikink).

\section{Finding the true and false vacua}\label{app:vacua}

\subsection{Finding the true vacuum}

The tensor $A$ defining a uniform iMPS \eqref{eq:iMPS} can be optimized to represent low-energy, translation-invariant states of gapped quantum spin chains using various algorithms. We use the nonlinear conjugate-gradient method described in \cite{milsted_2013_phi4} and implemented in the \emph{evoMPS} package \cite{milsted_evomps} to find a uniform iMPS that approximately describe the ground states of gapped quantum spin chains. We denote the optimized iMPS tensor $\Tvac$.

\subsection{Finding the false vacuum}
\label{app:falsevac}

We explain how we obtain an iMPS representation of the false vacuum in practice in App.~\ref{app:falsevac_grad} below. In this section, we consider the nature of the false vacuum more generally.

For the broken $\mathbb{Z}_2$ symmetry of the Ising-like chain in the main text, the false vacuum $|\overline{\Omega}\rangle$ is a state that has opposite spin orientation to the true vacuum $|\Omega\rangle$. It should also be \emph{like} a vacuum, in that it should be a spatially uniform, approximately static state near a local energetic minimum with respect to some constraint, such as locality.

A candidate state is the ``flipped'' vacuum
\begin{equation} \label{eq:flipvac}
  \left(\prod_j X_j \right)|\Omega\rangle.    
\end{equation}
It is spatially uniform and typically close to an energetic minimum in the following sense: If we apply a finite string $\prod_{j=1}^L X_j$ of length $L$, the change in energy will be \emph{positive} for small values of $L$, becoming negative only after the $\mathcal{O}(2hL)$ energy lost by replacing false vacuum with true vacuum on $L$ sites is larger than the $\mathcal{O}(1)$ energy penalty of spin anti-alignment at the boundaries. However, the flipped vacuum is generally \emph{not} close to being an eigenstate in case of a nonzero symmetry-breaking field parameter $h$ and therefore is not suitably static. Nevertheless, one might begin with the flipped vacuum and attempt to bring it closer to a false-vacuum eigenstate by lowering the energy, for example via imaginary time evolution:
\begin{equation}
    |\overline{\Omega}\rangle = e^{-\tau H} \left(\prod_j X_j \right) |\Omega\rangle.
\end{equation}

A problem with this approach is that, since we are not in a true energetic minimum, imaginary time evolution will ultimately take us back to the \emph{true} vacuum $|\Omega\rangle$.
At finite system sizes, this corresponds to a nonzero inner product between the flipped vacuum and the vacuum. Let us consider the $\mathbb{Z}_2$-broken Ising Hamiltonian ($\lambda=0$, $h > 0$) at finite system size $N$. Although in our simulations we work directly in the thermodynamic limit $N\rightarrow \infty$ using iMPS, finite $N$ is more convenient for the following calculation. We will see that the key result is independent of $N$. If we perturb around the bare theory of $g = 0$, we find
\begin{equation}
    \langle \Omega | \left(\prod_j X_j \right) | \Omega \rangle = 0 + \mathcal{O}(g^N),
\end{equation}
where $g \ll 1$. Overlaps $\langle E_i | \left(\prod_j X_j \right) | \Omega \rangle$ with energy eigenstates $|E_i\rangle$ that are close to the true vacuum (e.g.\ low-energy excitations) are also exponentially suppressed.

A simplified model allows us to estimate the timescale for ``decay'' to the true vacuum under imaginary time evolution. Take $|\psi\rangle := \psi_\Omega |\Omega\rangle + \sum_i \psi_i |\overline{E}_i\rangle$, where $\langle\psi|\psi\rangle = 1$ and $|\overline{E}_i\rangle$ represents an eigenstate in the false-vacuum ``sector'', i.e.\ with a flipped spin orientation versus $|\Omega\rangle$. This will be our model for the flipped vacuum $\left(\prod_j X_j \right) |\Omega\rangle$. From our perturbative calculation, we take $|\psi_\Omega| \approx g^N$, so that $\sum_i |\psi_i|^2 \approx 1 - g^{2N}$. Imaginary-time evolution give us
\begin{equation}
    e^{-\tau H} |\psi\rangle = e^{-\tau E_\Omega} \psi_\Omega |\Omega\rangle + \sum_i e^{-\tau \overline{E}_i} \psi_i |\overline{E}_i\rangle.
\end{equation}
Now we take $\overline{E}_i - E_\Omega \sim 2Nh$, since $|\overline{E}_i\rangle$ are flipped states which suffer an extensive energy penalty compared to $|\Omega\rangle$. The relative contribution of the vacuum after a time $\tau$ is then
\begin{equation}
    g^N e^{\tau 2Nh},
\end{equation}
which goes to 1 at $\tau_\Omega$, independently of $N$:
\begin{equation}
    \tau_\Omega = -\frac{1}{2h} \log g.
\end{equation}
Hence, for small $h$, one must evolve for a ``long'' time to see a significant vacuum contribution. For sufficiently large $\tau$ still satisfying $\tau \ll \tau_\Omega$, assuming initial occupancy and energetic separation of the $|\overline{E}_0\rangle$ state, we end up with:
\begin{equation}
e^{-\tau H} |\psi\rangle \approx e^{-\tau E_\Omega} \psi_\Omega |\Omega\rangle + e^{-\tau \overline{E}_0} \psi_0 |\overline{E}_0\rangle,
\end{equation}
where $|\overline{E}_0\rangle$ is a hypothetical lowest-energy contribution from the false-vacuum ``sector''. This picture is supported by numerical observations in which performing some imaginary-time evolution on $\left(\prod_j X_j \right) |\Omega\rangle$ quickly results in something that is (numerically) approximately an eigenstate.

\subsection{Finding an iMPS for the false vacuum}
\label{app:falsevac_grad}

To find an iMPS for the false vacuum, we begin with an iMPS approximation of the \emph{flipped vacuum} \eqref{eq:flipvac}, obtained from the iMPS approximation of the true vacuum.
Instead of using imaginary time evolution to reduce the energy of this state, as considered in the previous section, we use the same conjugate-gradient optimization method used to find the true vacuum \cite{milsted_evomps}. Like imaginary time evolution, such variational methods should eventually take the flipped state to the true vacuum state. In practice, however, we observe that for small symmetry-breaking fields $|h| \ll 1$ this does not happen. Instead, the state converges to a false-vacuum iMPS (parameterized by a tensor we denote $\Tfvac$) that is numerically indistinguishable from an energy eigenstate. This may be because of the limited available numerical precision\footnote{We use a double-precision floating-point representation, although the effective precision may be lower due to inversion of small Schmidt coefficients \cite{haegeman_unifying_2016}.}, which could preclude accurate representation of the gradient components that would lead to the true vacuum.

\section{MPS quasiparticle states}
\label{app:ex_basis}

We use a Bloch-state approach to represent low-energy excitations \cite{rommer_1997, haegeman_2012_excite, haegeman_2013_el_ex}. A localized quasiparticle state is constructed from vacuum tensors $A_L$ and $A_R$, which remain constant, together with an ``excitation tensor'' $B$ that can be chosen to represent different excitations:
\begin{multline} \label{eq:MPS_local_excite}
    |\phi_j(A_L, B, A_R)\rangle := \\
    \sum_{\vec{s}} v_L^\dagger \left( \prod_{i=-\infty}^{j-1} A^{s_i}_L \right)
    B^{s_j} \left( \prod_{k=j+1}^{\infty} A^{s_k}_R \right) v_R |\vec{s}\rangle.
\end{multline}
This ansatz can represent topological excitations, in case $A_L$ and $A_R$ refer to different vacua, as well as nontopological excitations, in case $A_L$ and $A_R$ represent the same vacuum state. We use the symbol $\phi$ to denote a generic excitation, and $\kappa$, $\overline{\kappa}$, or $\mu$ to refer to kinks, antikinks, or mesons specifically. For example, the kink states $|\kappa_j\rangle$ of the main text have $A_L = \Tvac$, $B = \Tkink$ and $A_R = \Tfvac$, while the meson states $|\mu_j\rangle$ have  $A_L = \Tvac$, $B = \Tmes$ and $A_R = \Tvac$. For tensor-network diagrams showing the parts of the tensor networks surrounding $B$ (for $|\kappa_j\rangle$ and $|\mu_j\rangle$) see Fig.~\ref{fig:states}. Due to exponential decay of correlations in the vacua represented by $A_L$ and $A_R$, the tensor $B$ represents a \emph{quasilocal} excitation and may affect expectation values across many lattice sites. In the following, we assume for simplicity that the MPS $|\phi_j\rangle$ have uniform bond dimension $D$.

Momentum eigenstates can be constructed as Fourier modes of the spatially localized excitations:
\begin{multline} \label{eq:MPS_excite}
    |\phi(A_L, B, A_R, p)\rangle := 
    \sum_j e^{\mathrm{i}pj} |\phi_j(A_L, B, A_R)\rangle.
\end{multline}
These momentum eigenstates enjoy a ``gauge'' freedom (parameter redundancy): the $B$ tensor may be transformed as
\begin{equation} \label{eq:Fmode_freedom}
    B^s \rightarrow B^s + A^s_L x - e^{-\mathrm{i}p} x A^s_R,
\end{equation}
where $x$ is a $D \times D$ matrix, without affecting the Fourier mode $|\phi(A_L, B, A_R, p)\rangle$. This freedom can be fixed in many ways. For example, the ``left orthogonality''\footnote{Called ``left gauge-fixing'' in \cite{haegeman_2012_excite}.} conditions \cite{haegeman_2012_excite, vanderstraeten_tangent_2019} are
\begin{equation} \label{eq:LGF}
    \vcenter{\hbox{\includegraphics[height=3.5em]{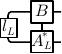}}} = 
    \langle l_L | \left( \sum_s B^{(s)} \otimes A_L^{(s)}{}^* \right) = 0,
\end{equation}
where $\langle l_L|$ is the dominant left eigenvector of the MPS transfer matrix of the left uniform bulk:
\begin{equation}
    \langle l_L | \left( \sum_s A_L^{(s)} \otimes A_L^{(s)}{}^* \right) = \langle l_L|.
\end{equation}
Similarly, the ``right orthogonality'' conditions are
\begin{equation} \label{eq:RGF}
    \vcenter{\hbox{\includegraphics[height=3.5em]{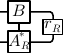}}} = 
    \left( \sum_s B^{(s)} \otimes A_R^{(s)}{}^* \right) |r_R\rangle = 0,
\end{equation}
where $|r_R\rangle$ is the dominant right eigenvector of the MPS transfer matrix of the right uniform bulk:
\begin{equation}
    \left( \sum_s A_R^{(s)} \otimes A_R^{(s)}{}^* \right) |r_R\rangle = |r_R\rangle.
\end{equation}
These conditions can be achieved for any initial tensor $B$ by transforming it with an appropriate choice of $x$ in \eqref{eq:Fmode_freedom}. Imposing either the left or right conditions, \eqref{eq:LGF} or \eqref{eq:RGF}, implies orthogonality of the position states: $\langle \phi_j(A_L, B, A_R)|\phi_k(A_L, B, A_R)\rangle = \delta_{jk}$. This is particularly convenient for working with the momentum eigenstates, as it greatly simplifies the computation of their inner products and expectation values \cite{haegeman_2012_excite}.

\subsection{Optimizing the excitation tensor $B$}
\label{app:ex_opt}

To find a tensor $B$ that accurately represents a \emph{particular} quasiparticle excitation, we use the methods of \cite{haegeman_2012_excite} with some modifications for dealing with the case in which one of $A_L$ and $A_R$ represents a \emph{false} vacuum (with a different energy density compared to the true vacuum). The basic idea is to project the Hamiltonian onto the ansatz space of momentum eigenstates $|\phi(A_L, B, A_R, p)\rangle$, resulting in an effective Hamiltonian for the tensor $B$ that can be solved using a standard sparse eigenvalue solver.

Note that, since $\langle\phi(A_L, B, A_R, p)|\phi(A_L, B, A_R, p')\rangle = \delta(p-p')$, it is natural to do this for a particular, chosen value of $p$. By solving for multiple eigenvalue-eigenvector pairs, a set of orthogonal tensors $B^{(a)}$ (the eigenvectors) can be found that accurately approximate several different low-energy excitations (labeled by the index $a$), as long as they are all below the two-particle threshold \cite{haegeman_2013_el_ex}. The eigenvalues are the energies of these excitations. By computing them for a range of $p$, one can obtain an approximate dispersion relation $E(p)$ for the quasiparticles in the system.

\begin{figure}
    (i) Ising: $g=0.8$, $h=0.007$
    \includegraphics[width=\linewidth]{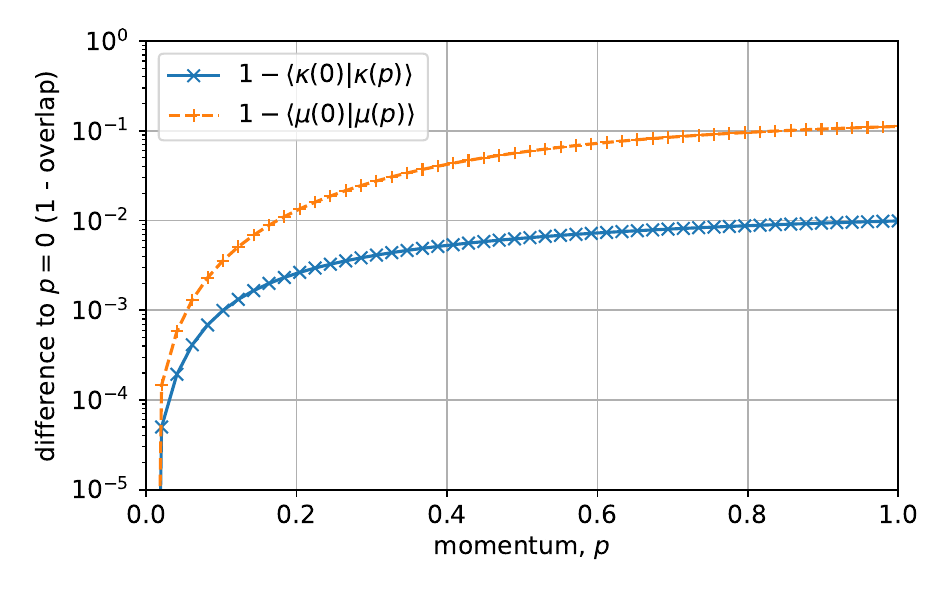}\\
    (ii) near TCI: $\lambda=0.41$, $g=0.98$, $h=0.001$\\
    \includegraphics[width=\linewidth]{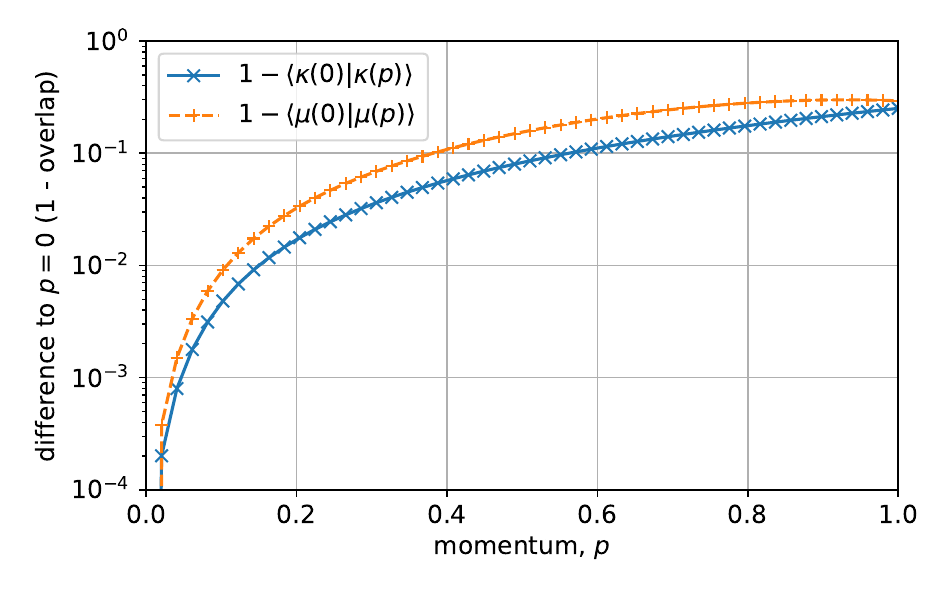}
    \caption{Error made ($1-\langle \phi_j(B) | \phi_j(B(p)) \rangle$) in ignoring the momentum dependence of the tensor $B$ used to construct MPS quasiparticle states, for both kink and meson excitations, for the Hamiltonian parameters used in simulations (i) and (ii) of the main text. The momentum-eigenstate freedom on $B(p)$ is fixed so that $\langle \phi_j(B) | \phi_k(B') \rangle = \delta_{jk}$.}
    \label{fig:momdep}
\end{figure}

Importantly, not only the energies, but also the optimized tensors $B^{(a)}(p)$, and hence the position states $|\phi_j(B)\rangle$, generally depend nontrivially on the value of $p$. This is illustrated in Fig.~\ref{fig:momdep}, which shows the error made in using a tensor $B^{(0)}(p=0)$ optimized for $p=0$, to represent the lowest-lying excitations at other momenta.

\subsubsection{Broken symmetry and kinks}
\label{app:broken_optimization}

In the presence of explicit symmetry breaking ($h \neq 0$), \emph{topological excitations} such as kinks and antikinks involve the false vacuum. The energy of a localized kink (or antikink) depends on its position, since different positions lead to different extensive contributions from the false vacuum\footnote{In an infinite system, as considered here, all kinks have \emph{infinite} energy with respect to the vacuum. However, the energy difference between two kinks is finite when their positions differ by a finite amount.}. As such, there are no energy-momentum eigenstates (of the Hamiltonian and momentum operators) corresponding to these excitations and we cannot find them by solving the effective Hamiltonian for the $B$ tensors considered above. Nevertheless, we expect there to be excitations that behave as quasiparticles subject to a confining force (which makes them accelerate). If one could somehow cancel the confining force, as if by accelerating at the same rate as the quasiparticle, the latter would appear to propagate freely.

With this picture in mind, we define a modified energy function
\begin{equation}
    \tilde E = \sum_{jk} e^{\mathrm{i}p(k-j)} \langle \phi_j | (H - \Delta E_j \; \mathds{1}) | \phi_{k}\rangle,
\end{equation}
where $\Delta E_j := \sum_{-\infty}^{j} e_L + \sum_{j+1}^{\infty} e_R$ and $e_L$, $e_R$ are the energy densities of the vacua parameterized by $A_L$ and $A_R$, respectively. Here, it is assumed that the momentum-eigenstate gauge-freedom on $B$, \eqref{eq:Fmode_freedom}, has been fixed so that $\langle \phi_j | \phi_k \rangle = \delta_{jk}$. With orthogonality of the position states, the identity term simply shifts the energy of the excitation in a position-dependent way, cancelling the position-dependent contribution due to the differing bulk energy densities. One can also write down a modified Hamiltonian
\begin{equation} \label{eq:tildeH}
    \tilde H = H - \sum_{j} P_j \Delta E_j,
\end{equation}
where $P_j$ is a projector onto the space of states spanned by $|\phi_j(B)\rangle$ for all $B$ satisfying the chosen orthogonality conditions: for such $B$, we thus have $P_j|\phi_k(B)\rangle = \delta_{jk} |\phi_k(B)\rangle$. We can then rewrite $\tilde E$ as 
\begin{equation}
    \tilde E = \langle\phi(A_L, B, A_R, p)| \tilde H |\phi(A_L, B, A_R, p)\rangle.
\end{equation}
We can thus optimize $B$ by computing eigenvalue-eigenvector pairs of $\tilde H$, after pushing it into the ansatz space, analogously to the symmetric case above.

\begin{figure}
    \includegraphics[width=\linewidth]{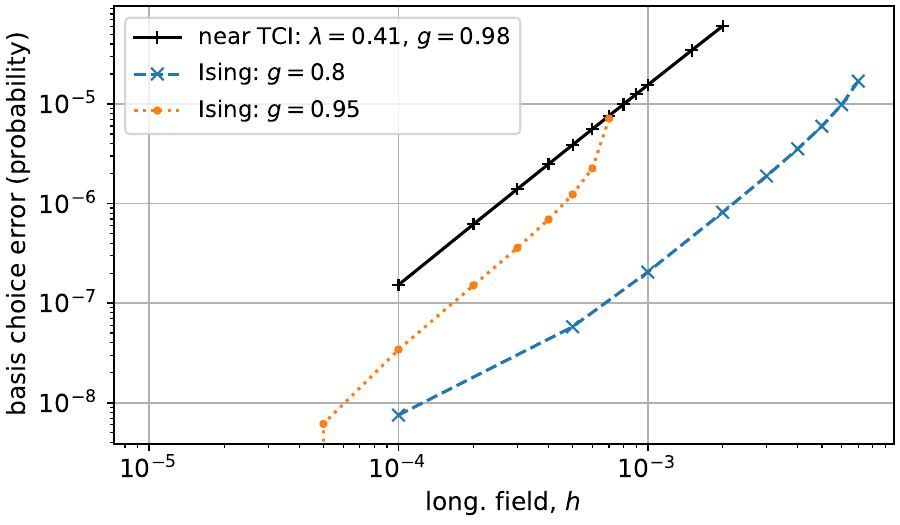}
    \caption{Estimated error (1 - infidelity per site) on momentum eigenstates $|\phi(B, p)\rangle$ for kink quasiparticles, due to the choice of orthogonality conditions on $B$ used to achieve $\langle \phi_j(B)|\phi_k(B)\rangle = \delta_{jk}$. Here we compare the left and right orthogonality conditions, \eqref{eq:LGF} and \eqref{eq:RGF}.}
    \label{fig:gauge_err}
\end{figure}

In this formulation it is manifest that the optimization procedure \emph{depends on the conditions used to achieve position-state orthogonality}, since different conditions will lead to different $P_j$ in \eqref{eq:tildeH}. In practice, we find that the difference this makes to the resulting optimized states $|\phi(A_L, B, A_R, p)\rangle$ is small: We consider the infidelity per site 
\begin{multline}
    1 - |\mathbb{Z}|^{-1} \left|\langle \phi(B_{LF}, p)|\phi(B_{RF}, p)\rangle \right| = \\
    1 - |\left|\langle \phi_j(B_{LF})|\phi(B_{RF}, p)\rangle \right|,
\end{multline}
where $B_{LF}$ and $B_{RF}$ are optimized for the same quasiparticle (and momentum $p$) using the \emph{left} and \emph{right} orthogonality conditions, \eqref{eq:LGF} and \eqref{eq:RGF}, respectively, and $|\mathbb{Z}|$ is the cardinality of the integers (accounting for the infinite norm of the momentum eigenstates). We find empirically the infidelity scales as $h^2$, where $h \ll 1$ is the symmetry-breaking parameter of the Hamiltonian, for the lowest-energy kink states and the Hamiltonian parameters considered in this paper. This dependence is shown in Fig.~\ref{fig:gauge_err}.

It may be possible to design improved optimization techniques that avoid this ambiguity. A prerequisite would be a cost function other than $\langle \tilde H \rangle$, presumably related to the stability of the quasiparticle wavepackets, that distinguishes usefully between the different choices that can be made in parameterizing $B$.

Another important observation is that, in \eqref{eq:tildeH}, we assume that the location $j$ of the $B$ tensor reliably indicates the position of the kink (or antikink). In fact, since the excitation described by $B$ is \emph{quasi}local, the location of the kink (defined as the point in space at which the spin expectation value crosses zero) may differ from $j$. Indeed, as shown below in Fig.~\ref{fig:kappa_j_spins}, there may be a relative shift of several lattice sites, depending on the choice of left or right orthogonality conditions on $B$. The shift will generally also depend on the momentum $p$, such that a dispersion relation $E(p)$ computed from the eigenvalues of $\tilde H$ should really be interpreted as a function of the $B$-tensor momentum, derived from the position $j$ of $B$, considered distinct from the kink momentum, derived from the kink position.

The momentum-dependent energy-shift due to these position shifts can be calculated by first computing the actual kink positions, relative to $j$, for each $|\phi_j(p)\rangle$, as a function of $p$. These shifts can then be multiplied by the false-vacuum excess energy density to compute the energy shift, which can in turn be used to ``correct'' the dispersion relation. This provides a more intuitive definition of the kink dispersion in the symmetry-broken setting. Note also, however, that since kink quasiparticles do not have a well-defined energy gap with respect to the vacuum state, these dispersion relations still cannot be used to compute particle-production thresholds. They could, however, be used to estimate the kink velocity $dE(p)/dp$, this being independent of energy shifts $E(p) \rightarrow E(p) + c$.

\subsection{Wavepackets}

Analogously to the momentum eigenstates of \eqref{eq:MPS_excite}, we can construct wavepackets from the localized quasiparticle states as 
\begin{multline} \label{eq:MPS_packet}
    |\phi(A_L, B, A_R, f)\rangle := 
    \sum_i f_i |\phi_i(A_L, B, A_R)\rangle,
\end{multline}
where in our simulations we choose $f_i$ to be a Gaussian centered at position $x$ with width $\sigma$. Importantly for our purposes, it is straightforward to turn such a wavepacket state into a \emph{single} MPS:
\begin{multline} \label{eq:MPS_packet_single}
    |\phi(A_L, B, A_R, f)\rangle = 
    \sum_{\vec{s}} w_L^\dagger \left( \prod_{i=-\infty}^{\infty} A_i^{s_i} \right) w_R |\vec{s}\rangle,
\end{multline}
where
\begin{equation}
    A_i^s := 
    \begin{pmatrix}
    A_L^s & f_i B^s\\
    0 & A_R^s
    \end{pmatrix}
\end{equation}
is a $2D \times 2D$ matrix, given that $A_L^s$, $A_R^s$, and $B^s$ are all $D \times D$ matrices.  We set the boundary conditions to be
\begin{align}
    w_L^\dagger &:= \left( v_L^\dagger, 0 \right) \\
    w_R &:= \begin{pmatrix}
    0 \\
    v_R
    \end{pmatrix}
\end{align}
for some generic choice of $v_L$ and $v_R$. If $|f_i|$ falls below some numerical threshold for all $i$ less than some $i_L$ and for all $i$ greater than some $i_R > i_L$, we can truncate it to zero and reduce the bond dimension to $D$ in those regions without introducing significant errors. This allows us to represent a truncated wavepacket in the thermodynamic limit using the nonuniform window ansatz \eqref{eq:iMPS_window}.

As discussed in the main text, this wavepacket construction \emph{ignores} any momentum dependence of the tensor $B$. While this introduces errors in the form of contributions from other excitations, as shown in Fig.~\ref{fig:momdep}, these become small for large $\sigma$, as indicated in Fig.~\ref{fig:err_vs_sig} below.

\subsection{Localized states and Bloch-state parameter redundancy}

The parameter redundancy, or ``gauge freedom'', on the $B$-tensors of the momentum eigenstates \eqref{eq:MPS_excite} \emph{does not} leave the localized states $|\phi_j(B)\rangle$ of \eqref{eq:MPS_local_excite} unchanged. These states, as well as the wavepacket states \eqref{eq:MPS_packet} built from them, depend on how these degrees of freedom are fixed. However, the procedure we use for choosing optimal $B$ tensors is based on momentum eigenstates and does not tell us how to optimally fix the remaining freedom.

That said, the impact of this choice on Gaussian wavepacket states must vanish in the limit $\sigma \rightarrow \infty$, where packets become momentum eigenstates. Hence we can reasonably expect the impact on wavepackets with finite width to become small as $\sigma$ increases. Since we already have a physical reason to choose broad wavepackets in our simulations (for slow wavepacket spread), this issue is not as severe as it may at first appear.

Nevertheless, we choose to use the \emph{reflection-symmetric} conditions of \cite{vandamme_realtime_2019a}, slightly adapted for the topologically nontrivial setting, to fix the remaining freedom on the $B$ tensors used to construct our initial bubble states. To be precise, we fix $B$ by choosing $x$ in \eqref{eq:Fmode_freedom} as
\begin{multline} \label{eq:SGF}
    x = \arg \min_{x'} \left( \left| \sum_s B(x')^{(s)} \otimes A_L^{(s)}{}^* \right|^2 \right. + \\
    \left. \left| \sum_s B(x')^{(s)} \otimes A_R^{(s)}{}^* \right|^2 \right).
\end{multline}
In terms of tensor networks, we can rewrite this as
\begin{equation}
    x = \arg \min_{x'} \left( \left| 
    \vcenter{\hbox{\includegraphics[height=3em]{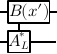}}}
    \right|^2 + 
    \left|
    \vcenter{\hbox{\includegraphics[height=3em]{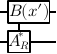}}}
    \right|^2 \right).
\end{equation}
Unlike the left and right orthogonality conditions, \eqref{eq:LGF} and \eqref{eq:RGF}, these conditions are manifestly symmetric under spatial reflections. The reflection-symmetric conditions can be formulated as an overdetermined linear-least-squares optimization problem and then solved using standard techniques.

\begin{figure}
    (i) Ising: $g=0.8$, $h=0.007$\\
    \includegraphics[width=0.95\linewidth]{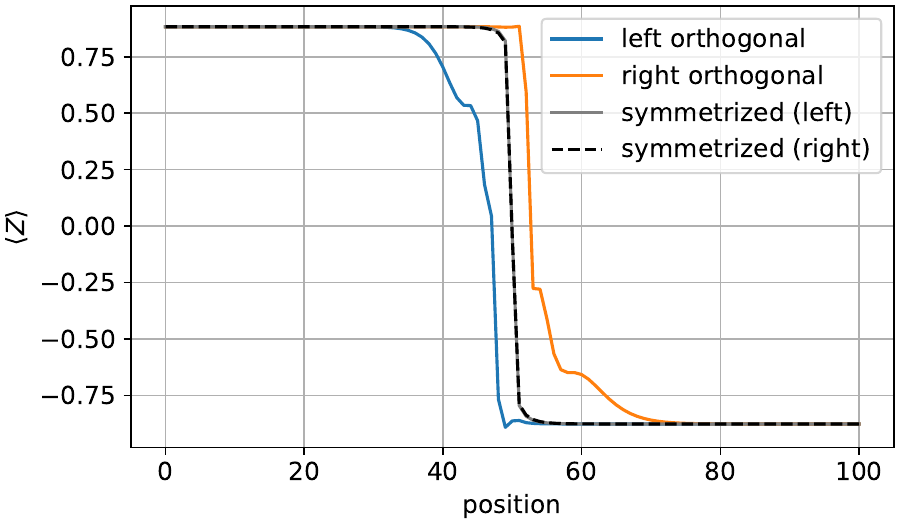}
    (ii) near TCI: $\lambda=0.41$, $g=0.98$, $h=0.001$\\
    \includegraphics[width=0.95\linewidth]{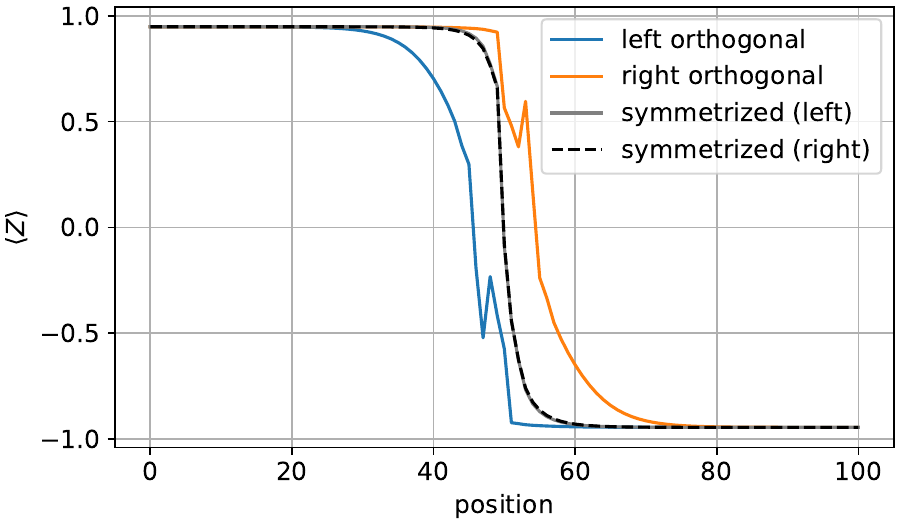}
    \caption{Spin expectation values of a kink position MPS $|\kappa_j\rangle$ for (i) the Ising model and (ii) close to the Tri-Critical-Ising (TCI) point. The bond dimension is $D=8$ for the Ising data, and $D=18$ for the TCI data. We plot the spins for various ways of fixing the momentum-eigenstate freedom: the left and right orthogonal conditions, \eqref{eq:LGF} or \eqref{eq:RGF}, and the reflection-symmetric conditions \eqref{eq:SGF}, beginning from a $B$ tensor optimized using either the left or right conditions (since this makes a small \emph{physical} difference to the result -- see App.~\ref{app:ex_opt}).}
    \label{fig:kappa_j_spins}
\end{figure}

See Fig.~\ref{fig:kappa_j_spins} for a comparison of the reflection-symmetric conditions to the left and right orthogonality conditions, in terms of the spin expectation values of localized topological states. We plot these results for $A_L$ and $A_R$ representing the two vacua of the Ising chain ($\lambda=0$, $g=0.8$) for both zero and nonzero longitudinal field strength $h$. The tensor $B$ is variationally optimized, as described above, so that the momentum eigenstate $|\phi(A_L, B, A_R, p=0)\rangle$ approximates the lowest-lying topological excitation. We note that, although the symmetrized states exhibit ``smoother'' spin expectation values in both cases, they are not perfectly symmetric in the case $h > 0$. The spatial asymmetry in the spin likely reflects the energetic asymmetry of topological states in this case. In both cases, there is certainly an aesthetic improvement to be had by imposing the symmetrization conditions, but it remains to be seen whether the symmetrized states are better representations of localized quasiparticles. To see that they are, we consider how well wavepackets built from them fit into the corresponding quasiparticle subspace.

\begin{figure}
    \centering
    \includegraphics[width=\linewidth]{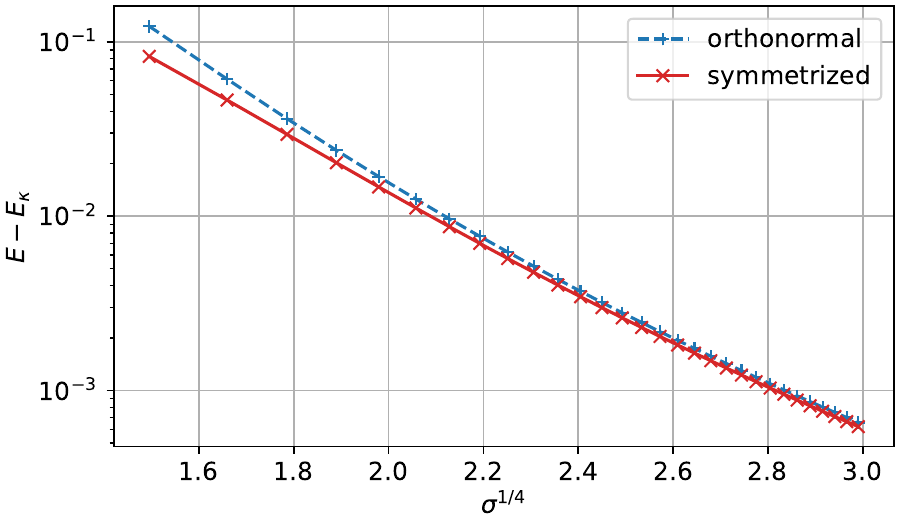}
    \caption{Single-kink wavepacket energy as a function of the width $\sigma$, for the $\mathbb{Z}_2$-symmetric Ising model $g=0.8, \lambda=0, h=0$ (in the SSB phase). We plot the energy (relative to the energy of the kink-quasiparticle momentum eigenstate) for two different ways of fixing the momentum-eigenstate freedom on the $B$ tensors used to construct the wavepacket state: an orthogonal choice $\langle\kappa_j|\kappa_k\rangle = \delta_{jk}$ (left and right conditions are equivalent when $h=0$) and the reflection-symmetrized choice \eqref{eq:SGF}. In both cases the MPS tensors used to construct the state are tuned to the wavepacket momentum $p=0$.}
    \label{fig:en_vs_sig}
\end{figure}

\begin{figure}
    \centering
    \includegraphics[width=\linewidth]{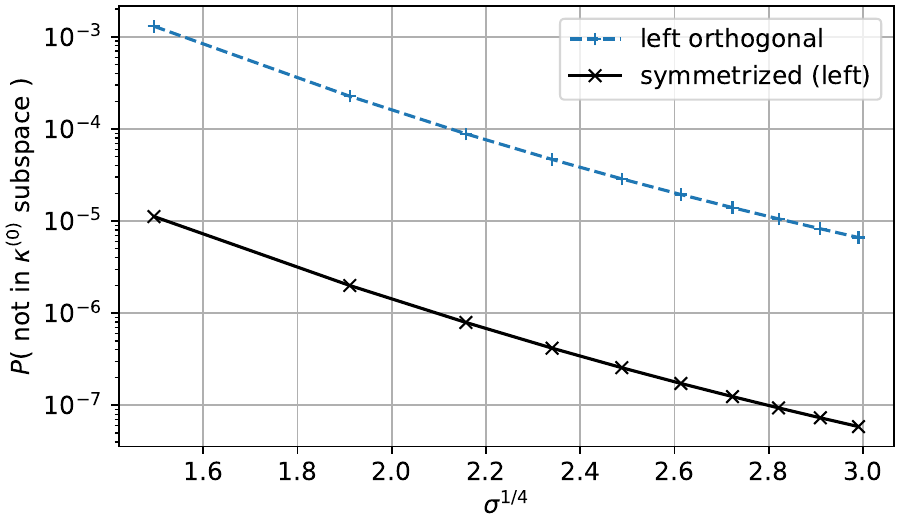}
    \caption{Portion of a single-kink wavepacket state outside of the single-kink subspace $\kappa^{(0)}$ as a function of the wavepacket width $\sigma$, for the $\mathbb{Z}_2$-broken Ising model $g=0.8$,  $\lambda=0$, $h=0.007$. We plot the error for two different ways of fixing the momentum-eigenstate freedom on the $B$ tensors used to construct the wavepacket: the left orthogonal choice $\langle\kappa_j|\kappa_k\rangle = \delta_{jk}$ and the reflection-symmetrized choice (with $B$ \emph{optimized} using the left orthogonal conditions). In both cases the MPS tensors used to construct the state are tuned to the wavepacket momentum, which is $p=0$. The projection into the $\kappa^{(0)}$ subspace uses $B$ tensors optimized using the left orthogonal conditions and fully accounts for momentum dependence via a Fourier analysis.}
    \label{fig:err_vs_sig} 
\end{figure}

In Fig.~\ref{fig:err_vs_sig}, we plot the portion of a kink wavepacket state (by probability) outside of the targeted kink-quasiparticle subspace for both the orthogonal and reflection-symmetric conditions. We explain how to carry out this kind of projection in App.~\ref{app:detection}. We observe that the symmetrized states result in a much more accurate wavepacket than the orthonormal states, by almost two orders of magnitude, confirming that the symmetrized choices are more than just aesthetically pleasing. In the symmetric Ising model ($h=0$), we can also compare the energy of wavepackets. In Fig.~\ref{fig:en_vs_sig}, we see that the kink wavepackets created with the symmetrized states have consistently lower energy, which indicates improved accuracy, since we are targeting the lowest-lying kink quasiparticle.

\subsection{Two-particle states}
\label{app:2particle}

To create false-vacuum bubbles, we need to combine two quasiparticles, namely a kink and an antikink. Two-particle states have the form
\begin{multline} \label{eq:MPS_local_kkbar}
    |\phi\phi_{jk}(A_L, B_L, A_C, B_R, A_R)\rangle := \sum_{\vec{s}} |\vec{s}\rangle \times\\
     v_L^\dagger \left( \prod_{i=-\infty}^{j-1} A^{s_i}_L \right)
    B_L^{s_j} \left( \prod_{l=i+1}^{k-1} A^{s_l}_C \right) 
    B_R^{s_k} \left( \prod_{m=l+1}^{\infty} A^{s_m}_R \right) v_R,
\end{multline}
where, compared to \eqref{eq:MPS_local_excite}, we now have a central (false) vacuum tensor $A_C$, as well as two excitation tensors, $B_L$ and $B_R$, instead of one. Analogously to the one-particle states, we use $\phi\phi$ to denote a generic pair of quasiparticles, specifying $\kappa\overline{\kappa}$ or $\mu\mu$ when we are discussing a kink-antikink pair or a meson pair specifically. Such states are illustrated in Fig.~\ref{fig:states}. We will assume that the quasiparticles are sufficiently well separated, so that interactions may be neglected and $B_L$ and $B_R$ can be held constant irrespective of the separation $d := k-j$. A sufficient condition for this to be justified, is that the reduced state for sites $i$ in between the two excitations $j < i < k$ reverts to that of the central vacuum MPS parameterized by $A_C$ for some range of $i$. If this happens for the reduced state on at least $r$ contiguous sites, where $r$ is the range of interactions in the Hamiltonian (for our model, $r=2$ when $\lambda=0$ and $r=3$ otherwise), this implies that the energy of the two-particle state, as a function of the separation $d$, is not affected by interaction between the particles at that location (only by differences in the ``vacuum'' energies). This means there can be no interaction energy.

This condition can be made more precise: we define the strength of interaction effects at location $i$ (with $j < i < k$) as the deviation of the left and right \emph{environment tensors} of the 2-particle state from the corresponding environment tensors of the central (false) vacuum MPS parameterized by $A_C$. Here, the environment at site $i$ is the tensor network for the reduced state on site $i$, excluding the tensors assigned to site $i$ itself. It naturally splits into left and right components, consisting of the tensors to the left and to the right of $i$, respectively. We compute the deviation for the left and right parts separately, as the norm of the difference between the central (false) vacuum environment and the 2-particle-state environments. For the left environment, we define
\begin{equation}
    \epsilon_L(i) := \left|\:\vcenter{\hbox{\includegraphics[height=3em]{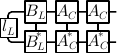}}} \dots \quad - \quad \vcenter{\hbox{\includegraphics[height=2em]{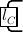}}}\: \right|,
\end{equation}
where the ellipsis indicates that the center transfer matrix should be repeated as many times as is necessary to reach site $i$ from the position of $B_L$. Similarly, for the right environment we define
\begin{equation}
    \epsilon_R(i) := \left|\: \dots \vcenter{\hbox{\includegraphics[height=3em]{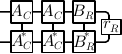}}} \quad - \quad \vcenter{\hbox{\includegraphics[height=2em]{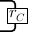}}}\: \right|.
\end{equation}
We define the magnitude of interaction effects at separation $d$ to be
\begin{equation} \label{eq:interaction_strength}
    \epsilon(d) := \min_{j < i < k} \left( \epsilon_L(i) + \epsilon_R(i) \right).
\end{equation}
This value is plotted, for the Hamiltonian parameters (ii) of the main text, in Fig.~\ref{fig:min_sep_tci} for a selection of low-energy quasiparticles.

Wavepackets can be constructed from the two-particle position states \eqref{eq:MPS_local_kkbar} analogously to the single-particle case \eqref{eq:MPS_packet}:
\begin{multline}
    |\phi\phi(A_L, B_L, A_C, B_R, A_R, f, g)\rangle := \\ \sum_{j<k} f_j g_k |\phi\phi_{jk}(A_L, B_L, A_C, B_R, A_R)\rangle,
\end{multline}
where, for our simulations, we choose the packet functions $f_j$ and $g_k$ to be Gaussians centered at $x_L$ and $x_R$, with momenta $p$ and $-p$, respectively. As in the single-particle case \eqref{eq:MPS_packet_single}, these wavepackets can be rewritten as a single MPS with bond dimension $2D$, where $D$ is the bond dimension of the vacua $A_L, A_C, A_R$. Note that, because position states are not defined for $k \ge j$, the wavepacket functions $f$ and $g$ are effectively truncated, leaving only the terms $j < k$, in this ansatz. Of course, if the wavepacket functions have negligible overlap, the effects of this truncation can themselves be neglected.

\section{Particle detection via quasiparticle basis states}
\label{app:detection}

As discussed in the main text, it is possible to use the optimized quasiparticle states and their two-particle combinations to estimate the particle content of a wavefunction $|\psi(t)\rangle$ as it evolves during simulation. For example, the inner product $\langle \kappa\overline{\kappa}_{jk}^{(a,b)}|\psi(t)\rangle$ is sensitive to the presence of a pair consisting of a type-$a$ kink quasiparticle at position $j$ and a type-$b$ antikink quasiparticle at position $k$.

The single-particle quasiparticle position states $|\phi_j\rangle$ of \eqref{eq:MPS_local_excite} can be made orthogonal by imposing either the left or right orthogonality conditions, \eqref{eq:LGF} or \eqref{eq:RGF}. For the two-particle states $|\phi\phi_{jk}^{(a,b)}\rangle$ of \eqref{eq:MPS_local_kkbar} (where $\phi\phi$ is generic notation for either a kink-antikink pair or a meson pair), we can achieve orthogonality by enforcing the left orthogonality conditions on the left excitation tensor, $B_L$, and the right orthogonality conditions on right excitation tensor, $B_R$. With these conditions, the two-particle states are orthogonal for one pair of species $a,b$
\begin{equation}
    \langle \phi\phi_{jk}^{(a,b)}|\phi\phi_{lm}^{(a,b)}\rangle = \delta_{jl} \delta_{km}.
\end{equation}
Furthermore, by limiting the subspace to states with separation $d := k-j$ large enough so that interaction effects are negligible (see App.~\ref{app:2particle}), we can achieve an approximately orthonormal basis across species as well as positions:
\begin{equation}
    \langle \phi\phi_{jk}^{(a,b)}|\phi\phi_{lm}^{(c,d)}\rangle \approx \delta_{jl} \delta_{km} \delta_{ac} \delta_{bd} \quad \forall \; k \gg j, \; m \gg l.
\end{equation}
If the left and right bulk vacuum tensors, $A_L$ and $A_R$ of the evolved state $|\psi(t)\rangle$, which in our simulations have the form \eqref{eq:iMPS_window}, match the left and right bulk tensors $A_L$ and $A_R$ in the two-particle states \eqref{eq:MPS_local_kkbar}, it is then straightforward to (approximately) project $|\psi(t)\rangle$ into the subspace spanned by these states.

However, we must take care when interpreting the overlaps of a wavefunction $|\psi(t)\rangle$ with quasiparticle position states such as $|\phi_j(B)\rangle$, as should be clear from the discussion of excitation tensors and quasiparticle position states above. In particular, the momentum-dependence of the $B$ tensors optimized to represent each quasiparticle, as well as the ambiguity in fixing the degrees of freedom \eqref{eq:Fmode_freedom} in $B$ that are not fixed by the optimization procedure (see App.~\ref{app:ex_opt}), make the interpretation of the overlap unclear unless the quasiparticle content of $|\psi(t)\rangle$ consists of broad spatial wavepackets, whose momentum support is focused around the momentum $p$ used to optimize the $B$ tensor. We next discuss two methods for avoiding these issues.

\subsection{Checking consistency in the projected wavefunction}

As described in the main text, one way to avoid issues with momentum-dependence and wavepacket breadth is to choose \emph{some} momentum $p$, optimize a tensor $B(p)$ at that momentum for the quasiparticle being targeted, then examine the overlaps $\psi_j := \langle \phi_j(B(p))|\psi(t)\rangle$ (or $\psi_{jk} :=\langle \phi\phi_{jk}(B(p), B'(p'))|\psi(t)\rangle$ for a two-particle basis). If the wavepacket width of the projected wavefunction $\psi_j$ is sufficiently large, and the momentum support (computed via Fourier transformation) sufficiently close to the chosen value of $p$, we know that the error made is small and can trust that the projection is accurately telling us about the quasiparticle content. We can quantify how broad the wavepacket must be, and how close the wavepacket momenta should be to $p$, via analyses such as those of Fig.~\ref{fig:momdep} and Fig.~\ref{fig:err_vs_sig}.

If the distribution of momenta in the wavepacket indicates a large error due to the choice of $p$ made while constructing the basis, it may be possible to iteratively \emph{tune} $p$ to achieve a better match.
This procedure fails, of course, if the wavepackets in $|\psi(t)\rangle$ are too narrow in position space (and hence too broad in momentum space) for the error to be kept small.

We use this procedure to compute the spin expectations values within each particle ``sector'' in Fig.~\ref{fig:spin_comps}, tuning the basis momenta $p_L$ and $p_R$ for two-particle bases $|\phi\phi_{jk}(B_L(p_L), B_R(p_R)\rangle$ to match the observed momenta.

\subsection{Fourier analysis}

To \emph{fully} account for the momentum-dependence of the quasiparticle states, and to \emph{eliminate} any issues due to the momentum-eigenstate ``gauge-freedom'' \eqref{eq:Fmode_freedom} on $B$ tensors, we can simply take overlaps with momentum eigenstates instead of with the position states. These states are exactly invariant under \eqref{eq:Fmode_freedom}, and the $B$ tensors used to construct them can be optimized for the momentum of the eigenstate to avoid momentum mismatch.

For example, to project onto a single-particle subspace at momentum $p$, one can compute $\langle \phi(p) | \psi(t)\rangle$, with $|\phi(p)\rangle$ from \eqref{eq:MPS_excite}. We can expand this overlap in terms of position states:
\begin{equation} \label{eq:mode_proj}
    \langle \phi(B(p), p) | \psi(t)\rangle = \sum_j e^{-\mathrm{i}pj} \langle \phi_j (B(p)) | \psi(t)\rangle,
\end{equation}
where $B(p)$ is an excitation tensor optimized to represent the quasiparticle $\phi$ at momentum $p$. This overlap can be computed in practice, despite the infinite sum over $j$, because the position of excitations in $|\psi(t)\rangle$ is limited to the nonuniform window of \eqref{eq:iMPS_window} in which the initial quasiparticles (comprising the bubble) were placed, so that there are only $\sim N_w$ nonzero position terms in this overlap.

To compute the projection onto the entire quasiparticle subspace, we must evaluate the integral
\begin{equation}
  \int_{-\pi}^\pi dp \; \langle \phi(p) | \psi(t)\rangle.    
\end{equation}
This can be done approximately by sampling, for example using a numerical integration scheme. In practice, we use the Fast Fourier-Transform (FFT) algorithm to transform the spatial components $\langle \phi_j(B(p)) | \psi(t)\rangle$ into a fixed sampling of momentum components at a resolution determined by the number of lattice sites $N_w$ summed over in \eqref{eq:mode_proj}.

We use this method to compute the projected single-kink wavefunction described in Fig.~\ref{fig:err_vs_sig}, to compute the kink-antikink scattering outcome probability in Fig.~\ref{fig:ising_error} of the main text, and to compute the various scattering outcome probabilities reported in Fig.~\ref{fig:spin_comps}.

\subsubsection{Efficient computation of the quasiparticle Fourier analysis}

The projection of a wavefunction $|\psi(t)\rangle$ onto many momentum modes is relatively computationally intensive, since for each momentum mode we must first compute a tuned excitation tensor $B(p)$, followed by a full set of position overlaps $\langle \phi_j(B(p))| \psi(t)\rangle$. In the case of single-particle states, the cost is $\mathcal{O}(N_w M)$, where $M$ is the number of momentum samples and $N_w$ is the nonuniform window size of $|\psi(t)\rangle$. For two-particle states, since we must consider cases in which the two particles have different momenta/positions, making the cost $\mathcal{O}(M^2 N_w^2)$. If $M \sim N_w$, and $N_w \sim 1000$, this may be prohibitive!

To reduce the cost, we can use the observation that the optimized excitation tensor $B(p)$, for a given quasiparticle, usually varies only slowly with the quasiparticle momentum $p$ by introducing a small momentum mismatch in a controlled way: We project the excitation tensors $B(p)$, optimized for each mode of momentum $p$ that we wish to sum over, onto a small basis of excitation tensors that capture the momentum dependence accurately across a wide range of $p$:
\begin{equation} \label{eq:B_basis}
    B(p) \approx \sum_\alpha b_\alpha(p) B_\alpha,
\end{equation}
for appropriately chosen coefficients $b_\alpha(p)$ and suitably chosen basis tensors $B_\alpha$. A suitable basis can be built by orthonormalizing (via a Gram-Schmidt procedure) a set of $B(p_\alpha)$ obtained at a selection of momenta (say, $p=-2,-1,0,1,2$). We find that $<10$ basis tensors is sufficient, for our chosen Hamiltonian parameters, to achieve an accuracy of $\sim 10^{-8}$ in \eqref{eq:B_basis}. Given such a basis, we then compute position-state overlaps only for the basis tensors, from which we can compute the projection of the wavefunction onto an arbitrary momentum mode while making only a small error.

For example, in the case of a two-particle basis, we first compute
$\psi_{jk;\alpha\beta} := \langle \phi\phi_{jk}(B_{L,\alpha}, B_{R,\beta}) | \psi(t)\rangle$. Then, using the coefficients of \eqref{eq:B_basis}, we approximate the overlap with the momentum-tuned position states as
\begin{equation}
  \langle \phi\phi_{jk}(B_L(p), B_R(p')) | \psi(t)\rangle \approx
  b_{L,\alpha}^*(p) b_{R,\beta}^*(p') \psi_{jk;\alpha\beta}.
\end{equation}
From here, the momentum-mode overlaps are but an FFT away.

\subsection{Other sources of error}

Even for the Fourier analysis, there are at least two sources of inaccuracy beyond the choice of iMPS bond dimensions in the vacua and the quality of the optimization procedures used to find the vacua and the excitations tensors $B$. First, there is the further ambiguity (see App.~\ref{app:broken_optimization}) in the case of topological excitations in the symmetry-broken setting ($h > 0$) owing to the dependence of the $B$-tensor optimization on an arbitrary choice of orthogonality conditions made during optimization. We do not currently know of a way to avoid this source of error. Fortunately, as illustrated in Fig~\ref{fig:gauge_err}, we have good evidence that it is small.

The other source of error comes from interaction effects, which are not captured properly by the two-particle states. As discussed above, one can choose the minimum separation of the two particles to avoid interaction, by throwing out position states where the ``interaction strength'' $\epsilon$, defined in \eqref{eq:interaction_strength}, rises above some threshold. In choosing the threshold, there is a tradeoff between capturing (potentially large) components of the wavefunction that have smaller separation, but likely incur some (possibly small) error due to interaction, and the magnitude of $\epsilon$, which is a conservative estimate of that error and is exponential in the separation. How to make this tradeoff optimally depends on the target wavefunction and the required precision of the projected wavefunction.

\begin{figure}
    \includegraphics[width=0.9\linewidth]{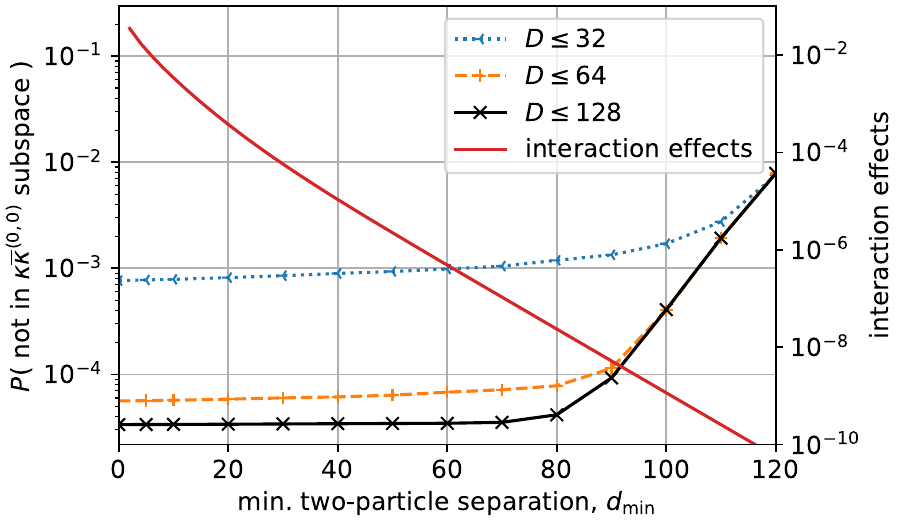}\\
    \caption{Portion of the evolved bubble wavefunction outside of the kink-antikink ``sector'' for simulation (i) of Fig.~\ref{fig:implots} ($\lambda=0$, $g=0.8$, $h=0.007$) at time $t = 240$, as a function of the minimum separation $d_{\min} := \min(k-j)$ permitted in the two-quasiparticles basis states ${|\kappa\overline{\kappa}_{j,k}^{(0,0)}\rangle}$. Probabilities are computed via a Fourier analysis, taking into account the momentum-dependence of the basis states. The basis error due to interaction effects is estimated using \eqref{eq:interaction_strength}.}
    \label{fig:min_sep_isg}
\end{figure}

\begin{figure}
    (a) $\kappa\overline{\kappa}^{(0,0)}$\\
    \includegraphics[width=0.9\linewidth]{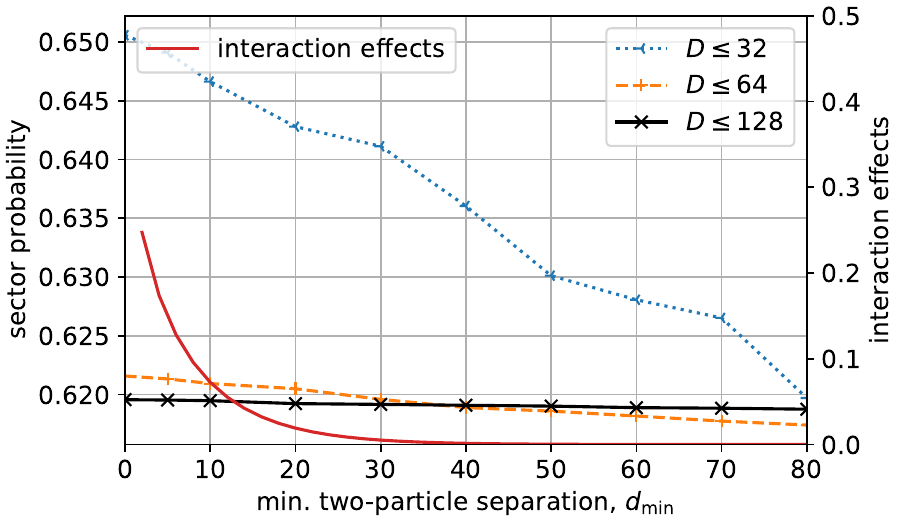}\\
    (b) $\kappa\overline{\kappa}^{(0,1)}$\\
    \includegraphics[width=0.9\linewidth]{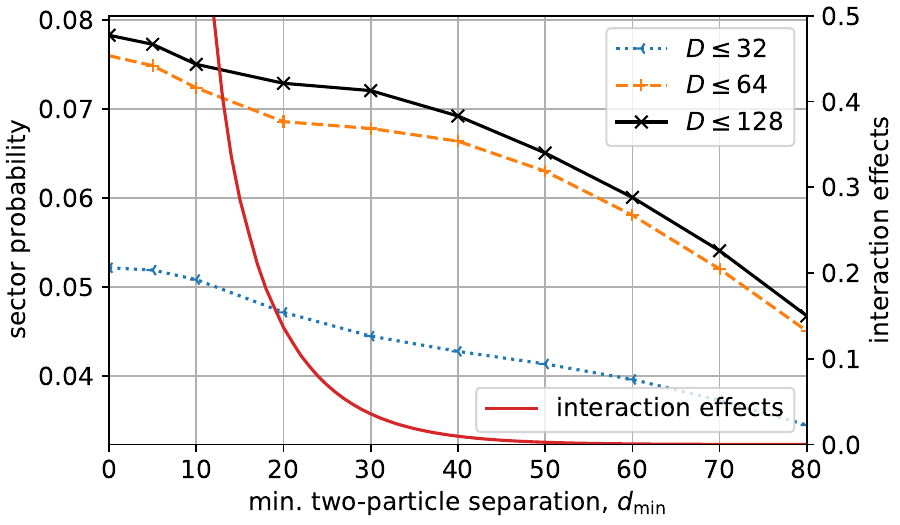}\\
    (c) $\mu\mu^{(0,0)}$\\
    \includegraphics[width=0.9\linewidth]{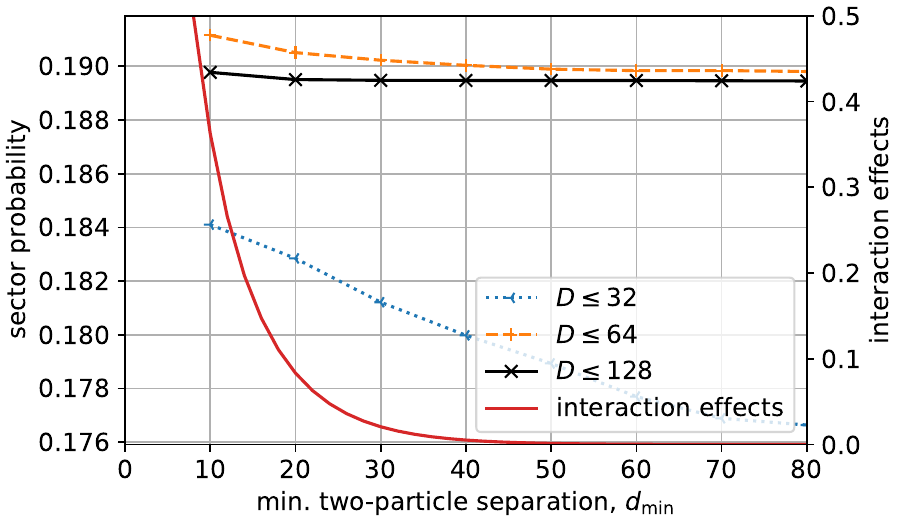}
    \caption{Quasiparticle ``sector'' projection probabilities for simulation (ii) of Fig.~\ref{fig:implots} ($\lambda=0.41$, $g=0.98$, $h=0.001$) at time $t\approx 426$, as a function of the minimum separation $d_{\min} := \min(k-j)$ permitted in the two-quasiparticles basis states ${|\cdot\cdot_{j,k}\rangle}$. Probabilities are computed via a Fourier analysis, taking into account the momentum-dependence of the basis states. The basis error due to interaction effects is estimated using \eqref{eq:interaction_strength}.}
    \label{fig:min_sep_tci}
\end{figure}

For our computations, we examined the dependence of the norm of the projected wavefunctions on the minimum separation $d_{\min}$ allowed in the two-particle basis states. Exemplary results are shown in Figs.~\ref{fig:min_sep_isg} and \ref{fig:min_sep_tci}, which also show the dependence of the norms on the simulation bond-dimension limit. In Fig.~\ref{fig:min_sep_isg}, we observe that, for the largest bond-dimension, the norm of the projected wavefunction is essentially constant for $d < 70$, despite the rising magnitude of interaction effects. This suggests that the wavefunction has negligible support at small separations (which we confirm via a Fourier analysis). We also observe that a minimum separation of $d_{\min} = 60$ is sufficient to keep $\epsilon < 10^{-6}$. Noting that the interaction strength is a property of the quasiparticle basis, and hence independent of time, we make the choice $d_{\min} = 60$ to avoid interaction effects when computing the data shown in Fig.~\ref{fig:ising_error} of the main text.

Fig.~\ref{fig:min_sep_tci} shows similar data for three quasiparticle-pair ``sectors'' for the Hamiltonian parameters (ii) of the main text. This data was used to estimate the scattering outcome probabilities shown in Fig.~\ref{fig:spin_comps} of the main text. We see that, for this target wavefunction $|\psi(t)\rangle$, interaction effects are not important for the $\kappa\overline{\kappa}^{(0,0)}$ and $\mu\overline{\mu}^{(0,0)}$ ``sectors''. However, they may influence the result at the level of $\sim 0.01$, possibly more, for $\kappa\overline{\kappa}^{(0,1)}$ (and, by reflection symmetry of the initial bubble state, $\kappa\overline{\kappa}^{(1,0)}$). The $\kappa\overline{\kappa}^{(0,1)}$ and $\kappa\overline{\kappa}^{(1,0)}$ results are likely to be more sensitive to interaction than the $\kappa\overline{\kappa}^{(0,0)}$ result because the former two ``sectors'' describe bubble states in which one of the walls (the kink or the antikink) is heavier than in the $\kappa\overline{\kappa}^{(0,0)}$ case. These ``lopsided'' bubbles will be smaller than a $\kappa\overline{\kappa}^{(0,0)}$ bubble at the same energy, leading to larger components of the wavefunction at small separations, where interaction effects are stronger.

\section{Evolving through time}
\label{app:tdvp}

To evolve an initial bubble iMPS in time, we define a window of $N_w$ lattice sites surrounding the bubble and allow the MPS tensors belonging to those sites to vary during the evolution, while keeping the rest of the MPS tensors fixed. In other words, we use the nonuniform window ansatz \eqref{eq:iMPS_window} with fixed bulk tensors $A_L$, $A_R$. To compute the evolved state, we use methods based on the Time-Dependent Variational Principle (TDVP), which is set out for this class of states in \cite{milsted_2013_sand} and implemented in the \emph{evoMPS} package \cite{milsted_evomps}.

The TDVP provides flow equations that describe the evolution of the MPS tensors needed to optimally approximate the evolution of the state by the Hamiltonian, given the constraint that the MPS bond dimension must remain \emph{fixed}. Various schemes can be used to integrate the flow equations: we use the popular Runge-Kutta 4/5 (RK4) numerical integrator (to directly integrate the global flow equations) as well as the ``projector-splitting'' (PS) integrator of \cite{haegeman_unifying_2016}. Since we want the MPS bond dimension to grow as needed (up to some maximum) as the entanglement of the state increases, we combine TDVP flow with techniques for increasing the bond dimension. In particular, we use the ``dynamical expansion'' scheme described in \cite{haegeman_2011} together with the RK4 integrator, as well as the two-site projector-splitting method of \cite{haegeman_unifying_2016}.

The PS method and the RK4 integrator (with dynamical expansion), despite having similar theoretical error rates for a given time-step size, behave differently in important ways. For a given step size, the PS method has a larger computational overhead per step, but has better numerical stability and precision since, unlike the ``traditional'' TDVP scheme of \cite{haegeman_2011}, it does not require the inversion of matrices with small eigenvalues.

We find that the RK4 scheme with dynamical expansion is too unstable to use reliably during the initial timesteps of our simulations, which begin with an MPS of relatively small bond dimension (at most twice the vacuum bond dimension). However, we find RK4 can be used successfully after performing a small number of initial steps using the two-site PS scheme. During these initial PS steps, the bond dimension increases significantly. Later in the evolution, once the bond dimension has stabilized, we find the much faster RK4 scheme is able to take over without significant impact on the results. During the evolution, we do allow the MPS bond dimension to grow beyond a predefined maximum value.

To better understand the effects of the integration scheme on our simulations, as well as the impact of the bond-dimension limit, we compute two quantities indicative of numerical error: the energy drift and the truncation error. Although the exact evolution of the quantum state conserves the energy, the imperfect integration of the TDVP flow equations, combined with the limited bond dimension, leads to a small energy drift. This drift is an indicator of error incurred more generally during the evolution. We estimate the truncation error -- the portion of the state by norm that is lost due to the bond dimension limit -- as the maximum value, taken over position, of the \emph{minimum} Schmidt coefficient (the $D$th-largest) for the left-right bipartition at that position. Since there are rarely large jumps in the Schmidt spectrum, this value provides a good estimate of the magnitude of the terms that cannot be represented due to the bond-dimension limit.

\begin{figure}
    \centering
    \includegraphics[width=0.95\linewidth]{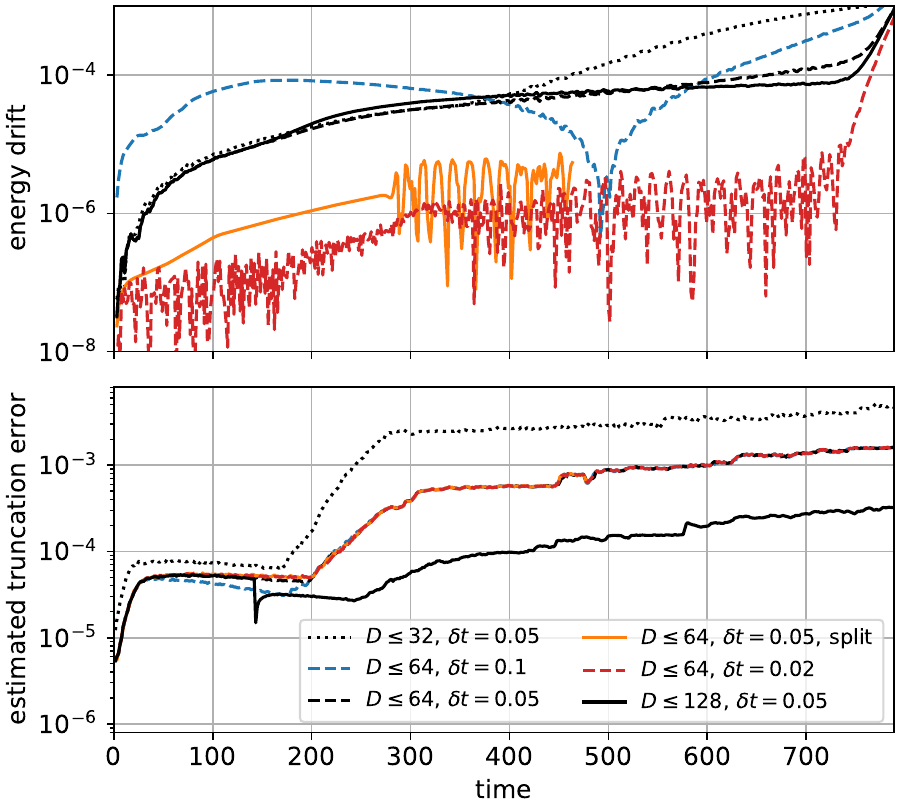}
    \caption{Evolution of the energy expectation value and MPS truncation error for simulation (ii) of the main text. We compare different maximum bond dimensions $D$ as well as different RK4 time-step sizes $\delta t$ (shown here in unscaled lattice Hamiltonian units, for $D\leq 64$), observing that the effects of the time step are most noticeable in the energy drift, whereas the bond-dimension most obviously affects the truncation error at around the time of the first collision ($t \approx 240$). The $D\leq 128$ simulation is initialized from $D\leq 64$, $\delta t = 0.05$ at $t\approx 142$. The uptick in the energy drift at $t\approx 750$ is due to unconfined wavepackets hitting the boundaries of the simulation window.}
    \label{fig:tci_num_err}
\end{figure}

\begin{figure}
    \centering
    \includegraphics[width=0.95\linewidth]{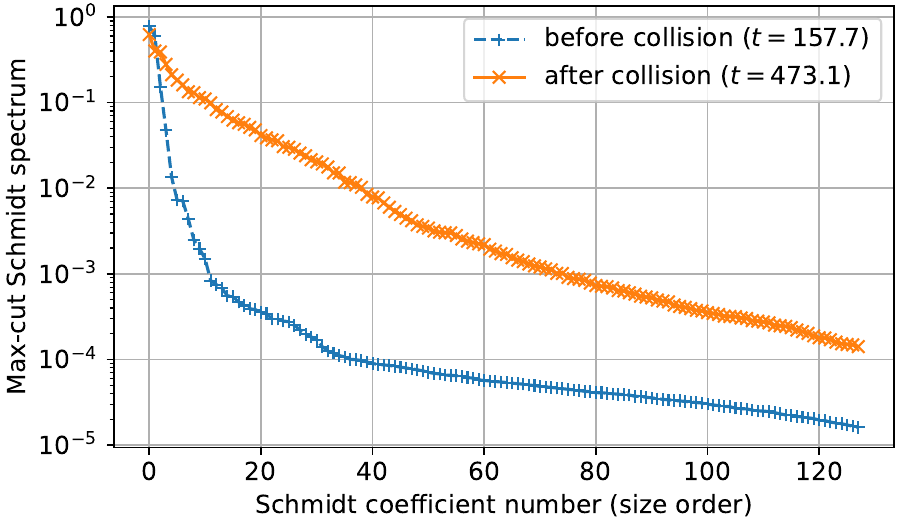}
    \caption{Schmidt spectra for the maximum-entropy cut before and after the first collision in simulation (ii) of the main text. The bond dimension is $128$.}
    \label{fig:schmidt_spec}
\end{figure}

As shown for simulation (ii) of the main text in Fig.~\ref{fig:tci_num_err}, the energy drift is particularly sensitive to the integration scheme and step size, whereas the truncation error is, unsurprisingly, most sensitive to the bond-dimension limit. We note that the truncation error jumps at the time of the first kink-antikink collision ($t\approx 250$), consistent with the entanglement jump observed in Fig.~\ref{fig:entropy_time} of the main text. Fig.~\ref{fig:schmidt_spec} provides a more detailed picture of the entanglement structure, showing the full entanglement (Schmidt) spectrum (up to truncation) before and after the first collision at the cut with the largest entanglement entropy.

The computational cost of simulating up to some fixed time $t$ scales as $\mathcal{O}\left(\frac{D^3}{\delta t}\right)$. We are therefore eventually forced to trade accuracy for computational cost. For this particular simulation, we judge $D\leq 128$ and $\delta t = 0.05$ to provide sufficient accuracy, while keeping the computational requirements manageable, to enable us to study the outcomes of at least the first kink-antikink collision event in detail. 

\section{Comparison with quench approaches}
\label{app:quenches}

We construct our initial false-vacuum bubble from individual kink and antikink quasiparticles, separated by a region of metastable false vacuum. Similar states can be constructed via a simpler approach: act on the uniform true vacuum $|\Omega\rangle$ with a suitable string operator that, in the case of the Ising-type model we study, flips all the spins over a range of sites:
\begin{equation} \label{eq:str}
    |S_{jk}\rangle := X_j X_{j+1} \dots X_{k-1} X_k |\Omega\rangle
\end{equation}
For small longitudinal field $h$, we know that flipping \emph{all} the spins gets us from the vacuum to a state \emph{close} to the false vacuum (see App.~\ref{app:falsevac}), so if $d:= k-j$ is sufficiently large, the reduced state will be close to that of the false vacuum in the middle of the flipped region and topological excitations will be created at the edges of the string. In general, these excitations will be a combination of many topological quasiparticles of varying energy, hence the walls of a bubble created in this way will be unstable to interactions between these quasiparticles. The walls are also highly localized in position and hence have large momentum uncertainty.

By smearing the edges of the string in space, their momenta can be focused. This results in states of the form
\begin{equation} \label{eq:str_pkts}
    |\Psi\rangle = \sum_{j<k} f_j(x_L, p_L) f_k(x_R, p_R) |S_{jk}\rangle,
\end{equation}
where $f_j$ and $f_k$ are wavepacket functions for the left and right edge, respectively. If we choose these to be Gaussian, as for the wavepackets in the main text, we get a bubble state similar to those used in the main text, except that the walls have undetermined quasiparticle content. 

\begin{figure}
    \centering
    \includegraphics[width=\linewidth]{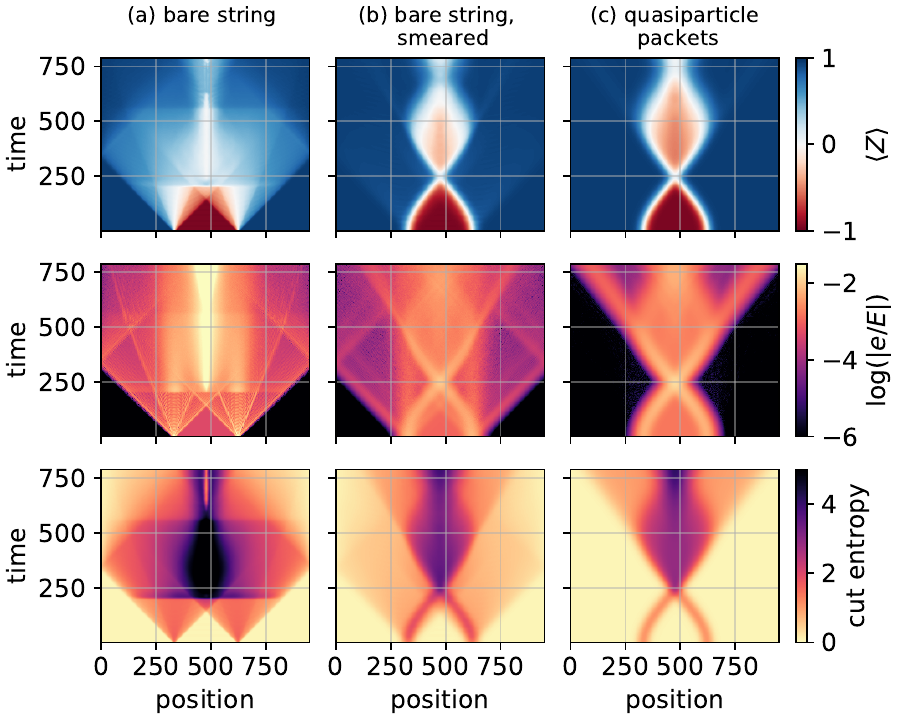}
    \caption{Evolution of spin and energy density expectation values and the cut entropy for parameter set (ii) of the main text ($\lambda=0.41$, $g=0.98$, $h=0.001$), with three different initial states. State (a) is prepared by acting on the vacuum with a string operator $\prod_{j=x_L}^{x_R-1} X_j$, which flips the spins to form a bubble-like state with energy $E/m_\mu = 8.69$. State (b) is similar to (a), but with the ends of the string smeared out using Gaussian packets of width $\sigma=40$, reducing the energy to $E/m_\mu = 4.02$. State (c) is the initial state discussed in the main text with $E/m_\mu = 2.62$, using quasiparticle wavepackets for the kinks and the false vacuum for the middle region. The evolution parameters are the same in all cases: The maximum bond dimension is $64$ and the RK4 step size is $\approx 0.08$ ($0.05$ in unscaled lattice Hamiltonian units). In (a), dramatic errors in the simulation occur at $t\approx 240$, indicating the difficulty of simulating these dynamics versus (b) and (c). In both (a) and (b), ballistic energy-spread emanating from the initial kinks indicates that they have complex quasiparticle content, resulting in immediate inelastic scattering. In contrast, the tuned quasiparticle kinks of (c) do not produce appreciable ballistic spread until the bubble walls have collided.}
    \label{fig:quench}
\end{figure}

In Fig.~\ref{fig:quench}, we compare the dynamics of three different initial states for the Hamiltonian parameters (ii) of the main text: (a) the fully localized string of \eqref{eq:str}, (b) the smeared string of \eqref{eq:str_pkts}, and (c) the tuned quasiparticle kink-antikink wavepackets used in the main text. We choose the same kink-antikink separation in all cases, and the same wavepacket widths in the latter two. The TDVP step size and the maximum bond dimension were also the same in all three cases. Simulations (a) and (b) both exhibit clear ballistic spread of energy from the initial bubble edges, indicating their instability, whereas simulation (c) only shows ballistic spread after the initial bubble walls have collided, consistent with the walls consisting of individual quasiparticles. 

\begin{figure}
    \centering
    \includegraphics[width=0.95\linewidth]{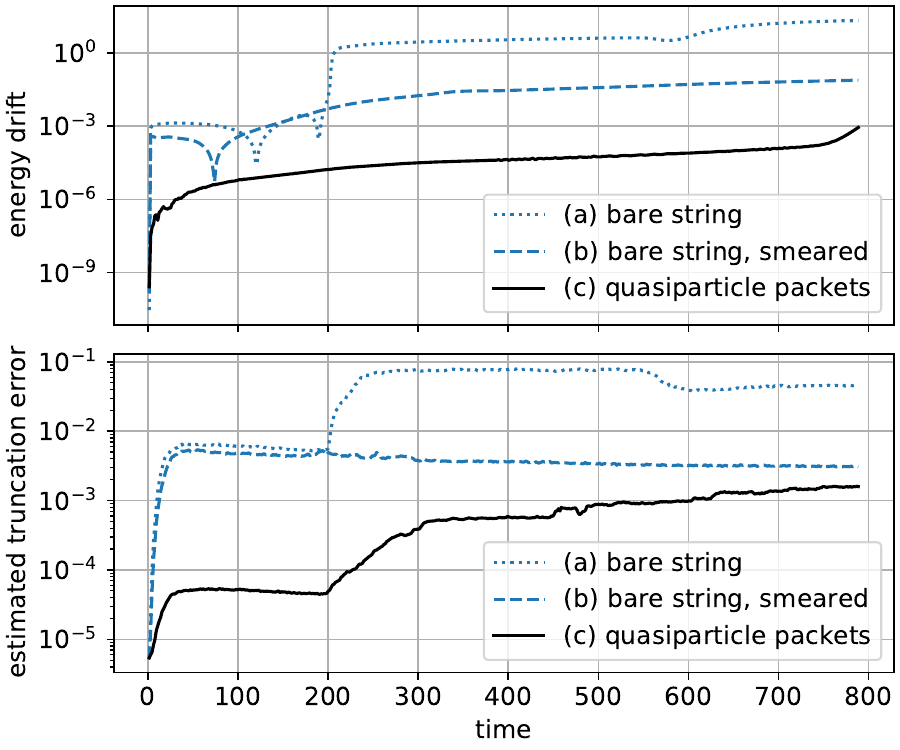}
    \caption{Evolution of the energy expectation value and MPS truncation error for the simulations of Fig.~\ref{fig:quench}. Energy drift ($|1-E(t)/ E(0)|$) indicates deviation from unitary evolution and results from restriction to a maximum bond dimension of $64$ as well as from numerical integration errors (RK4 step size $\approx 0.08$). Truncation error (estimated as the maximum over cuts of the smallest Schmidt coefficient) results from the limited bond dimension and increases as entanglement is produced.}
    \label{fig:quench_num_err}
\end{figure}

It is noteworthy that the evolution of state (a) encounters catastrophic numerical errors at around $t=240$, unlike simulations (b) and (c), suggesting that scenario (a) is much harder to simulate accurately\footnote{These errors also occur if the PS integrator is used throughout the evolution, rather than switching to RK4 after some entanglement has built up.}. In Fig.~\ref{fig:quench_num_err} we show the energy drift and estimated truncation error (smallest retained Schmidt coefficient for the most entangled cut) for the same three simulations. This data clearly indicates that simulation (c) is easiest to simulate, since both the energy drift and the truncation error remain smaller with the same evolution parameters. The large truncation error at early times in cases (a) and (b) is consistent with a large amount of entanglement being generated early on, coming from the multiple excitations created at the ends of the string operator.

\section{Velocity and Bloch oscillations}
\label{app:velocity}

\subsection{Single-kink evolution}

In case of explicit symmetry breaking ($h > 0$), an isolated kink wavepacket, with true vacuum on one side and false vacuum on the other, will accelerate toward the false vacuum, absorbing the excess energy density of the latter. On the lattice, however, the kinetic energy cannot increase indefinitely. Instead, as discussed in the main text, the kink begins to undergo \emph{Bloch oscillations}, eventually decelerating and reversing its direction of travel, as shown in Fig~\ref{fig:1kink_pos_mom_main} for Ising model parameters. By projecting into the $\kappa$ basis of single kinks (see App.~\ref{app:detection}), we can easily compute the kink position and momentum for such a simulation. As shown in Fig.~\ref{fig:1kink_pos_mom_main}, the momentum increases linearly with time, making it easy to understand the evolution of the wavepacket position via the group velocity, which is given by $dE(p)/dp$, where $E(p)$ is the quasiparticle dispersion relation. In Fig.~\ref{fig:1kink_vel_main} we show the velocity derived from the position compared to the velocity derived from the momentum (via the dispersion relation).

\subsection{Bubble evolution}

\begin{figure}
    \centering
    \includegraphics[width=0.95\linewidth]{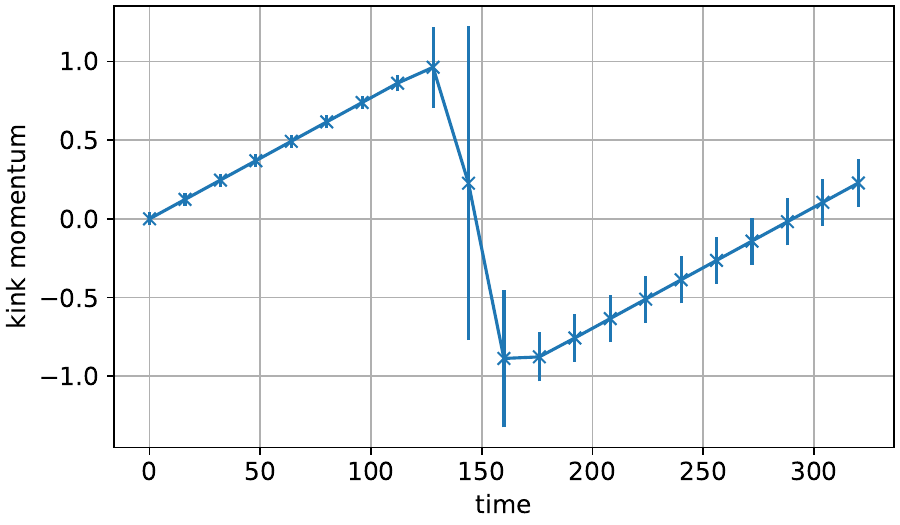}
    \caption{Momentum of the kink in simulation (i) of the main text (Ising), computed from the projected wavefunction in the $\kappa\overline{\kappa}^{(0,0)}$ basis. The bond dimension is $128$. Since the $\kappa\overline{\kappa}^{(0,0)}$ basis does not accurately capture the state when the kink is very close to the antikink, this data is not complete during the collision ($t\approx 140$). The error bars indicate the standard deviation of the momentum distribution.}
    \label{fig:kinkmom_isg}
\end{figure}

In the case of a bubble state, the initial kink and antikink behave as their isolated counterparts, accelerating into the false vacuum until they near each other and interact. Their momenta increase linearly until the collision, as shown in lattice units ($-\pi < p \le \pi$) for the Ising model (parameter set (i) of the main text) in Fig.~\ref{fig:kinkmom_isg}. It is interesting to note that the momentum \emph{variance} is significantly larger after the collision than it is before, indicative of an (in this case elastic) interaction. 

\begin{figure}
    \centering
    \includegraphics[width=0.95\linewidth]{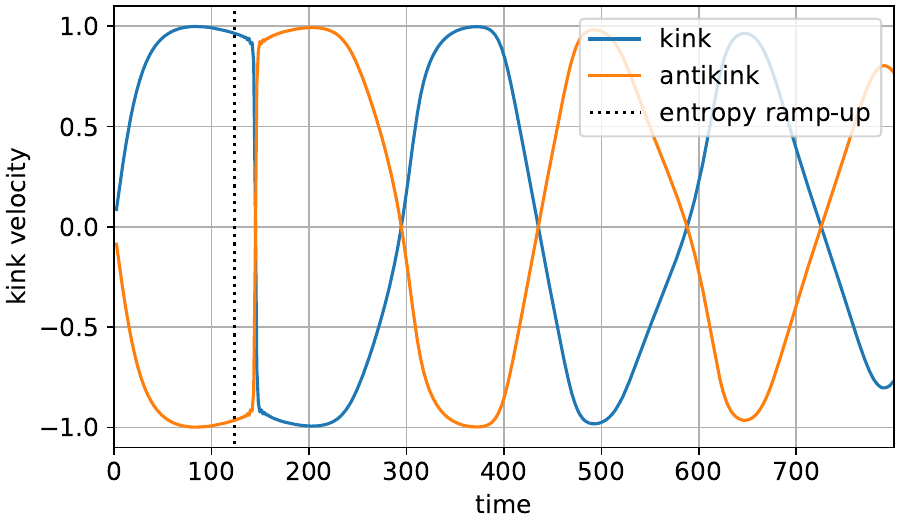}
    \caption{Kink and antikink wavepacket velocity for the Ising model, simulation (i) of the main text, computed as the finite difference of the interpolated position of the 0-intercept of $\langle Z_j \rangle$. The bond dimension is $128$. Note that the data is only likely to be accurate up to $t\approx 800$, as suggested by Fig.~\ref{fig:entropy_time}. The onset of the first collision is indicated by the dotted line, which is the time at which the maximum cut entropy begins to grow rapidly.}
    \label{fig:vel_isg}
\end{figure}

The velocity of the kink and antikink, defined here as the velocity of the point in space at which the (interpolated) spin expectation value crosses zero, evolve as shown in Fig.~\ref{fig:vel_isg}, in accordance with the dispersion relation of the kink quasiparticle excitation. In this simulation, the kink achieves its maximum velocity well before the collision, and begins to decelerate as part of a Bloch oscillation. The pre-collision velocity can be kept from reaching its maximum by reducing the kink-antikink separation in the initial bubble state, as we have confirmed with other simulations.

\begin{figure}
    \centering
    \includegraphics[width=0.95\linewidth]{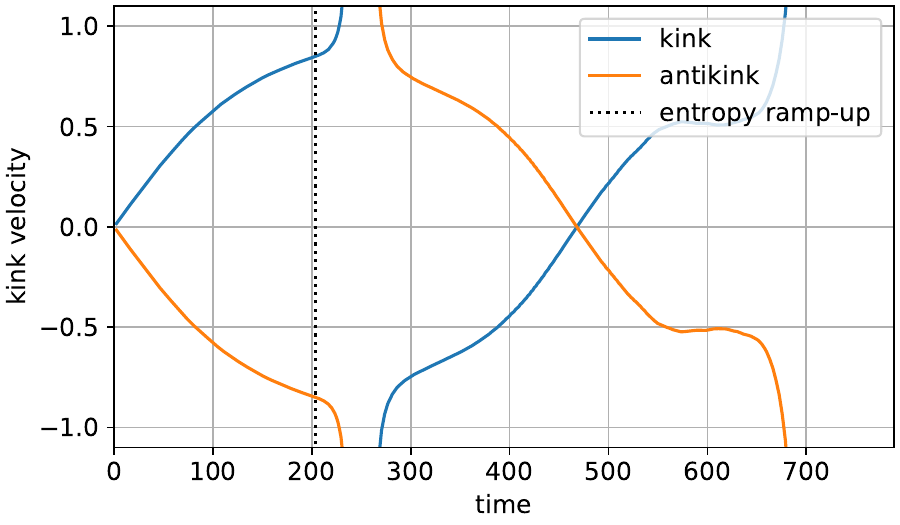}
    \caption{Kink and antikink wavepacket velocity for simulation (ii), computed as the finite difference of the interpolated position of the 0-intercept of $\langle Z_j \rangle$. The bond dimension is $128$. The interpretation of the 0-intercept as the position breaks down both during and, to some extent, after the collision: During the collision, the zero intercept disappears altogether as the kink and antikink merge and all spin expectation values are $>0$. After the collision, there are in this case (see Fig.~\ref{fig:spin_comps}) at least two different bubble ``branches'' of the wavefunction, both contributing to the spin expectation values. The onset of the first collision is indicated by the dotted line, which is the time at which the maximum cut entropy begins to grow rapidly.}
    \label{fig:vel_tci}
\end{figure}

In Fig.~\ref{fig:vel_tci}, we show the kink and antikink velocity evolution for simulation (ii) of the main text. In this case, we have set the initial kink-antikink separation so that the velocity does not reach a maximum prior to collision (indicating that Bloch oscillations have not yet begun).

\section{Zero longitudinal field}
\label{app:zero_field}

Here we examine the behavior of ``bubbles'' (kink-antikink pairs) when we set the longitudinal field $h=0$. Our model with $\lambda \ne 0$ is not integrable, even if we turn off the longitudinal magnetic field $h$. Generically, we should therefore expect to observe inelastic kink-antikink collisions.

We prepare bubble states, for the Hamiltonian parameters $g=0.9$, $\lambda=0.3$, $h=0$, in which the kink and antikink have initial momentum $p$ and $-p$, respectively. By varying $p$, we can choose the total energy to be either above or below the threshold for quasiparticle pair production, which we numerically estimate to be $2m_{\mu} = 1.88$ (relative to the vacuum energy).

\begin{figure}
  \centering
  \includegraphics[width=0.9\linewidth]{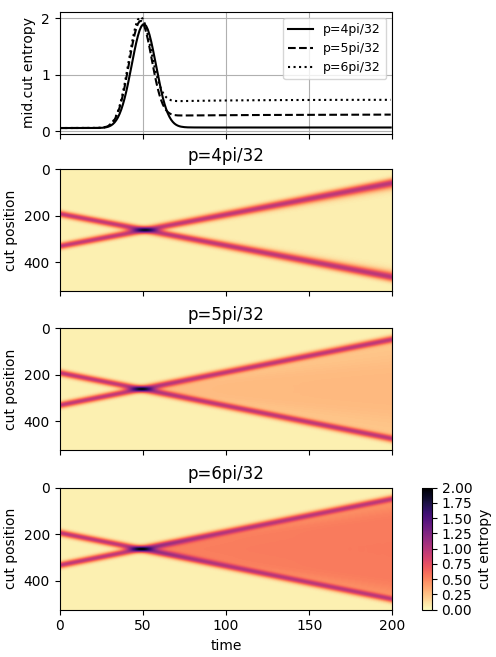}
  \caption{\label{fig:zeroh_ent} Cut entropy for kink-antikink collisions, in the absence of a longitudinal field, with initial kink momentum $p$ and antikink momentum $-p$, for three different values of $p$. The Hamiltonian parameters are $g=0.9$, $\lambda=0.3$, $h=0$, and the wavepacket width is $\sigma=19.0$. The vacuum bond dimension is $D=14$, with a limit $D\le 64$ imposed during evolution. The integration time-step size was $\delta t =0.05$ in unscaled lattice Hamiltonian units.}
\end{figure}

In Fig.~\ref{fig:zeroh_ent}, we show the cut entropy as a function of space and time and, separately for clarity, the time-dependence of the cut entropy at the midpoint between the quasiparticle wavepackets. We choose three different initial momenta: $p=\frac{4\pi}{32}$, $p=\frac{5\pi}{32}$, and $p=\frac{6\pi}{32}$, corresponding to bubble energies of $E=1.65$, $E=1.92$, and $E=2.20$, respectively. We observe that the post-collision mid-chain entropy returns to its vacuum value for $p=\frac{4\pi}{32}$, suggesting a \emph{trivial} scattering event (see App.~\ref{app:elastic_ent}). For $p=\frac{5\pi}{32}$ and $p=\frac{6\pi}{32}$, we observe a residual entropy surplus after the collision, suggesting \emph{nontrivial} scattering. Since the onset of this extra entropy contribution coincides with the energy crossing the two-meson threshold $2m_{\mu}$, it is likely due to an increasing probability of meson pair production.

It is interesting to note that, if we set $h>0$ while keeping the other Hamiltonian parameters the same, we observe \emph{nontrivial}, albeit elastic, scattering of kink-antikink pairs when the energy is below the two-meson threshold. Turning off the longitudinal field appears to turn off this nontrivial elastic contribution, so that kinks and antikinks scatter trivially.

\section{Entanglement generated by an elastic collision}
\label{app:elastic_ent}

Here we consider how two single-particle wavepackets become entangled when they scatter elastically in one spatial dimension. The entanglement arises from the momentum dependence of the scattering phase shift.

For this analysis we ignore lattice effects and consider two distinguishable particles $A$ and $B$ propagating in the continuum. A pure state expanded in the momentum basis has the form
\begin{align}
|\psi \rangle  = \int \mathrm{d}p \,\mathrm{d}q ~\psi(p,q)|p{\rangle _A}|q{\rangle _B}~,
\end{align}
with the normalization
\begin{align}
\int \mathrm{d}p \,\mathrm{d}q~ |\psi(p,q){|^2} = 1~.
\end{align}
We assume the initial state factorizes as the product $\psi(p,q) = \psi_A(p)\psi_B(q)$ of two widely separated wavepackets, but after elastic scattering, the wave packets becomes correlated due to the momentum dependent phase shift $\phi(p,q)$:
\begin{align}
\psi(p,q) = {\psi_A}(p){\psi_B}(q){e^{i\phi (p,q)}}~.
\end{align}
Tracing out  particle $B$, we obtain the reduced density matrix for particle $A$
\begin{align}
&{\rho _A} = \int \mathrm{d}{p_1}\,\mathrm{d}{p_2}\left| {{p_1}} \right\rangle {\rho _A}\left( {{p_1},{p_2}} \right)\left\langle {{p_2}} \right|~,\nonumber\\
&{\rho _A}\left( {{p_1},{p_2}} \right) = \int \mathrm{d}q ~\psi\left( {{p_1},q} \right){\psi^*}\left( {{p_2},q} \right)~,
\end{align}
where $*$ denotes complex conjugation. 

To quantify the entanglement of particles $A$ and $B$, we compute the R\'enyi entropies of $\rho_A$,
\begin{equation}
    S_n = \frac{1}{1-n}\log_2 \rho_A^n~,
\end{equation}
where 
\begin{widetext}
\begin{align}\label{eq:rhoA-integral}
   \textrm{tr} \rho_A^n & =  
   \int \mathrm{d}p_1 \mathrm{d}p_2 \dots \mathrm{d}p_n ~\mathrm{d}q_1 \mathrm{d}q_2 \dots \mathrm{d}q_n~
   |\psi_A(p_1)|^2|\psi_A(p_2)|^2\dots |\psi_A(p_n)|^2~  |\psi_B(q_1)|^2|\psi_B(q_2)|^2\dots|\psi_B(q_n)|^2 \notag\\
   &\exp\left[i\left( \phi(p_1,q_1) - \phi(p_2,q_1) +\phi(p_2, q_2)-\phi(p_3, q_2)\dots + \phi(p_n,q_n)-\phi(p_1,q_n)\right)\right] ~.
\end{align}
\end{widetext}
Now suppose that the wave packets for particles $A$ and $B$ are Gaussian: 
\begin{align}
&{\left| {{\psi_A}(p)} \right|^2} = \frac{1}{{\sqrt {2\pi } {\Delta _A}}}{e^{ - {{(p - \bar p)}^2}/2\Delta _A^2}}~, \nonumber\\
&{\left| {{\psi_B}(q)} \right|^2} = \frac{1}{{\sqrt {2\pi } {\Delta _B}}}{e^{ - {{(q - \bar q)}^2}/2\Delta _B^2}}~.
\end{align}
If the phase shift were slowly varying over the range in $p$ and $q$ where the wave packets have significant support, we could approximate $\textrm{tr} \rho_A^n$ by expanding $\phi(p,q)$ to quadratic order about $(\bar p, \bar q)$. But in that case the scattered wave packets are only slightly entangled. In order to do an analytic computation, we will assume that $\phi(p,q)$ is exactly quadratic even if the phase shift varies rapidly. Then the only term that matters is the cross term
\begin{equation}
    \phi(p,q) = \phi_2 p q + \dots~,
    \end{equation}
because the exponential of the other terms factorizes into a function of $p$ times a function of $q$, which does not contribute to the entanglement of particles $A$ and $B$.

By evaluating a Gaussian integral, we find
\begin{align}\label{eq:renyiproduct}
&{\mathop{\rm tr}\nolimits} \rho _A^n = {\left[ {\prod\limits_{k = 1}^{n - 1} {\left( {1 + 4{\alpha ^2}{{\sin }^2}\left( {\frac{{k\pi }}{n}} \right)} \right)} } \right]^{ - 1/2}}~,\nonumber\\
&{S_n} = \frac{1}{{2(n - 1)}}\sum\limits_{k = 1}^{n - 1} {\log_2 \left( {1 + 4{\alpha ^2}{{\sin }^2}\left( {\frac{{k\pi }}{n}} \right)} \right)} ~,
\end{align}
where
\begin{align}
\alpha=\phi_{2} \Delta_{A} \Delta_{B}~.
\end{align}
As we anticipated, for $\left |\alpha \right|\gg 1$ the phase shift is rapidly varying and the entanglement is substantial. 
Using the formula
\begin{align}
   \prod_{k=1}^{n-1} 4\sin^2 \left(\frac{k \pi }{n}\right) = n^2
\end{align}
we find
\begin{align}
     \textrm tr\rho_A^n = \frac{1}{n|\alpha|^{n-1}}\left(1 + O(\alpha^{-2})\right).
\end{align}
We can extract the large-$\alpha$ behavior of the von Neumann entropy by taking the limit
\begin{align}\label{eq:von-neumann-large-alpha}
    S_1 &= \mathop{\rm tr}\rho_A\log_2\rho_A =\lim_{n\to 1} \frac{1}{1-n}\log_2 \textrm{tr}\rho_A^n\notag\\
    &\approx\lim_{n\to 1} \frac{(n-1)\log_2|\alpha| + \log_2 n}{n-1}= \log_2 (e |\alpha|).
\end{align}
The entanglement entropy of particles $A$ and $B$, after the elastic scattering event, scales like $\log|\alpha|$; therefore we expect to need a bond dimension scaling like $|\alpha|$ to simulate the scattering process accurately using an MPS approximation. 

In fact, by invoking properties of Chebyshev polynomials, the product over $k$ in \eqref{eq:renyiproduct} can be evaluated explicitly, yielding \cite{tang_private}
\begin{align}
{S_n} = \frac{1}{{2(n - 1)}}\left[ 1 + 2n{{\log }_2}\left| \alpha  \right| + {{\log }_2}(\cosh (bn) - 1) \right]~,
\end{align}
where
\begin{align}
b \equiv \arccosh \left( {1 + \frac{1}{{2{{\left| \alpha  \right|}^2}}}} \right)~.
\end{align}
In the limit $n\to 1$  we find
\begin{align}
S_1 = {\log _2}\left| \alpha  \right| + \frac{b}{2} \left(1 + 4{{\left| \alpha  \right|}^2}\right)^{1/2} \log_2 e ~;
\end{align}
taking the large-$\alpha$ limit using $\arccosh(1+x) = \sqrt{2x}\left(1 + O(x) \right)$, we recover \eqref{eq:von-neumann-large-alpha}.

\bibliography{bubble}

\begin{thebibliography}{86}%
\makeatletter
\providecommand \@ifxundefined [1]{%
 \@ifx{#1\undefined}
}%
\providecommand \@ifnum [1]{%
 \ifnum #1\expandafter \@firstoftwo
 \else \expandafter \@secondoftwo
 \fi
}%
\providecommand \@ifx [1]{%
 \ifx #1\expandafter \@firstoftwo
 \else \expandafter \@secondoftwo
 \fi
}%
\providecommand \natexlab [1]{#1}%
\providecommand \enquote  [1]{``#1''}%
\providecommand \bibnamefont  [1]{#1}%
\providecommand \bibfnamefont [1]{#1}%
\providecommand \citenamefont [1]{#1}%
\providecommand \href@noop [0]{\@secondoftwo}%
\providecommand \href [0]{\begingroup \@sanitize@url \@href}%
\providecommand \@href[1]{\@@startlink{#1}\@@href}%
\providecommand \@@href[1]{\endgroup#1\@@endlink}%
\providecommand \@sanitize@url [0]{\catcode `\\12\catcode `\$12\catcode
  `\&12\catcode `\#12\catcode `\^12\catcode `\_12\catcode `\%12\relax}%
\providecommand \@@startlink[1]{}%
\providecommand \@@endlink[0]{}%
\providecommand \url  [0]{\begingroup\@sanitize@url \@url }%
\providecommand \@url [1]{\endgroup\@href {#1}{\urlprefix }}%
\providecommand \urlprefix  [0]{URL }%
\providecommand \Eprint [0]{\href }%
\providecommand \doibase [0]{https://doi.org/}%
\providecommand \selectlanguage [0]{\@gobble}%
\providecommand \bibinfo  [0]{\@secondoftwo}%
\providecommand \bibfield  [0]{\@secondoftwo}%
\providecommand \translation [1]{[#1]}%
\providecommand \BibitemOpen [0]{}%
\providecommand \bibitemStop [0]{}%
\providecommand \bibitemNoStop [0]{.\EOS\space}%
\providecommand \EOS [0]{\spacefactor3000\relax}%
\providecommand \BibitemShut  [1]{\csname bibitem#1\endcsname}%
\let\auto@bib@innerbib\@empty
\bibitem [{\citenamefont {Coleman}(1977)}]{Coleman:1977py}%
  \BibitemOpen
  \bibfield  {author} {\bibinfo {author} {\bibfnamefont {S.~R.}\ \bibnamefont
  {Coleman}},\ }\bibfield  {title} {\bibinfo {title} {{The Fate of the False
  Vacuum. 1. Semiclassical Theory}},\ }\href
  {https://doi.org/10.1103/PhysRevD.16.1248} {\bibfield  {journal} {\bibinfo
  {journal} {Phys. Rev. D}\ }\textbf {\bibinfo {volume} {15}},\ \bibinfo
  {pages} {2929} (\bibinfo {year} {1977})},\ \bibinfo {note} {[Erratum: Phys.
  Rev. D 16, 1248(E) (1977)]}\BibitemShut {NoStop}%
\bibitem [{\citenamefont {Callan}\ and\ \citenamefont
  {Coleman}(1977)}]{Callan:1977pt}%
  \BibitemOpen
  \bibfield  {author} {\bibinfo {author} {\bibfnamefont {C.~G.}\ \bibnamefont
  {Callan}, \bibfnamefont {Jr.}}\ and\ \bibinfo {author} {\bibfnamefont
  {S.~R.}\ \bibnamefont {Coleman}},\ }\bibfield  {title} {\bibinfo {title}
  {{The Fate of the False Vacuum. 2. First Quantum Corrections}},\ }\href
  {https://doi.org/10.1103/PhysRevD.16.1762} {\bibfield  {journal} {\bibinfo
  {journal} {Phys. Rev. D}\ }\textbf {\bibinfo {volume} {16}},\ \bibinfo
  {pages} {1762} (\bibinfo {year} {1977})}\BibitemShut {NoStop}%
\bibitem [{\citenamefont {Turner}\ and\ \citenamefont
  {Wilczek}(1982)}]{Turner:1982xj}%
  \BibitemOpen
  \bibfield  {author} {\bibinfo {author} {\bibfnamefont {M.~S.}\ \bibnamefont
  {Turner}}\ and\ \bibinfo {author} {\bibfnamefont {F.}~\bibnamefont
  {Wilczek}},\ }\bibfield  {title} {\bibinfo {title} {{Might our vacuum be
  metastable?}},\ }\href {https://doi.org/10.1038/298633a0} {\bibfield
  {journal} {\bibinfo  {journal} {Nature}\ }\textbf {\bibinfo {volume} {298}},\
  \bibinfo {pages} {633} (\bibinfo {year} {1982})}\BibitemShut {NoStop}%
\bibitem [{\citenamefont {Coleman}\ and\ \citenamefont
  {De~Luccia}(1980)}]{Coleman:1980aw}%
  \BibitemOpen
  \bibfield  {author} {\bibinfo {author} {\bibfnamefont {S.~R.}\ \bibnamefont
  {Coleman}}\ and\ \bibinfo {author} {\bibfnamefont {F.}~\bibnamefont
  {De~Luccia}},\ }\bibfield  {title} {\bibinfo {title} {{Gravitational Effects
  on and of Vacuum Decay}},\ }\href {https://doi.org/10.1103/PhysRevD.21.3305}
  {\bibfield  {journal} {\bibinfo  {journal} {Phys. Rev. D}\ }\textbf {\bibinfo
  {volume} {21}},\ \bibinfo {pages} {3305} (\bibinfo {year}
  {1980})}\BibitemShut {NoStop}%
\bibitem [{\citenamefont {Markkanen}\ \emph {et~al.}(2018)\citenamefont
  {Markkanen}, \citenamefont {Rajantie},\ and\ \citenamefont
  {Stopyra}}]{markkanen_cosmological_2018}%
  \BibitemOpen
  \bibfield  {author} {\bibinfo {author} {\bibfnamefont {T.}~\bibnamefont
  {Markkanen}}, \bibinfo {author} {\bibfnamefont {A.}~\bibnamefont
  {Rajantie}},\ and\ \bibinfo {author} {\bibfnamefont {S.}~\bibnamefont
  {Stopyra}},\ }\bibfield  {title} {\bibinfo {title} {{Cosmological Aspects of
  Higgs Vacuum Metastability}},\ }\href
  {https://doi.org/10.3389/fspas.2018.00040} {\bibfield  {journal} {\bibinfo
  {journal} {Front. Astron. Space Sci.}\ }\textbf {\bibinfo {volume} {5}},\
  \bibinfo {pages} {40} (\bibinfo {year} {2018})},\ \Eprint
  {https://arxiv.org/abs/1809.06923} {arXiv:1809.06923 [astro-ph.CO]}
  \BibitemShut {NoStop}%
\bibitem [{\citenamefont {Giblin}\ \emph {et~al.}(2010)\citenamefont {Giblin},
  \citenamefont {Hui}, \citenamefont {Lim},\ and\ \citenamefont
  {Yang}}]{giblin_2010}%
  \BibitemOpen
  \bibfield  {author} {\bibinfo {author} {\bibfnamefont {J.~T.}\ \bibnamefont
  {Giblin}}, \bibinfo {author} {\bibfnamefont {L.}~\bibnamefont {Hui}},
  \bibinfo {author} {\bibfnamefont {E.~A.}\ \bibnamefont {Lim}},\ and\ \bibinfo
  {author} {\bibfnamefont {I.-S.}\ \bibnamefont {Yang}},\ }\bibfield  {title}
  {\bibinfo {title} {How to run through walls: Dynamics of bubble and soliton
  collisions},\ }\href {https://doi.org/10.1103/physrevd.82.045019} {\bibfield
  {journal} {\bibinfo  {journal} {Phys. Rev. D}\ }\textbf {\bibinfo {volume}
  {82}},\ \bibinfo {pages} {045019} (\bibinfo {year} {2010})},\ \Eprint
  {https://arxiv.org/abs/1005.3493} {arXiv:1005.3493} \BibitemShut {NoStop}%
\bibitem [{\citenamefont {Amin}\ \emph {et~al.}(2013)\citenamefont {Amin},
  \citenamefont {Lim},\ and\ \citenamefont {Yang}}]{amin_2013}%
  \BibitemOpen
  \bibfield  {author} {\bibinfo {author} {\bibfnamefont {M.~A.}\ \bibnamefont
  {Amin}}, \bibinfo {author} {\bibfnamefont {E.~A.}\ \bibnamefont {Lim}},\ and\
  \bibinfo {author} {\bibfnamefont {I.-S.}\ \bibnamefont {Yang}},\ }\bibfield
  {title} {\bibinfo {title} {A scattering theory of ultrarelativistic
  solitons},\ }\href {https://doi.org/10.1103/physrevd.88.105024} {\bibfield
  {journal} {\bibinfo  {journal} {Phys. Rev. D}\ }\textbf {\bibinfo {volume}
  {88}},\ \bibinfo {pages} {105024} (\bibinfo {year} {2013})},\ \Eprint
  {https://arxiv.org/abs/1308.0606} {arXiv:1308.0606} \BibitemShut {NoStop}%
\bibitem [{\citenamefont {Durr}\ \emph {et~al.}(2008)\citenamefont {Durr} \emph
  {et~al.}}]{durr_2008}%
  \BibitemOpen
  \bibfield  {author} {\bibinfo {author} {\bibfnamefont {S.}~\bibnamefont
  {Durr}} \emph {et~al.},\ }\bibfield  {title} {\bibinfo {title} {{Ab-Initio
  Determination of Light Hadron Masses}},\ }\href
  {https://doi.org/10.1126/science.1163233} {\bibfield  {journal} {\bibinfo
  {journal} {Science}\ }\textbf {\bibinfo {volume} {322}},\ \bibinfo {pages}
  {1224} (\bibinfo {year} {2008})},\ \Eprint {https://arxiv.org/abs/0906.3599}
  {arXiv:0906.3599 [hep-lat]} \BibitemShut {NoStop}%
\bibitem [{\citenamefont {Kanwar}\ and\ \citenamefont
  {Wagman}(2021)}]{kanwar_2021}%
  \BibitemOpen
  \bibfield  {author} {\bibinfo {author} {\bibfnamefont {G.}~\bibnamefont
  {Kanwar}}\ and\ \bibinfo {author} {\bibfnamefont {M.~L.}\ \bibnamefont
  {Wagman}},\ }\bibfield  {title} {\bibinfo {title} {Real-time lattice gauge
  theory actions: Unitarity, convergence, and path integral contour
  deformations},\ }\href {https://doi.org/10.1103/physrevd.104.014513}
  {\bibfield  {journal} {\bibinfo  {journal} {Phys. Rev. D}\ }\textbf {\bibinfo
  {volume} {104}},\ \bibinfo {pages} {014513} (\bibinfo {year} {2021})},\
  \Eprint {https://arxiv.org/abs/2103.02602} {arXiv:2103.02602} \BibitemShut
  {NoStop}%
\bibitem [{\citenamefont {Milsted}\ \emph
  {et~al.}(2013{\natexlab{a}})\citenamefont {Milsted}, \citenamefont
  {Haegeman},\ and\ \citenamefont {Osborne}}]{milsted_2013_phi4}%
  \BibitemOpen
  \bibfield  {author} {\bibinfo {author} {\bibfnamefont {A.}~\bibnamefont
  {Milsted}}, \bibinfo {author} {\bibfnamefont {J.}~\bibnamefont {Haegeman}},\
  and\ \bibinfo {author} {\bibfnamefont {T.~J.}\ \bibnamefont {Osborne}},\
  }\bibfield  {title} {\bibinfo {title} {{Matrix product states and variational
  methods applied to critical quantum field theory}},\ }\href
  {https://doi.org/10.1103/PhysRevD.88.085030} {\bibfield  {journal} {\bibinfo
  {journal} {Phys. Rev. D}\ }\textbf {\bibinfo {volume} {88}},\ \bibinfo
  {pages} {085030} (\bibinfo {year} {2013}{\natexlab{a}})},\ \Eprint
  {https://arxiv.org/abs/1302.5582} {arXiv:1302.5582 [hep-lat]} \BibitemShut
  {NoStop}%
\bibitem [{\citenamefont {K\"uhn}\ \emph {et~al.}(2015)\citenamefont {K\"uhn},
  \citenamefont {Cirac},\ and\ \citenamefont
  {Ba\~nuls}}]{kuhn_nonabelian_2015}%
  \BibitemOpen
  \bibfield  {author} {\bibinfo {author} {\bibfnamefont {S.}~\bibnamefont
  {K\"uhn}}, \bibinfo {author} {\bibfnamefont {J.~I.}\ \bibnamefont {Cirac}},\
  and\ \bibinfo {author} {\bibfnamefont {M.~C.}\ \bibnamefont {Ba\~nuls}},\
  }\bibfield  {title} {\bibinfo {title} {{Non-Abelian string breaking phenomena
  with Matrix Product States}},\ }\href
  {https://doi.org/10.1007/JHEP07(2015)130} {\bibfield  {journal} {\bibinfo
  {journal} {JHEP}\ }\textbf {\bibinfo {volume} {07}},\ \bibinfo {pages}
  {130}},\ \Eprint {https://arxiv.org/abs/1505.04441} {arXiv:1505.04441
  [hep-lat]} \BibitemShut {NoStop}%
\bibitem [{\citenamefont {Pichler}\ \emph {et~al.}(2016)\citenamefont
  {Pichler}, \citenamefont {Dalmonte}, \citenamefont {Rico}, \citenamefont
  {Zoller},\ and\ \citenamefont {Montangero}}]{pichler_realtime_2016}%
  \BibitemOpen
  \bibfield  {author} {\bibinfo {author} {\bibfnamefont {T.}~\bibnamefont
  {Pichler}}, \bibinfo {author} {\bibfnamefont {M.}~\bibnamefont {Dalmonte}},
  \bibinfo {author} {\bibfnamefont {E.}~\bibnamefont {Rico}}, \bibinfo {author}
  {\bibfnamefont {P.}~\bibnamefont {Zoller}},\ and\ \bibinfo {author}
  {\bibfnamefont {S.}~\bibnamefont {Montangero}},\ }\bibfield  {title}
  {\bibinfo {title} {{Real-time Dynamics in U(1) Lattice Gauge Theories with
  Tensor Networks}},\ }\href {https://doi.org/10.1103/PhysRevX.6.011023}
  {\bibfield  {journal} {\bibinfo  {journal} {Phys. Rev. X}\ }\textbf {\bibinfo
  {volume} {6}},\ \bibinfo {pages} {011023} (\bibinfo {year} {2016})},\ \Eprint
  {https://arxiv.org/abs/1505.04440} {arXiv:1505.04440 [cond-mat.quant-gas]}
  \BibitemShut {NoStop}%
\bibitem [{\citenamefont {Buyens}\ \emph {et~al.}(2017)\citenamefont {Buyens},
  \citenamefont {Haegeman}, \citenamefont {Hebenstreit}, \citenamefont
  {Verstraete},\ and\ \citenamefont {Van~Acoleyen}}]{buyens_realtime_2017}%
  \BibitemOpen
  \bibfield  {author} {\bibinfo {author} {\bibfnamefont {B.}~\bibnamefont
  {Buyens}}, \bibinfo {author} {\bibfnamefont {J.}~\bibnamefont {Haegeman}},
  \bibinfo {author} {\bibfnamefont {F.}~\bibnamefont {Hebenstreit}}, \bibinfo
  {author} {\bibfnamefont {F.}~\bibnamefont {Verstraete}},\ and\ \bibinfo
  {author} {\bibfnamefont {K.}~\bibnamefont {Van~Acoleyen}},\ }\bibfield
  {title} {\bibinfo {title} {{Real-time simulation of the Schwinger effect with
  Matrix Product States}},\ }\href {https://doi.org/10.1103/PhysRevD.96.114501}
  {\bibfield  {journal} {\bibinfo  {journal} {Phys. Rev. D}\ }\textbf {\bibinfo
  {volume} {96}},\ \bibinfo {pages} {114501} (\bibinfo {year} {2017})},\
  \Eprint {https://arxiv.org/abs/1612.00739} {arXiv:1612.00739 [hep-lat]}
  \BibitemShut {NoStop}%
\bibitem [{\citenamefont {Chanda}\ \emph {et~al.}(2020)\citenamefont {Chanda},
  \citenamefont {Zakrzewski}, \citenamefont {Lewenstein},\ and\ \citenamefont
  {Tagliacozzo}}]{chanda_confinement_2020}%
  \BibitemOpen
  \bibfield  {author} {\bibinfo {author} {\bibfnamefont {T.}~\bibnamefont
  {Chanda}}, \bibinfo {author} {\bibfnamefont {J.}~\bibnamefont {Zakrzewski}},
  \bibinfo {author} {\bibfnamefont {M.}~\bibnamefont {Lewenstein}},\ and\
  \bibinfo {author} {\bibfnamefont {L.}~\bibnamefont {Tagliacozzo}},\
  }\bibfield  {title} {\bibinfo {title} {{Confinement and lack of
  thermalization after quenches in the bosonic Schwinger model}},\ }\href
  {https://doi.org/10.1103/PhysRevLett.124.180602} {\bibfield  {journal}
  {\bibinfo  {journal} {Phys. Rev. Lett.}\ }\textbf {\bibinfo {volume} {124}},\
  \bibinfo {pages} {180602} (\bibinfo {year} {2020})},\ \Eprint
  {https://arxiv.org/abs/1909.12657} {arXiv:1909.12657 [cond-mat.stat-mech]}
  \BibitemShut {NoStop}%
\bibitem [{\citenamefont {Magnifico}\ \emph {et~al.}(2020)\citenamefont
  {Magnifico}, \citenamefont {Dalmonte}, \citenamefont {Facchi}, \citenamefont
  {Pascazio}, \citenamefont {Pepe},\ and\ \citenamefont
  {Ercolessi}}]{magnifico_real_2020}%
  \BibitemOpen
  \bibfield  {author} {\bibinfo {author} {\bibfnamefont {G.}~\bibnamefont
  {Magnifico}}, \bibinfo {author} {\bibfnamefont {M.}~\bibnamefont {Dalmonte}},
  \bibinfo {author} {\bibfnamefont {P.}~\bibnamefont {Facchi}}, \bibinfo
  {author} {\bibfnamefont {S.}~\bibnamefont {Pascazio}}, \bibinfo {author}
  {\bibfnamefont {F.~V.}\ \bibnamefont {Pepe}},\ and\ \bibinfo {author}
  {\bibfnamefont {E.}~\bibnamefont {Ercolessi}},\ }\bibfield  {title} {\bibinfo
  {title} {{Real Time Dynamics and Confinement in the $\mathbb{Z}_{n}$
  Schwinger-Weyl lattice model for 1+1 QED}},\ }\href
  {https://doi.org/10.22331/q-2020-06-15-281} {\bibfield  {journal} {\bibinfo
  {journal} {Quantum}\ }\textbf {\bibinfo {volume} {4}},\ \bibinfo {pages}
  {281} (\bibinfo {year} {2020})},\ \Eprint {https://arxiv.org/abs/1909.04821}
  {arXiv:1909.04821 [quant-ph]} \BibitemShut {NoStop}%
\bibitem [{\citenamefont {Jordan}\ \emph {et~al.}(2012)\citenamefont {Jordan},
  \citenamefont {Lee},\ and\ \citenamefont {Preskill}}]{jordan_quantum_2012}%
  \BibitemOpen
  \bibfield  {author} {\bibinfo {author} {\bibfnamefont {S.~P.}\ \bibnamefont
  {Jordan}}, \bibinfo {author} {\bibfnamefont {K.~S.}\ \bibnamefont {Lee}},\
  and\ \bibinfo {author} {\bibfnamefont {J.}~\bibnamefont {Preskill}},\
  }\bibfield  {title} {\bibinfo {title} {{Quantum Algorithms for Quantum Field
  Theories}},\ }\href {https://doi.org/10.1126/science.1217069} {\bibfield
  {journal} {\bibinfo  {journal} {Science}\ }\textbf {\bibinfo {volume}
  {336}},\ \bibinfo {pages} {1130} (\bibinfo {year} {2012})},\ \Eprint
  {https://arxiv.org/abs/1111.3633} {arXiv:1111.3633 [quant-ph]} \BibitemShut
  {NoStop}%
\bibitem [{\citenamefont {Jordan}\ \emph {et~al.}(2014)\citenamefont {Jordan},
  \citenamefont {Lee},\ and\ \citenamefont {Preskill}}]{jordan_quantum_2014}%
  \BibitemOpen
  \bibfield  {author} {\bibinfo {author} {\bibfnamefont {S.~P.}\ \bibnamefont
  {Jordan}}, \bibinfo {author} {\bibfnamefont {K.~S.}\ \bibnamefont {Lee}},\
  and\ \bibinfo {author} {\bibfnamefont {J.}~\bibnamefont {Preskill}},\
  }\bibfield  {title} {\bibinfo {title} {{Quantum Computation of Scattering in
  Scalar Quantum Field Theories}},\ }\href@noop {} {\bibfield  {journal}
  {\bibinfo  {journal} {Quant. Inf. Comput.}\ }\textbf {\bibinfo {volume}
  {14}},\ \bibinfo {pages} {1014} (\bibinfo {year} {2014})},\ \Eprint
  {https://arxiv.org/abs/1112.4833} {arXiv:1112.4833 [hep-th]} \BibitemShut
  {NoStop}%
\bibitem [{\citenamefont {Mezzacapo}\ \emph {et~al.}(2015)\citenamefont
  {Mezzacapo}, \citenamefont {Rico}, \citenamefont {Sab\'\i{}n}, \citenamefont
  {Egusquiza}, \citenamefont {Lamata},\ and\ \citenamefont
  {Solano}}]{mezzacapo_nonabelian_2015}%
  \BibitemOpen
  \bibfield  {author} {\bibinfo {author} {\bibfnamefont {A.}~\bibnamefont
  {Mezzacapo}}, \bibinfo {author} {\bibfnamefont {E.}~\bibnamefont {Rico}},
  \bibinfo {author} {\bibfnamefont {C.}~\bibnamefont {Sab\'\i{}n}}, \bibinfo
  {author} {\bibfnamefont {I.~L.}\ \bibnamefont {Egusquiza}}, \bibinfo {author}
  {\bibfnamefont {L.}~\bibnamefont {Lamata}},\ and\ \bibinfo {author}
  {\bibfnamefont {E.}~\bibnamefont {Solano}},\ }\bibfield  {title} {\bibinfo
  {title} {{Non-Abelian $SU(2)$ Lattice Gauge Theories in Superconducting
  Circuits}},\ }\href {https://doi.org/10.1103/PhysRevLett.115.240502}
  {\bibfield  {journal} {\bibinfo  {journal} {Phys. Rev. Lett.}\ }\textbf
  {\bibinfo {volume} {115}},\ \bibinfo {pages} {240502} (\bibinfo {year}
  {2015})},\ \Eprint {https://arxiv.org/abs/1505.04720} {arXiv:1505.04720
  [quant-ph]} \BibitemShut {NoStop}%
\bibitem [{\citenamefont {Martinez}\ \emph {et~al.}(2016)\citenamefont
  {Martinez} \emph {et~al.}}]{martinez_realtime_2016}%
  \BibitemOpen
  \bibfield  {author} {\bibinfo {author} {\bibfnamefont {E.}~\bibnamefont
  {Martinez}} \emph {et~al.},\ }\bibfield  {title} {\bibinfo {title}
  {{Real-time dynamics of lattice gauge theories with a few-qubit quantum
  computer}},\ }\href {https://doi.org/10.1038/nature18318} {\bibfield
  {journal} {\bibinfo  {journal} {Nature}\ }\textbf {\bibinfo {volume} {534}},\
  \bibinfo {pages} {516} (\bibinfo {year} {2016})},\ \Eprint
  {https://arxiv.org/abs/1605.04570} {arXiv:1605.04570 [quant-ph]} \BibitemShut
  {NoStop}%
\bibitem [{\citenamefont {Zohar}\ \emph {et~al.}(2017)\citenamefont {Zohar},
  \citenamefont {Farace}, \citenamefont {Reznik},\ and\ \citenamefont
  {Cirac}}]{zohar_digital_2017}%
  \BibitemOpen
  \bibfield  {author} {\bibinfo {author} {\bibfnamefont {E.}~\bibnamefont
  {Zohar}}, \bibinfo {author} {\bibfnamefont {A.}~\bibnamefont {Farace}},
  \bibinfo {author} {\bibfnamefont {B.}~\bibnamefont {Reznik}},\ and\ \bibinfo
  {author} {\bibfnamefont {J.~I.}\ \bibnamefont {Cirac}},\ }\bibfield  {title}
  {\bibinfo {title} {{Digital quantum simulation of $\mathbb{Z}_2$ lattice
  gauge theories with dynamical fermionic matter}},\ }\href
  {https://doi.org/10.1103/PhysRevLett.118.070501} {\bibfield  {journal}
  {\bibinfo  {journal} {Phys. Rev. Lett.}\ }\textbf {\bibinfo {volume} {118}},\
  \bibinfo {pages} {070501} (\bibinfo {year} {2017})},\ \Eprint
  {https://arxiv.org/abs/1607.03656} {arXiv:1607.03656} \BibitemShut {NoStop}%
\bibitem [{\citenamefont {Muschik}\ \emph {et~al.}(2017)\citenamefont
  {Muschik}, \citenamefont {Heyl}, \citenamefont {Martinez}, \citenamefont
  {Monz}, \citenamefont {Schindler}, \citenamefont {Vogell}, \citenamefont
  {Dalmonte}, \citenamefont {Hauke}, \citenamefont {Blatt},\ and\ \citenamefont
  {Zoller}}]{muschik_u1_2017}%
  \BibitemOpen
  \bibfield  {author} {\bibinfo {author} {\bibfnamefont {C.}~\bibnamefont
  {Muschik}}, \bibinfo {author} {\bibfnamefont {M.}~\bibnamefont {Heyl}},
  \bibinfo {author} {\bibfnamefont {E.}~\bibnamefont {Martinez}}, \bibinfo
  {author} {\bibfnamefont {T.}~\bibnamefont {Monz}}, \bibinfo {author}
  {\bibfnamefont {P.}~\bibnamefont {Schindler}}, \bibinfo {author}
  {\bibfnamefont {B.}~\bibnamefont {Vogell}}, \bibinfo {author} {\bibfnamefont
  {M.}~\bibnamefont {Dalmonte}}, \bibinfo {author} {\bibfnamefont
  {P.}~\bibnamefont {Hauke}}, \bibinfo {author} {\bibfnamefont
  {R.}~\bibnamefont {Blatt}},\ and\ \bibinfo {author} {\bibfnamefont
  {P.}~\bibnamefont {Zoller}},\ }\bibfield  {title} {\bibinfo {title} {{U(1)
  Wilson lattice gauge theories in digital quantum simulators}},\ }\href
  {https://doi.org/10.1088/1367-2630/aa89ab} {\bibfield  {journal} {\bibinfo
  {journal} {New J. Phys.}\ }\textbf {\bibinfo {volume} {19}},\ \bibinfo
  {pages} {103020} (\bibinfo {year} {2017})},\ \Eprint
  {https://arxiv.org/abs/1612.08653} {arXiv:1612.08653 [quant-ph]} \BibitemShut
  {NoStop}%
\bibitem [{\citenamefont {Bender}\ \emph {et~al.}(2018)\citenamefont {Bender},
  \citenamefont {Zohar}, \citenamefont {Farace},\ and\ \citenamefont
  {Cirac}}]{bender_digital_2018}%
  \BibitemOpen
  \bibfield  {author} {\bibinfo {author} {\bibfnamefont {J.}~\bibnamefont
  {Bender}}, \bibinfo {author} {\bibfnamefont {E.}~\bibnamefont {Zohar}},
  \bibinfo {author} {\bibfnamefont {A.}~\bibnamefont {Farace}},\ and\ \bibinfo
  {author} {\bibfnamefont {J.~I.}\ \bibnamefont {Cirac}},\ }\bibfield  {title}
  {\bibinfo {title} {{Digital quantum simulation of lattice gauge theories in
  three spatial dimensions}},\ }\href
  {https://doi.org/10.1088/1367-2630/aadb71} {\bibfield  {journal} {\bibinfo
  {journal} {New J. Phys.}\ }\textbf {\bibinfo {volume} {20}},\ \bibinfo
  {pages} {093001} (\bibinfo {year} {2018})},\ \Eprint
  {https://arxiv.org/abs/1804.02082} {arXiv:1804.02082 [quant-ph]} \BibitemShut
  {NoStop}%
\bibitem [{\citenamefont {Preskill}(2018)}]{preskill_simulating_2018}%
  \BibitemOpen
  \bibfield  {author} {\bibinfo {author} {\bibfnamefont {J.}~\bibnamefont
  {Preskill}},\ }\bibfield  {title} {\bibinfo {title} {{Simulating quantum
  field theory with a quantum computer}},\ }\href
  {https://doi.org/10.22323/1.334.0024} {\bibfield  {journal} {\bibinfo
  {journal} {PoS}\ }\textbf {\bibinfo {volume} {LATTICE2018}},\ \bibinfo
  {pages} {024} (\bibinfo {year} {2018})},\ \Eprint
  {https://arxiv.org/abs/1811.10085} {arXiv:1811.10085 [hep-lat]} \BibitemShut
  {NoStop}%
\bibitem [{\citenamefont {Lamm}\ \emph {et~al.}(2019)\citenamefont {Lamm},
  \citenamefont {Lawrence},\ and\ \citenamefont
  {Yamauchi}}]{lamm_general_2019}%
  \BibitemOpen
  \bibfield  {author} {\bibinfo {author} {\bibfnamefont {H.}~\bibnamefont
  {Lamm}}, \bibinfo {author} {\bibfnamefont {S.}~\bibnamefont {Lawrence}},\
  and\ \bibinfo {author} {\bibfnamefont {Y.}~\bibnamefont {Yamauchi}} (\bibinfo
  {collaboration} {NuQS}),\ }\bibfield  {title} {\bibinfo {title} {{General
  Methods for Digital Quantum Simulation of Gauge Theories}},\ }\href
  {https://doi.org/10.1103/PhysRevD.100.034518} {\bibfield  {journal} {\bibinfo
   {journal} {Phys. Rev. D}\ }\textbf {\bibinfo {volume} {100}},\ \bibinfo
  {pages} {034518} (\bibinfo {year} {2019})},\ \Eprint
  {https://arxiv.org/abs/1903.08807} {arXiv:1903.08807 [hep-lat]} \BibitemShut
  {NoStop}%
\bibitem [{\citenamefont {Kreshchuk}\ \emph {et~al.}(2020)\citenamefont
  {Kreshchuk}, \citenamefont {Kirby}, \citenamefont {Goldstein}, \citenamefont
  {Beauchemin},\ and\ \citenamefont {Love}}]{kreshchuk_quantum_2020}%
  \BibitemOpen
  \bibfield  {author} {\bibinfo {author} {\bibfnamefont {M.}~\bibnamefont
  {Kreshchuk}}, \bibinfo {author} {\bibfnamefont {W.~M.}\ \bibnamefont
  {Kirby}}, \bibinfo {author} {\bibfnamefont {G.}~\bibnamefont {Goldstein}},
  \bibinfo {author} {\bibfnamefont {H.}~\bibnamefont {Beauchemin}},\ and\
  \bibinfo {author} {\bibfnamefont {P.~J.}\ \bibnamefont {Love}},\ }\href@noop
  {} {\bibinfo {title} {Quantum simulation of quantum field theory in the
  light-front formulation}} (\bibinfo {year} {2020}),\ \Eprint
  {https://arxiv.org/abs/2002.04016} {arXiv:2002.04016 [quant-ph]} \BibitemShut
  {NoStop}%
\bibitem [{\citenamefont {Liu}\ and\ \citenamefont {Xin}(2020)}]{Liu:2020eoa}%
  \BibitemOpen
  \bibfield  {author} {\bibinfo {author} {\bibfnamefont {J.}~\bibnamefont
  {Liu}}\ and\ \bibinfo {author} {\bibfnamefont {Y.}~\bibnamefont {Xin}},\
  }\bibfield  {title} {\bibinfo {title} {{Quantum simulation of quantum field
  theories as quantum chemistry}},\ }\href
  {https://doi.org/10.1007/JHEP12(2020)011} {\bibfield  {journal} {\bibinfo
  {journal} {JHEP}\ }\textbf {\bibinfo {volume} {12}},\ \bibinfo {pages}
  {011}},\ \Eprint {https://arxiv.org/abs/2004.13234} {arXiv:2004.13234
  [hep-th]} \BibitemShut {NoStop}%
\bibitem [{\citenamefont {Du}\ \emph {et~al.}(2020)\citenamefont {Du},
  \citenamefont {Vary}, \citenamefont {Zhao},\ and\ \citenamefont
  {Zuo}}]{du_quantum_2020}%
  \BibitemOpen
  \bibfield  {author} {\bibinfo {author} {\bibfnamefont {W.}~\bibnamefont
  {Du}}, \bibinfo {author} {\bibfnamefont {J.~P.}\ \bibnamefont {Vary}},
  \bibinfo {author} {\bibfnamefont {X.}~\bibnamefont {Zhao}},\ and\ \bibinfo
  {author} {\bibfnamefont {W.}~\bibnamefont {Zuo}},\ }\href@noop {} {\bibinfo
  {title} {{Quantum Simulation of Nuclear Inelastic Scattering}}} (\bibinfo
  {year} {2020}),\ \Eprint {https://arxiv.org/abs/2006.01369} {arXiv:2006.01369
  [nucl-th]} \BibitemShut {NoStop}%
\bibitem [{\citenamefont {Farrelly}\ and\ \citenamefont
  {Streich}(2020)}]{farrelly_discretizing_2020}%
  \BibitemOpen
  \bibfield  {author} {\bibinfo {author} {\bibfnamefont {T.}~\bibnamefont
  {Farrelly}}\ and\ \bibinfo {author} {\bibfnamefont {J.}~\bibnamefont
  {Streich}},\ }\href@noop {} {\bibinfo {title} {{Discretizing quantum field
  theories for quantum simulation}}} (\bibinfo {year} {2020}),\ \Eprint
  {https://arxiv.org/abs/2002.02643} {arXiv:2002.02643 [quant-ph]} \BibitemShut
  {NoStop}%
\bibitem [{\citenamefont {Shaw}\ \emph {et~al.}(2020)\citenamefont {Shaw},
  \citenamefont {Lougovski}, \citenamefont {Stryker},\ and\ \citenamefont
  {Wiebe}}]{shaw_quantum_2020}%
  \BibitemOpen
  \bibfield  {author} {\bibinfo {author} {\bibfnamefont {A.~F.}\ \bibnamefont
  {Shaw}}, \bibinfo {author} {\bibfnamefont {P.}~\bibnamefont {Lougovski}},
  \bibinfo {author} {\bibfnamefont {J.~R.}\ \bibnamefont {Stryker}},\ and\
  \bibinfo {author} {\bibfnamefont {N.}~\bibnamefont {Wiebe}},\ }\bibfield
  {title} {\bibinfo {title} {{Quantum Algorithms for Simulating the Lattice
  Schwinger Model}},\ }\href {https://doi.org/10.22331/q-2020-08-10-306}
  {\bibfield  {journal} {\bibinfo  {journal} {Quantum}\ }\textbf {\bibinfo
  {volume} {4}},\ \bibinfo {pages} {306} (\bibinfo {year} {2020})},\ \Eprint
  {https://arxiv.org/abs/2002.11146} {arXiv:2002.11146 [quant-ph]} \BibitemShut
  {NoStop}%
\bibitem [{\citenamefont {Klco}\ \emph {et~al.}(2020)\citenamefont {Klco},
  \citenamefont {Savage},\ and\ \citenamefont {Stryker}}]{klco_su2_2020}%
  \BibitemOpen
  \bibfield  {author} {\bibinfo {author} {\bibfnamefont {N.}~\bibnamefont
  {Klco}}, \bibinfo {author} {\bibfnamefont {M.~J.}\ \bibnamefont {Savage}},\
  and\ \bibinfo {author} {\bibfnamefont {J.~R.}\ \bibnamefont {Stryker}},\
  }\bibfield  {title} {\bibinfo {title} {{SU(2) non-Abelian gauge field theory
  in one dimension on digital quantum computers}},\ }\href
  {https://doi.org/10.1103/PhysRevD.101.074512} {\bibfield  {journal} {\bibinfo
   {journal} {Phys. Rev. D}\ }\textbf {\bibinfo {volume} {101}},\ \bibinfo
  {pages} {074512} (\bibinfo {year} {2020})},\ \Eprint
  {https://arxiv.org/abs/1908.06935} {arXiv:1908.06935 [quant-ph]} \BibitemShut
  {NoStop}%
\bibitem [{\citenamefont {Vovrosh}\ and\ \citenamefont
  {Knolle}(2020)}]{vovrosh2020confinement}%
  \BibitemOpen
  \bibfield  {author} {\bibinfo {author} {\bibfnamefont {J.}~\bibnamefont
  {Vovrosh}}\ and\ \bibinfo {author} {\bibfnamefont {J.}~\bibnamefont
  {Knolle}},\ }\href@noop {} {\bibinfo {title} {Confinement and entanglement
  dynamics on a digital quantum computer}} (\bibinfo {year} {2020}),\ \Eprint
  {https://arxiv.org/abs/2001.03044} {arXiv:2001.03044 [cond-mat.str-el]}
  \BibitemShut {NoStop}%
\bibitem [{\citenamefont {Li}\ and\ \citenamefont {Liu}(2020)}]{Li:2020kbv}%
  \BibitemOpen
  \bibfield  {author} {\bibinfo {author} {\bibfnamefont {Y.-Z.}\ \bibnamefont
  {Li}}\ and\ \bibinfo {author} {\bibfnamefont {J.}~\bibnamefont {Liu}},\
  }\href@noop {} {\bibinfo {title} {{On Quantum Simulation Of Cosmic
  Inflation}}} (\bibinfo {year} {2020}),\ \Eprint
  {https://arxiv.org/abs/2009.10921} {arXiv:2009.10921 [quant-ph]} \BibitemShut
  {NoStop}%
\bibitem [{\citenamefont {Gonz\'alez-Cuadra}\ \emph {et~al.}(2017)\citenamefont
  {Gonz\'alez-Cuadra}, \citenamefont {Zohar},\ and\ \citenamefont
  {Cirac}}]{gonzalez-cuadra_quantum_2017}%
  \BibitemOpen
  \bibfield  {author} {\bibinfo {author} {\bibfnamefont {D.}~\bibnamefont
  {Gonz\'alez-Cuadra}}, \bibinfo {author} {\bibfnamefont {E.}~\bibnamefont
  {Zohar}},\ and\ \bibinfo {author} {\bibfnamefont {J.~I.}\ \bibnamefont
  {Cirac}},\ }\bibfield  {title} {\bibinfo {title} {{Quantum Simulation of the
  Abelian-Higgs Lattice Gauge Theory with Ultracold Atoms}},\ }\href
  {https://doi.org/10.1088/1367-2630/aa6f37} {\bibfield  {journal} {\bibinfo
  {journal} {New J. Phys.}\ }\textbf {\bibinfo {volume} {19}},\ \bibinfo
  {pages} {063038} (\bibinfo {year} {2017})},\ \Eprint
  {https://arxiv.org/abs/1702.05492} {arXiv:1702.05492 [quant-ph]} \BibitemShut
  {NoStop}%
\bibitem [{\citenamefont {Surace}\ \emph {et~al.}(2020)\citenamefont {Surace},
  \citenamefont {Mazza}, \citenamefont {Giudici}, \citenamefont {Lerose},
  \citenamefont {Gambassi},\ and\ \citenamefont
  {Dalmonte}}]{surace_lattice_2020}%
  \BibitemOpen
  \bibfield  {author} {\bibinfo {author} {\bibfnamefont {F.~M.}\ \bibnamefont
  {Surace}}, \bibinfo {author} {\bibfnamefont {P.~P.}\ \bibnamefont {Mazza}},
  \bibinfo {author} {\bibfnamefont {G.}~\bibnamefont {Giudici}}, \bibinfo
  {author} {\bibfnamefont {A.}~\bibnamefont {Lerose}}, \bibinfo {author}
  {\bibfnamefont {A.}~\bibnamefont {Gambassi}},\ and\ \bibinfo {author}
  {\bibfnamefont {M.}~\bibnamefont {Dalmonte}},\ }\bibfield  {title} {\bibinfo
  {title} {{Lattice gauge theories and string dynamics in Rydberg atom quantum
  simulators}},\ }\href {https://doi.org/10.1103/PhysRevX.10.021041} {\bibfield
   {journal} {\bibinfo  {journal} {Phys. Rev. X}\ }\textbf {\bibinfo {volume}
  {10}},\ \bibinfo {pages} {021041} (\bibinfo {year} {2020})},\ \Eprint
  {https://arxiv.org/abs/1902.09551} {arXiv:1902.09551 [cond-mat.quant-gas]}
  \BibitemShut {NoStop}%
\bibitem [{\citenamefont {Surace}\ and\ \citenamefont
  {Lerose}(2020)}]{surace_scattering_2020}%
  \BibitemOpen
  \bibfield  {author} {\bibinfo {author} {\bibfnamefont {F.~M.}\ \bibnamefont
  {Surace}}\ and\ \bibinfo {author} {\bibfnamefont {A.}~\bibnamefont
  {Lerose}},\ }\href@noop {} {\bibinfo {title} {{Scattering of mesons in
  quantum simulators}}} (\bibinfo {year} {2020}),\ \Eprint
  {https://arxiv.org/abs/2011.10583} {arXiv:2011.10583 [cond-mat.quant-gas]}
  \BibitemShut {NoStop}%
\bibitem [{\citenamefont {Notarnicola}\ \emph {et~al.}(2020)\citenamefont
  {Notarnicola}, \citenamefont {Collura},\ and\ \citenamefont
  {Montangero}}]{notarnicola_realtimedynamics_2020}%
  \BibitemOpen
  \bibfield  {author} {\bibinfo {author} {\bibfnamefont {S.}~\bibnamefont
  {Notarnicola}}, \bibinfo {author} {\bibfnamefont {M.}~\bibnamefont
  {Collura}},\ and\ \bibinfo {author} {\bibfnamefont {S.}~\bibnamefont
  {Montangero}},\ }\bibfield  {title} {\bibinfo {title} {{Real-time-dynamics
  quantum simulation of (1+1)-dimensional lattice QED with Rydberg atoms}},\
  }\href {https://doi.org/10.1103/PhysRevResearch.2.013288} {\bibfield
  {journal} {\bibinfo  {journal} {Phys. Rev. Res.}\ }\textbf {\bibinfo {volume}
  {2}},\ \bibinfo {pages} {013288} (\bibinfo {year} {2020})},\ \Eprint
  {https://arxiv.org/abs/1907.12579} {arXiv:1907.12579 [cond-mat.quant-gas]}
  \BibitemShut {NoStop}%
\bibitem [{\citenamefont {Davoudi}\ \emph {et~al.}(2020)\citenamefont
  {Davoudi}, \citenamefont {Hafezi}, \citenamefont {Monroe}, \citenamefont
  {Pagano}, \citenamefont {Seif},\ and\ \citenamefont
  {Shaw}}]{davoudi_analog_2020}%
  \BibitemOpen
  \bibfield  {author} {\bibinfo {author} {\bibfnamefont {Z.}~\bibnamefont
  {Davoudi}}, \bibinfo {author} {\bibfnamefont {M.}~\bibnamefont {Hafezi}},
  \bibinfo {author} {\bibfnamefont {C.}~\bibnamefont {Monroe}}, \bibinfo
  {author} {\bibfnamefont {G.}~\bibnamefont {Pagano}}, \bibinfo {author}
  {\bibfnamefont {A.}~\bibnamefont {Seif}},\ and\ \bibinfo {author}
  {\bibfnamefont {A.}~\bibnamefont {Shaw}},\ }\bibfield  {title} {\bibinfo
  {title} {{Towards analog quantum simulations of lattice gauge theories with
  trapped ions}},\ }\href {https://doi.org/10.1103/PhysRevResearch.2.023015}
  {\bibfield  {journal} {\bibinfo  {journal} {Phys. Rev. Res.}\ }\textbf
  {\bibinfo {volume} {2}},\ \bibinfo {pages} {023015} (\bibinfo {year}
  {2020})},\ \Eprint {https://arxiv.org/abs/1908.03210} {arXiv:1908.03210
  [quant-ph]} \BibitemShut {NoStop}%
\bibitem [{\citenamefont {Celi}\ \emph {et~al.}(2020)\citenamefont {Celi},
  \citenamefont {Vermersch}, \citenamefont {Viyuela}, \citenamefont {Pichler},
  \citenamefont {Lukin},\ and\ \citenamefont {Zoller}}]{celi_emerging_2020}%
  \BibitemOpen
  \bibfield  {author} {\bibinfo {author} {\bibfnamefont {A.}~\bibnamefont
  {Celi}}, \bibinfo {author} {\bibfnamefont {B.}~\bibnamefont {Vermersch}},
  \bibinfo {author} {\bibfnamefont {O.}~\bibnamefont {Viyuela}}, \bibinfo
  {author} {\bibfnamefont {H.}~\bibnamefont {Pichler}}, \bibinfo {author}
  {\bibfnamefont {M.~D.}\ \bibnamefont {Lukin}},\ and\ \bibinfo {author}
  {\bibfnamefont {P.}~\bibnamefont {Zoller}},\ }\bibfield  {title} {\bibinfo
  {title} {{Emerging Two-Dimensional Gauge Theories in Rydberg Configurable
  Arrays}},\ }\href {https://doi.org/10.1103/PhysRevX.10.021057} {\bibfield
  {journal} {\bibinfo  {journal} {Phys. Rev. X}\ }\textbf {\bibinfo {volume}
  {10}},\ \bibinfo {pages} {021057} (\bibinfo {year} {2020})},\ \Eprint
  {https://arxiv.org/abs/1907.03311} {arXiv:1907.03311 [quant-ph]} \BibitemShut
  {NoStop}%
\bibitem [{\citenamefont {Ba\~nuls}\ \emph {et~al.}(2020)\citenamefont
  {Ba\~nuls} \emph {et~al.}}]{banuls_simulating_2020}%
  \BibitemOpen
  \bibfield  {author} {\bibinfo {author} {\bibfnamefont {M.}~\bibnamefont
  {Ba\~nuls}} \emph {et~al.},\ }\bibfield  {title} {\bibinfo {title}
  {{Simulating Lattice Gauge Theories within Quantum Technologies}},\ }\href
  {https://doi.org/10.1140/epjd/e2020-100571-8} {\bibfield  {journal} {\bibinfo
   {journal} {Eur. Phys. J. D}\ }\textbf {\bibinfo {volume} {74}},\ \bibinfo
  {pages} {165} (\bibinfo {year} {2020})},\ \Eprint
  {https://arxiv.org/abs/1911.00003} {arXiv:1911.00003 [quant-ph]} \BibitemShut
  {NoStop}%
\bibitem [{\citenamefont {Fannes}\ \emph {et~al.}(1992)\citenamefont {Fannes},
  \citenamefont {Nachtergaele},\ and\ \citenamefont {Werner}}]{fannes_1992}%
  \BibitemOpen
  \bibfield  {author} {\bibinfo {author} {\bibfnamefont {M.}~\bibnamefont
  {Fannes}}, \bibinfo {author} {\bibfnamefont {B.}~\bibnamefont
  {Nachtergaele}},\ and\ \bibinfo {author} {\bibfnamefont {R.~F.}\ \bibnamefont
  {Werner}},\ }\bibfield  {title} {\bibinfo {title} {Finitely correlated states
  on quantum spin chains},\ }\href {https://doi.org/10.1007/BF02099178}
  {\bibfield  {journal} {\bibinfo  {journal} {Commun. Math. Phys.}\ }\textbf
  {\bibinfo {volume} {144}},\ \bibinfo {pages} {443} (\bibinfo {year}
  {1992})}\BibitemShut {NoStop}%
\bibitem [{\citenamefont {Rommer}\ and\ \citenamefont
  {Ostlund}(1997)}]{rommer_1997}%
  \BibitemOpen
  \bibfield  {author} {\bibinfo {author} {\bibfnamefont {S.}~\bibnamefont
  {Rommer}}\ and\ \bibinfo {author} {\bibfnamefont {S.}~\bibnamefont
  {Ostlund}},\ }\bibfield  {title} {\bibinfo {title} {{Class of ansatz wave
  functions for one-dimensional spin systems and their relation to the density
  matrix renormalization group}},\ }\href
  {https://doi.org/10.1103/PhysRevB.55.2164} {\bibfield  {journal} {\bibinfo
  {journal} {Phys. Rev. B}\ }\textbf {\bibinfo {volume} {55}},\ \bibinfo
  {pages} {2164} (\bibinfo {year} {1997})},\ \Eprint
  {https://arxiv.org/abs/cond-mat/9606213} {arXiv:cond-mat/9606213}
  \BibitemShut {NoStop}%
\bibitem [{\citenamefont {Sugihara}(2004)}]{sugihara_2004}%
  \BibitemOpen
  \bibfield  {author} {\bibinfo {author} {\bibfnamefont {T.}~\bibnamefont
  {Sugihara}},\ }\bibfield  {title} {\bibinfo {title} {{Density matrix
  renormalization group in a two-dimensional lambda phi4 Hamiltonian lattice
  model}},\ }\href {https://doi.org/10.1088/1126-6708/2004/05/007} {\bibfield
  {journal} {\bibinfo  {journal} {JHEP}\ }\textbf {\bibinfo {volume} {05}},\
  \bibinfo {pages} {007}},\ \Eprint {https://arxiv.org/abs/hep-lat/0403008}
  {arXiv:hep-lat/0403008} \BibitemShut {NoStop}%
\bibitem [{\citenamefont {Liu}\ \emph {et~al.}(2021)\citenamefont {Liu},
  \citenamefont {Preskill},\ and\ \citenamefont
  {\c{S}ahino\u{g}lu}}]{burak_upcoming}%
  \BibitemOpen
  \bibfield  {author} {\bibinfo {author} {\bibfnamefont {J.}~\bibnamefont
  {Liu}}, \bibinfo {author} {\bibfnamefont {J.}~\bibnamefont {Preskill}},\ and\
  \bibinfo {author} {\bibfnamefont {B.}~\bibnamefont {\c{S}ahino\u{g}lu}},\
  }\href@noop {} {\bibfield  {journal} {\bibinfo  {journal} {upcoming}\ }
  (\bibinfo {year} {2021})}\BibitemShut {NoStop}%
\bibitem [{\citenamefont {Friedan}\ \emph {et~al.}(1984)\citenamefont
  {Friedan}, \citenamefont {Qiu},\ and\ \citenamefont
  {Shenker}}]{friedan_conformal_1984}%
  \BibitemOpen
  \bibfield  {author} {\bibinfo {author} {\bibfnamefont {D.}~\bibnamefont
  {Friedan}}, \bibinfo {author} {\bibfnamefont {Z.-a.}\ \bibnamefont {Qiu}},\
  and\ \bibinfo {author} {\bibfnamefont {S.~H.}\ \bibnamefont {Shenker}},\
  }\bibfield  {title} {\bibinfo {title} {{Conformal Invariance, Unitarity and
  Two-Dimensional Critical Exponents}},\ }\href
  {https://doi.org/10.1103/PhysRevLett.52.1575} {\bibfield  {journal} {\bibinfo
   {journal} {Phys. Rev. Lett.}\ }\textbf {\bibinfo {volume} {52}},\ \bibinfo
  {pages} {1575} (\bibinfo {year} {1984})}\BibitemShut {NoStop}%
\bibitem [{\citenamefont {Lerose}\ \emph {et~al.}(2020)\citenamefont {Lerose},
  \citenamefont {Surace}, \citenamefont {Mazza}, \citenamefont {Perfetto},
  \citenamefont {Collura},\ and\ \citenamefont
  {Gambassi}}]{lerose_quasilocalized_2020}%
  \BibitemOpen
  \bibfield  {author} {\bibinfo {author} {\bibfnamefont {A.}~\bibnamefont
  {Lerose}}, \bibinfo {author} {\bibfnamefont {F.~M.}\ \bibnamefont {Surace}},
  \bibinfo {author} {\bibfnamefont {P.~P.}\ \bibnamefont {Mazza}}, \bibinfo
  {author} {\bibfnamefont {G.}~\bibnamefont {Perfetto}}, \bibinfo {author}
  {\bibfnamefont {M.}~\bibnamefont {Collura}},\ and\ \bibinfo {author}
  {\bibfnamefont {A.}~\bibnamefont {Gambassi}},\ }\bibfield  {title} {\bibinfo
  {title} {{Quasilocalized dynamics from confinement of quantum excitations}},\
  }\href {https://doi.org/10.1103/PhysRevB.102.041118} {\bibfield  {journal}
  {\bibinfo  {journal} {Phys. Rev. B}\ }\textbf {\bibinfo {volume} {102}},\
  \bibinfo {pages} {041118(R)} (\bibinfo {year} {2020})},\ \Eprint
  {https://arxiv.org/abs/1911.07877} {arXiv:1911.07877 [cond-mat.stat-mech]}
  \BibitemShut {NoStop}%
\bibitem [{\citenamefont {Fonseca}\ and\ \citenamefont
  {Zamolodchikov}(2001)}]{fonseca_ising_2001}%
  \BibitemOpen
  \bibfield  {author} {\bibinfo {author} {\bibfnamefont {P.}~\bibnamefont
  {Fonseca}}\ and\ \bibinfo {author} {\bibfnamefont {A.}~\bibnamefont
  {Zamolodchikov}},\ }\href@noop {} {\bibinfo {title} {{Ising field theory in a
  magnetic field: Analytic properties of the free energy}}} (\bibinfo {year}
  {2001}),\ \Eprint {https://arxiv.org/abs/hep-th/0112167}
  {arXiv:hep-th/0112167} \BibitemShut {NoStop}%
\bibitem [{\citenamefont {Nagele}\ \emph {et~al.}(2019)\citenamefont {Nagele},
  \citenamefont {Cejudo}, \citenamefont {Byrnes},\ and\ \citenamefont
  {Kleban}}]{Nagele_2019}%
  \BibitemOpen
  \bibfield  {author} {\bibinfo {author} {\bibfnamefont {C.}~\bibnamefont
  {Nagele}}, \bibinfo {author} {\bibfnamefont {J.~E.}\ \bibnamefont {Cejudo}},
  \bibinfo {author} {\bibfnamefont {T.}~\bibnamefont {Byrnes}},\ and\ \bibinfo
  {author} {\bibfnamefont {M.}~\bibnamefont {Kleban}},\ }\bibfield  {title}
  {\bibinfo {title} {Flux unwinding in the lattice schwinger model},\ }\href
  {https://doi.org/10.1103/physrevd.99.094501} {\bibfield  {journal} {\bibinfo
  {journal} {Phys. Rev. D}\ }\textbf {\bibinfo {volume} {99}},\ \bibinfo
  {pages} {094501} (\bibinfo {year} {2019})},\ \Eprint
  {https://arxiv.org/abs/1811.03096} {arXiv:1811.03096 [hep-th]} \BibitemShut
  {NoStop}%
\bibitem [{\citenamefont {Karpov}\ \emph {et~al.}(2020)\citenamefont {Karpov},
  \citenamefont {Zhu}, \citenamefont {Heller},\ and\ \citenamefont
  {Heyl}}]{Karpov:2020pqe}%
  \BibitemOpen
  \bibfield  {author} {\bibinfo {author} {\bibfnamefont {P.}~\bibnamefont
  {Karpov}}, \bibinfo {author} {\bibfnamefont {G.}~\bibnamefont {Zhu}},
  \bibinfo {author} {\bibfnamefont {M.}~\bibnamefont {Heller}},\ and\ \bibinfo
  {author} {\bibfnamefont {M.}~\bibnamefont {Heyl}},\ }\href@noop {} {\bibinfo
  {title} {{Spatiotemporal dynamics of particle collisions in quantum spin
  chains}}} (\bibinfo {year} {2020}),\ \Eprint
  {https://arxiv.org/abs/2011.11624} {arXiv:2011.11624 [cond-mat.quant-gas]}
  \BibitemShut {NoStop}%
\bibitem [{\citenamefont {Van~Damme}\ \emph {et~al.}(2021)\citenamefont
  {Van~Damme}, \citenamefont {Vanderstraeten}, \citenamefont {De~Nardis},
  \citenamefont {Haegeman},\ and\ \citenamefont
  {Verstraete}}]{vandamme_realtime_2019a}%
  \BibitemOpen
  \bibfield  {author} {\bibinfo {author} {\bibfnamefont {M.}~\bibnamefont
  {Van~Damme}}, \bibinfo {author} {\bibfnamefont {L.}~\bibnamefont
  {Vanderstraeten}}, \bibinfo {author} {\bibfnamefont {J.}~\bibnamefont
  {De~Nardis}}, \bibinfo {author} {\bibfnamefont {J.}~\bibnamefont
  {Haegeman}},\ and\ \bibinfo {author} {\bibfnamefont {F.}~\bibnamefont
  {Verstraete}},\ }\bibfield  {title} {\bibinfo {title} {Real-time scattering
  of interacting quasiparticles in quantum spin chains},\ }\href
  {https://doi.org/10.1103/PhysRevResearch.3.013078} {\bibfield  {journal}
  {\bibinfo  {journal} {Phys. Rev. Research}\ }\textbf {\bibinfo {volume}
  {3}},\ \bibinfo {pages} {013078} (\bibinfo {year} {2021})},\ \Eprint
  {https://arxiv.org/abs/1907.02474} {arXiv:1907.02474 [cond-mat.str-el]}
  \BibitemShut {NoStop}%
\bibitem [{\citenamefont {Vlijm}\ \emph {et~al.}(2015)\citenamefont {Vlijm},
  \citenamefont {Ganahl}, \citenamefont {Fioretto}, \citenamefont {Brockmann},
  \citenamefont {Haque}, \citenamefont {Evertz},\ and\ \citenamefont
  {Caux}}]{vlijm_quasisoliton_2015}%
  \BibitemOpen
  \bibfield  {author} {\bibinfo {author} {\bibfnamefont {R.}~\bibnamefont
  {Vlijm}}, \bibinfo {author} {\bibfnamefont {M.}~\bibnamefont {Ganahl}},
  \bibinfo {author} {\bibfnamefont {D.}~\bibnamefont {Fioretto}}, \bibinfo
  {author} {\bibfnamefont {M.}~\bibnamefont {Brockmann}}, \bibinfo {author}
  {\bibfnamefont {M.}~\bibnamefont {Haque}}, \bibinfo {author} {\bibfnamefont
  {H.~G.}\ \bibnamefont {Evertz}},\ and\ \bibinfo {author} {\bibfnamefont
  {J.-S.}\ \bibnamefont {Caux}},\ }\bibfield  {title} {\bibinfo {title}
  {Quasi-soliton scattering in quantum spin chains},\ }\href
  {https://doi.org/10.1103/PhysRevB.92.214427} {\bibfield  {journal} {\bibinfo
  {journal} {Phys. Rev. B}\ }\textbf {\bibinfo {volume} {92}},\ \bibinfo
  {pages} {214427} (\bibinfo {year} {2015})},\ \Eprint
  {https://arxiv.org/abs/1507.08624} {arXiv:1507.08624 [cond-mat.str-el]}
  \BibitemShut {NoStop}%
\bibitem [{\citenamefont {O'Brien}\ and\ \citenamefont
  {Fendley}(2018)}]{obrien_lattice_2018}%
  \BibitemOpen
  \bibfield  {author} {\bibinfo {author} {\bibfnamefont {E.}~\bibnamefont
  {O'Brien}}\ and\ \bibinfo {author} {\bibfnamefont {P.}~\bibnamefont
  {Fendley}},\ }\bibfield  {title} {\bibinfo {title} {{Lattice supersymmetry
  and order-disorder coexistence in the tricritical Ising model}},\ }\href
  {https://doi.org/10.1103/PhysRevLett.120.206403} {\bibfield  {journal}
  {\bibinfo  {journal} {Phys. Rev. Lett.}\ }\textbf {\bibinfo {volume} {120}},\
  \bibinfo {pages} {206403} (\bibinfo {year} {2018})},\ \Eprint
  {https://arxiv.org/abs/1712.06662} {arXiv:1712.06662 [cond-mat.stat-mech]}
  \BibitemShut {NoStop}%
\bibitem [{\citenamefont {Sannomiya}\ and\ \citenamefont
  {Katsura}(2019)}]{sannomiya_supersymmetry_2019}%
  \BibitemOpen
  \bibfield  {author} {\bibinfo {author} {\bibfnamefont {N.}~\bibnamefont
  {Sannomiya}}\ and\ \bibinfo {author} {\bibfnamefont {H.}~\bibnamefont
  {Katsura}},\ }\bibfield  {title} {\bibinfo {title} {{Supersymmetry Breaking
  and Nambu-Goldstone Fermions in Interacting Majorana Chains}},\ }\href
  {https://doi.org/10.1103/PhysRevD.99.045002} {\bibfield  {journal} {\bibinfo
  {journal} {Phys. Rev. D}\ }\textbf {\bibinfo {volume} {99}},\ \bibinfo
  {pages} {045002} (\bibinfo {year} {2019})},\ \Eprint
  {https://arxiv.org/abs/1712.01148} {arXiv:1712.01148 [cond-mat.str-el]}
  \BibitemShut {NoStop}%
\bibitem [{\citenamefont {Fonseca}\ and\ \citenamefont
  {Zamolodchikov}(2006)}]{fonseca_ising_2006}%
  \BibitemOpen
  \bibfield  {author} {\bibinfo {author} {\bibfnamefont {P.}~\bibnamefont
  {Fonseca}}\ and\ \bibinfo {author} {\bibfnamefont {A.}~\bibnamefont
  {Zamolodchikov}},\ }\href@noop {} {\bibinfo {title} {{Ising spectroscopy. I.
  Mesons at T \ensuremath{<} T(c)}}} (\bibinfo {year} {2006}),\ \Eprint
  {https://arxiv.org/abs/hep-th/0612304} {arXiv:hep-th/0612304} \BibitemShut
  {NoStop}%
\bibitem [{\citenamefont {Mussardo}\ and\ \citenamefont
  {Takacs}(2009)}]{mussardo_effective_2009}%
  \BibitemOpen
  \bibfield  {author} {\bibinfo {author} {\bibfnamefont {G.}~\bibnamefont
  {Mussardo}}\ and\ \bibinfo {author} {\bibfnamefont {G.}~\bibnamefont
  {Takacs}},\ }\bibfield  {title} {\bibinfo {title} {{Effective potentials and
  kink spectra in non-integrable perturbed conformal field theories}},\ }\href
  {https://doi.org/10.1088/1751-8113/42/30/304022} {\bibfield  {journal}
  {\bibinfo  {journal} {J. Phys. A}\ }\textbf {\bibinfo {volume} {42}},\
  \bibinfo {pages} {304022} (\bibinfo {year} {2009})},\ \Eprint
  {https://arxiv.org/abs/0901.3537} {arXiv:0901.3537 [hep-th]} \BibitemShut
  {NoStop}%
\bibitem [{\citenamefont
  {Zamolodchikov}(1989{\natexlab{a}})}]{zamolodchikov_integrable_1989}%
  \BibitemOpen
  \bibfield  {author} {\bibinfo {author} {\bibfnamefont {A.}~\bibnamefont
  {Zamolodchikov}},\ }\bibfield  {title} {\bibinfo {title} {{Integrable field
  theory from conformal field theory}},\ }\href
  {https://doi.org/10.1016/B978-0-12-385342-4.50022-6} {\bibfield  {journal}
  {\bibinfo  {journal} {Adv. Stud. Pure Math.}\ }\textbf {\bibinfo {volume}
  {19}},\ \bibinfo {pages} {641} (\bibinfo {year}
  {1989}{\natexlab{a}})}\BibitemShut {NoStop}%
\bibitem [{\citenamefont
  {Zamolodchikov}(1989{\natexlab{b}})}]{zamolodchikov_integrals_1989a}%
  \BibitemOpen
  \bibfield  {author} {\bibinfo {author} {\bibfnamefont {A.}~\bibnamefont
  {Zamolodchikov}},\ }\bibfield  {title} {\bibinfo {title} {{Integrals of
  Motion and S Matrix of the (Scaled) T=T(c) Ising Model with Magnetic
  Field}},\ }\href {https://doi.org/10.1142/S0217751X8900176X} {\bibfield
  {journal} {\bibinfo  {journal} {Int. J. Mod. Phys. A}\ }\textbf {\bibinfo
  {volume} {4}},\ \bibinfo {pages} {4235} (\bibinfo {year}
  {1989}{\natexlab{b}})}\BibitemShut {NoStop}%
\bibitem [{\citenamefont {Delfino}\ \emph {et~al.}(1996)\citenamefont
  {Delfino}, \citenamefont {Mussardo},\ and\ \citenamefont
  {Simonetti}}]{delfino_nonintegrable_1996}%
  \BibitemOpen
  \bibfield  {author} {\bibinfo {author} {\bibfnamefont {G.}~\bibnamefont
  {Delfino}}, \bibinfo {author} {\bibfnamefont {G.}~\bibnamefont {Mussardo}},\
  and\ \bibinfo {author} {\bibfnamefont {P.}~\bibnamefont {Simonetti}},\
  }\bibfield  {title} {\bibinfo {title} {{Nonintegrable quantum field theories
  as perturbations of certain integrable models}},\ }\href
  {https://doi.org/10.1016/0550-3213(96)00265-9} {\bibfield  {journal}
  {\bibinfo  {journal} {Nucl. Phys. B}\ }\textbf {\bibinfo {volume} {473}},\
  \bibinfo {pages} {469} (\bibinfo {year} {1996})},\ \Eprint
  {https://arxiv.org/abs/hep-th/9603011} {arXiv:hep-th/9603011} \BibitemShut
  {NoStop}%
\bibitem [{\citenamefont {Mussardo}(2011)}]{mussardo_integrability_2011}%
  \BibitemOpen
  \bibfield  {author} {\bibinfo {author} {\bibfnamefont {G.}~\bibnamefont
  {Mussardo}},\ }\bibfield  {title} {\bibinfo {title} {{Integrability,
  Non-integrability and confinement}},\ }\href
  {https://doi.org/10.1088/1742-5468/2011/01/P01002} {\bibfield  {journal}
  {\bibinfo  {journal} {J. Stat. Mech.}\ }\textbf {\bibinfo {volume} {1101}},\
  \bibinfo {pages} {P01002} (\bibinfo {year} {2011})},\ \Eprint
  {https://arxiv.org/abs/1010.5519} {arXiv:1010.5519 [cond-mat.stat-mech]}
  \BibitemShut {NoStop}%
\bibitem [{\citenamefont {McCoy}\ and\ \citenamefont
  {Wu}(1978)}]{mccoy_twodimensional_1978}%
  \BibitemOpen
  \bibfield  {author} {\bibinfo {author} {\bibfnamefont {B.~M.}\ \bibnamefont
  {McCoy}}\ and\ \bibinfo {author} {\bibfnamefont {T.~T.}\ \bibnamefont {Wu}},\
  }\bibfield  {title} {\bibinfo {title} {{Two-dimensional Ising Field Theory in
  a Magnetic Field: Breakup of the Cut in the Two Point Function}},\ }\href
  {https://doi.org/10.1103/PhysRevD.18.1259} {\bibfield  {journal} {\bibinfo
  {journal} {Phys. Rev. D}\ }\textbf {\bibinfo {volume} {18}},\ \bibinfo
  {pages} {1259} (\bibinfo {year} {1978})}\BibitemShut {NoStop}%
\bibitem [{\citenamefont {Rakovszky}\ \emph {et~al.}(2016)\citenamefont
  {Rakovszky}, \citenamefont {Mestyán}, \citenamefont {Collura}, \citenamefont
  {Kormos},\ and\ \citenamefont {Takács}}]{Rakovszky_2016}%
  \BibitemOpen
  \bibfield  {author} {\bibinfo {author} {\bibfnamefont {T.}~\bibnamefont
  {Rakovszky}}, \bibinfo {author} {\bibfnamefont {M.}~\bibnamefont {Mestyán}},
  \bibinfo {author} {\bibfnamefont {M.}~\bibnamefont {Collura}}, \bibinfo
  {author} {\bibfnamefont {M.}~\bibnamefont {Kormos}},\ and\ \bibinfo {author}
  {\bibfnamefont {G.}~\bibnamefont {Takács}},\ }\bibfield  {title} {\bibinfo
  {title} {Hamiltonian truncation approach to quenches in the ising field
  theory},\ }\href {https://doi.org/10.1016/j.nuclphysb.2016.08.024} {\bibfield
   {journal} {\bibinfo  {journal} {Nucl. Phys. B}\ }\textbf {\bibinfo {volume}
  {911}},\ \bibinfo {pages} {805–845} (\bibinfo {year} {2016})},\ \Eprint
  {https://arxiv.org/abs/1607.01068} {arXiv:1607.01068} \BibitemShut {NoStop}%
\bibitem [{\citenamefont {Kormos}\ \emph {et~al.}(2017)\citenamefont {Kormos},
  \citenamefont {Collura}, \citenamefont {Tak{\'a}cs},\ and\ \citenamefont
  {Calabrese}}]{kormos_real_2017}%
  \BibitemOpen
  \bibfield  {author} {\bibinfo {author} {\bibfnamefont {M.}~\bibnamefont
  {Kormos}}, \bibinfo {author} {\bibfnamefont {M.}~\bibnamefont {Collura}},
  \bibinfo {author} {\bibfnamefont {G.}~\bibnamefont {Tak{\'a}cs}},\ and\
  \bibinfo {author} {\bibfnamefont {P.}~\bibnamefont {Calabrese}},\ }\bibfield
  {title} {\bibinfo {title} {Real-time confinement following a quantum quench
  to a non-integrable model},\ }\href {https://doi.org/10.1038/nphys3934}
  {\bibfield  {journal} {\bibinfo  {journal} {Nature Physics}\ }\textbf
  {\bibinfo {volume} {13}},\ \bibinfo {pages} {246} (\bibinfo {year} {2017})},\
  \Eprint {https://arxiv.org/abs/1604.03571} {arXiv:1604.03571
  [cond-mat.stat-mech]} \BibitemShut {NoStop}%
\bibitem [{\citenamefont {Lin}\ and\ \citenamefont
  {Motrunich}(2017)}]{lin_quasiparticle_2017}%
  \BibitemOpen
  \bibfield  {author} {\bibinfo {author} {\bibfnamefont {C.-J.}\ \bibnamefont
  {Lin}}\ and\ \bibinfo {author} {\bibfnamefont {O.~I.}\ \bibnamefont
  {Motrunich}},\ }\bibfield  {title} {\bibinfo {title} {Quasiparticle
  explanation of the weak-thermalization regime under quench in a nonintegrable
  quantum spin chain},\ }\href {https://doi.org/10.1103/PhysRevA.95.023621}
  {\bibfield  {journal} {\bibinfo  {journal} {Phys. Rev. A}\ }\textbf {\bibinfo
  {volume} {95}},\ \bibinfo {pages} {023621} (\bibinfo {year} {2017})},\
  \Eprint {https://arxiv.org/abs/1610.04287} {arXiv:1610.04287} \BibitemShut
  {NoStop}%
\bibitem [{\citenamefont {Hódsági}\ \emph {et~al.}(2018)\citenamefont
  {Hódsági}, \citenamefont {Kormos},\ and\ \citenamefont
  {Takács}}]{hodsagi_quench_2018}%
  \BibitemOpen
  \bibfield  {author} {\bibinfo {author} {\bibfnamefont {K.}~\bibnamefont
  {Hódsági}}, \bibinfo {author} {\bibfnamefont {M.}~\bibnamefont {Kormos}},\
  and\ \bibinfo {author} {\bibfnamefont {G.}~\bibnamefont {Takács}},\
  }\bibfield  {title} {\bibinfo {title} {Quench dynamics of the ising field
  theory in a magnetic field},\ }\href
  {https://doi.org/10.21468/scipostphys.5.3.027} {\bibfield  {journal}
  {\bibinfo  {journal} {SciPost Phys.}\ }\textbf {\bibinfo {volume} {5}},\
  (\bibinfo {year} {2018})},\ \Eprint {https://arxiv.org/abs/1803.01158}
  {arXiv:1803.01158} \BibitemShut {NoStop}%
\bibitem [{\citenamefont {Castro-Alvaredo}\ \emph {et~al.}(2020)\citenamefont
  {Castro-Alvaredo}, \citenamefont {Lencs\'es}, \citenamefont {Sz\'ecs\'enyi},\
  and\ \citenamefont {Viti}}]{castro-alvaredo_entanglement_2020}%
  \BibitemOpen
  \bibfield  {author} {\bibinfo {author} {\bibfnamefont {O.~A.}\ \bibnamefont
  {Castro-Alvaredo}}, \bibinfo {author} {\bibfnamefont {M.}~\bibnamefont
  {Lencs\'es}}, \bibinfo {author} {\bibfnamefont {I.~M.}\ \bibnamefont
  {Sz\'ecs\'enyi}},\ and\ \bibinfo {author} {\bibfnamefont {J.}~\bibnamefont
  {Viti}},\ }\bibfield  {title} {\bibinfo {title} {{Entanglement Oscillations
  near a Quantum Critical Point}},\ }\href
  {https://doi.org/10.1103/PhysRevLett.124.230601} {\bibfield  {journal}
  {\bibinfo  {journal} {Phys. Rev. Lett.}\ }\textbf {\bibinfo {volume} {124}},\
  \bibinfo {pages} {230601} (\bibinfo {year} {2020})},\ \Eprint
  {https://arxiv.org/abs/2001.10007} {arXiv:2001.10007 [cond-mat.stat-mech]}
  \BibitemShut {NoStop}%
\bibitem [{\citenamefont {James}\ \emph {et~al.}(2019)\citenamefont {James},
  \citenamefont {Konik},\ and\ \citenamefont
  {Robinson}}]{james_nonthermal_2019}%
  \BibitemOpen
  \bibfield  {author} {\bibinfo {author} {\bibfnamefont {A.~J.~A.}\
  \bibnamefont {James}}, \bibinfo {author} {\bibfnamefont {R.~M.}\ \bibnamefont
  {Konik}},\ and\ \bibinfo {author} {\bibfnamefont {N.~J.}\ \bibnamefont
  {Robinson}},\ }\bibfield  {title} {\bibinfo {title} {Nonthermal states
  arising from confinement in one and two dimensions},\ }\href
  {https://doi.org/10.1103/PhysRevLett.122.130603} {\bibfield  {journal}
  {\bibinfo  {journal} {Phys. Rev. Lett.}\ }\textbf {\bibinfo {volume} {122}},\
  \bibinfo {pages} {130603} (\bibinfo {year} {2019})},\ \Eprint
  {https://arxiv.org/abs/1804.09990} {arXiv:1804.09990 [cond-mat.stat-mech]}
  \BibitemShut {NoStop}%
\bibitem [{\citenamefont {Wurtz}\ and\ \citenamefont
  {Polkovnikov}(2020)}]{wurtz_emergent_2020}%
  \BibitemOpen
  \bibfield  {author} {\bibinfo {author} {\bibfnamefont {J.}~\bibnamefont
  {Wurtz}}\ and\ \bibinfo {author} {\bibfnamefont {A.}~\bibnamefont
  {Polkovnikov}},\ }\bibfield  {title} {\bibinfo {title} {Emergent conservation
  laws and nonthermal states in the mixed-field ising model},\ }\href
  {https://doi.org/10.1103/PhysRevB.101.195138} {\bibfield  {journal} {\bibinfo
   {journal} {Phys. Rev. B}\ }\textbf {\bibinfo {volume} {101}},\ \bibinfo
  {pages} {195138} (\bibinfo {year} {2020})},\ \Eprint
  {https://arxiv.org/abs/2002.08969} {arXiv:2002.08969 [cond-mat.str-el]}
  \BibitemShut {NoStop}%
\bibitem [{\citenamefont {Lassig}\ \emph {et~al.}(1991)\citenamefont {Lassig},
  \citenamefont {Mussardo},\ and\ \citenamefont {Cardy}}]{lassig_scaling_1991}%
  \BibitemOpen
  \bibfield  {author} {\bibinfo {author} {\bibfnamefont {M.}~\bibnamefont
  {Lassig}}, \bibinfo {author} {\bibfnamefont {G.}~\bibnamefont {Mussardo}},\
  and\ \bibinfo {author} {\bibfnamefont {J.~L.}\ \bibnamefont {Cardy}},\
  }\bibfield  {title} {\bibinfo {title} {{The scaling region of the tricritical
  Ising model in two-dimensions}},\ }\href
  {https://doi.org/10.1016/0550-3213(91)90206-D} {\bibfield  {journal}
  {\bibinfo  {journal} {Nucl. Phys. B}\ }\textbf {\bibinfo {volume} {348}},\
  \bibinfo {pages} {591} (\bibinfo {year} {1991})}\BibitemShut {NoStop}%
\bibitem [{\citenamefont {Lepori}\ \emph {et~al.}(2008)\citenamefont {Lepori},
  \citenamefont {Mussardo},\ and\ \citenamefont {Toth}}]{lepori_particle_2008}%
  \BibitemOpen
  \bibfield  {author} {\bibinfo {author} {\bibfnamefont {L.}~\bibnamefont
  {Lepori}}, \bibinfo {author} {\bibfnamefont {G.}~\bibnamefont {Mussardo}},\
  and\ \bibinfo {author} {\bibfnamefont {G.~Z.}\ \bibnamefont {Toth}},\
  }\bibfield  {title} {\bibinfo {title} {{The particle spectrum of the
  Tricritical Ising Model with spin reversal symmetric perturbations}},\ }\href
  {https://doi.org/10.1088/1742-5468/2008/09/P09004} {\bibfield  {journal}
  {\bibinfo  {journal} {J. Stat. Mech.}\ }\textbf {\bibinfo {volume} {0809}},\
  \bibinfo {pages} {P09004} (\bibinfo {year} {2008})},\ \Eprint
  {https://arxiv.org/abs/0806.4715} {arXiv:0806.4715 [hep-th]} \BibitemShut
  {NoStop}%
\bibitem [{\citenamefont {Vidal}(2003)}]{vidal_2003}%
  \BibitemOpen
  \bibfield  {author} {\bibinfo {author} {\bibfnamefont {G.}~\bibnamefont
  {Vidal}},\ }\bibfield  {title} {\bibinfo {title} {Efficient classical
  simulation of slightly entangled quantum computations},\ }\href
  {https://doi.org/10.1103/physrevlett.91.147902} {\bibfield  {journal}
  {\bibinfo  {journal} {Phys. Rev. Lett.}\ }\textbf {\bibinfo {volume} {91}},\
  \bibinfo {pages} {147902} (\bibinfo {year} {2003})},\ \Eprint
  {https://arxiv.org/abs/quant-ph/0301063} {arXiv:quant-ph/0301063}
  \BibitemShut {NoStop}%
\bibitem [{\citenamefont {Vidal}(2004)}]{vidal_2004}%
  \BibitemOpen
  \bibfield  {author} {\bibinfo {author} {\bibfnamefont {G.}~\bibnamefont
  {Vidal}},\ }\bibfield  {title} {\bibinfo {title} {Efficient simulation of
  one-dimensional quantum many-body systems},\ }\href
  {https://doi.org/10.1103/PhysRevLett.93.040502} {\bibfield  {journal}
  {\bibinfo  {journal} {Phys. Rev. Lett.}\ }\textbf {\bibinfo {volume} {93}},\
  \bibinfo {pages} {040502} (\bibinfo {year} {2004})},\ \Eprint
  {https://arxiv.org/abs/quant-ph/0310089} {arXiv:quant-ph/0310089}
  \BibitemShut {NoStop}%
\bibitem [{\citenamefont {Verstraete}\ and\ \citenamefont
  {Cirac}(2006)}]{verstraete_2006}%
  \BibitemOpen
  \bibfield  {author} {\bibinfo {author} {\bibfnamefont {F.}~\bibnamefont
  {Verstraete}}\ and\ \bibinfo {author} {\bibfnamefont {J.~I.}\ \bibnamefont
  {Cirac}},\ }\bibfield  {title} {\bibinfo {title} {Matrix product states
  represent ground states faithfully},\ }\href
  {https://doi.org/10.1103/PhysRevB.73.094423} {\bibfield  {journal} {\bibinfo
  {journal} {Phys. Rev. B}\ }\textbf {\bibinfo {volume} {73}},\ \bibinfo
  {pages} {094423} (\bibinfo {year} {2006})},\ \Eprint
  {https://arxiv.org/abs/cond-mat/0505140} {arXiv:cond-mat/0505140}
  \BibitemShut {NoStop}%
\bibitem [{\citenamefont {Hastings}(2007)}]{hastings2007}%
  \BibitemOpen
  \bibfield  {author} {\bibinfo {author} {\bibfnamefont {M.~B.}\ \bibnamefont
  {Hastings}},\ }\bibfield  {title} {\bibinfo {title} {An area law for
  one-dimensional quantum systems},\ }\href
  {https://doi.org/10.1088/1742-5468/2007/08/p08024} {\bibfield  {journal}
  {\bibinfo  {journal} {J. Stat. Mech.}\ }\textbf {\bibinfo {volume} {2007}},\
  \bibinfo {pages} {P08024–P08024} (\bibinfo {year} {2007})},\ \Eprint
  {https://arxiv.org/abs/0705.2024} {arXiv:0705.2024} \BibitemShut {NoStop}%
\bibitem [{\citenamefont {Haegeman}\ \emph {et~al.}(2011)\citenamefont
  {Haegeman}, \citenamefont {Cirac}, \citenamefont {Osborne}, \citenamefont
  {Pizorn}, \citenamefont {Verschelde},\ and\ \citenamefont
  {Verstraete}}]{haegeman_2011}%
  \BibitemOpen
  \bibfield  {author} {\bibinfo {author} {\bibfnamefont {J.}~\bibnamefont
  {Haegeman}}, \bibinfo {author} {\bibfnamefont {J.~I.}\ \bibnamefont {Cirac}},
  \bibinfo {author} {\bibfnamefont {T.~J.}\ \bibnamefont {Osborne}}, \bibinfo
  {author} {\bibfnamefont {I.}~\bibnamefont {Pizorn}}, \bibinfo {author}
  {\bibfnamefont {H.}~\bibnamefont {Verschelde}},\ and\ \bibinfo {author}
  {\bibfnamefont {F.}~\bibnamefont {Verstraete}},\ }\bibfield  {title}
  {\bibinfo {title} {{Time-Dependent Variational Principle for Quantum
  Lattices}},\ }\href {https://doi.org/10.1103/PhysRevLett.107.070601}
  {\bibfield  {journal} {\bibinfo  {journal} {Phys. Rev. Lett.}\ }\textbf
  {\bibinfo {volume} {107}},\ \bibinfo {pages} {070601} (\bibinfo {year}
  {2011})},\ \Eprint {https://arxiv.org/abs/1103.0936} {arXiv:1103.0936
  [cond-mat.str-el]} \BibitemShut {NoStop}%
\bibitem [{\citenamefont {Haegeman}\ \emph {et~al.}(2012)\citenamefont
  {Haegeman}, \citenamefont {Pirvu}, \citenamefont {Weir}, \citenamefont
  {Cirac}, \citenamefont {Osborne}, \citenamefont {Verschelde},\ and\
  \citenamefont {Verstraete}}]{haegeman_2012_excite}%
  \BibitemOpen
  \bibfield  {author} {\bibinfo {author} {\bibfnamefont {J.}~\bibnamefont
  {Haegeman}}, \bibinfo {author} {\bibfnamefont {B.}~\bibnamefont {Pirvu}},
  \bibinfo {author} {\bibfnamefont {D.~J.}\ \bibnamefont {Weir}}, \bibinfo
  {author} {\bibfnamefont {J.~I.}\ \bibnamefont {Cirac}}, \bibinfo {author}
  {\bibfnamefont {T.~J.}\ \bibnamefont {Osborne}}, \bibinfo {author}
  {\bibfnamefont {H.}~\bibnamefont {Verschelde}},\ and\ \bibinfo {author}
  {\bibfnamefont {F.}~\bibnamefont {Verstraete}},\ }\bibfield  {title}
  {\bibinfo {title} {Variational matrix product ansatz for dispersion
  relations},\ }\href {https://doi.org/10.1103/PhysRevB.85.100408} {\bibfield
  {journal} {\bibinfo  {journal} {Phys. Rev. B}\ }\textbf {\bibinfo {volume}
  {85}},\ \bibinfo {pages} {100408(R)} (\bibinfo {year} {2012})},\ \Eprint
  {https://arxiv.org/abs/1103.2286} {arXiv:1103.2286 [quant-ph]} \BibitemShut
  {NoStop}%
\bibitem [{\citenamefont {Haegeman}\ \emph
  {et~al.}(2013{\natexlab{a}})\citenamefont {Haegeman}, \citenamefont
  {Michalakis}, \citenamefont {Nachtergaele}, \citenamefont {Osborne},
  \citenamefont {Schuch},\ and\ \citenamefont
  {Verstraete}}]{haegeman_2013_el_ex}%
  \BibitemOpen
  \bibfield  {author} {\bibinfo {author} {\bibfnamefont {J.}~\bibnamefont
  {Haegeman}}, \bibinfo {author} {\bibfnamefont {S.}~\bibnamefont
  {Michalakis}}, \bibinfo {author} {\bibfnamefont {B.}~\bibnamefont
  {Nachtergaele}}, \bibinfo {author} {\bibfnamefont {T.~J.}\ \bibnamefont
  {Osborne}}, \bibinfo {author} {\bibfnamefont {N.}~\bibnamefont {Schuch}},\
  and\ \bibinfo {author} {\bibfnamefont {F.}~\bibnamefont {Verstraete}},\
  }\bibfield  {title} {\bibinfo {title} {Elementary excitations in gapped
  quantum spin systems},\ }\href
  {https://doi.org/10.1103/physrevlett.111.080401} {\bibfield  {journal}
  {\bibinfo  {journal} {Phys. Rev. Lett.}\ }\textbf {\bibinfo {volume} {111}},\
  \bibinfo {pages} {080401} (\bibinfo {year} {2013}{\natexlab{a}})},\ \Eprint
  {https://arxiv.org/abs/1305.2176} {arXiv:1305.2176 [quant-ph]} \BibitemShut
  {NoStop}%
\bibitem [{\citenamefont {Milsted}\ \emph
  {et~al.}(2013{\natexlab{b}})\citenamefont {Milsted}, \citenamefont
  {Haegeman}, \citenamefont {Osborne},\ and\ \citenamefont
  {Verstraete}}]{milsted_2013_sand}%
  \BibitemOpen
  \bibfield  {author} {\bibinfo {author} {\bibfnamefont {A.}~\bibnamefont
  {Milsted}}, \bibinfo {author} {\bibfnamefont {J.}~\bibnamefont {Haegeman}},
  \bibinfo {author} {\bibfnamefont {T.~J.}\ \bibnamefont {Osborne}},\ and\
  \bibinfo {author} {\bibfnamefont {F.}~\bibnamefont {Verstraete}},\ }\bibfield
   {title} {\bibinfo {title} {Variational matrix product ansatz for nonuniform
  dynamics in the thermodynamic limit},\ }\href
  {https://doi.org/10.1103/PhysRevB.88.155116} {\bibfield  {journal} {\bibinfo
  {journal} {Phys. Rev. B}\ }\textbf {\bibinfo {volume} {88}},\ \bibinfo
  {pages} {155116} (\bibinfo {year} {2013}{\natexlab{b}})},\ \Eprint
  {https://arxiv.org/abs/1207.0691} {arXiv:1207.0691 [cond-mat.str-el]}
  \BibitemShut {NoStop}%
\bibitem [{\citenamefont {Vanderstraeten}\ \emph {et~al.}(2015)\citenamefont
  {Vanderstraeten}, \citenamefont {Verstraete},\ and\ \citenamefont
  {Haegeman}}]{vanderstraeten_2015}%
  \BibitemOpen
  \bibfield  {author} {\bibinfo {author} {\bibfnamefont {L.}~\bibnamefont
  {Vanderstraeten}}, \bibinfo {author} {\bibfnamefont {F.}~\bibnamefont
  {Verstraete}},\ and\ \bibinfo {author} {\bibfnamefont {J.}~\bibnamefont
  {Haegeman}},\ }\bibfield  {title} {\bibinfo {title} {Scattering particles in
  quantum spin chains},\ }\href {https://doi.org/10.1103/PhysRevB.92.125136}
  {\bibfield  {journal} {\bibinfo  {journal} {Phys. Rev. B}\ }\textbf {\bibinfo
  {volume} {92}},\ \bibinfo {pages} {125136} (\bibinfo {year} {2015})},\
  \Eprint {https://arxiv.org/abs/1506.01008} {arXiv:1506.01008
  [cond-mat.str-el]} \BibitemShut {NoStop}%
\bibitem [{\citenamefont {Haegeman}\ \emph
  {et~al.}(2013{\natexlab{b}})\citenamefont {Haegeman}, \citenamefont
  {Osborne},\ and\ \citenamefont {Verstraete}}]{haegeman_2013_post}%
  \BibitemOpen
  \bibfield  {author} {\bibinfo {author} {\bibfnamefont {J.}~\bibnamefont
  {Haegeman}}, \bibinfo {author} {\bibfnamefont {T.~J.}\ \bibnamefont
  {Osborne}},\ and\ \bibinfo {author} {\bibfnamefont {F.}~\bibnamefont
  {Verstraete}},\ }\bibfield  {title} {\bibinfo {title} {Post-matrix product
  state methods: To tangent space and beyond},\ }\href
  {https://doi.org/10.1103/physrevb.88.075133} {\bibfield  {journal} {\bibinfo
  {journal} {Phys. Rev. B}\ }\textbf {\bibinfo {volume} {88}},\ \bibinfo
  {pages} {075133} (\bibinfo {year} {2013}{\natexlab{b}})},\ \Eprint
  {https://arxiv.org/abs/1305.1894} {arXiv:1305.1894 [quant-ph]} \BibitemShut
  {NoStop}%
\bibitem [{\citenamefont {Haegeman}\ \emph {et~al.}(2016)\citenamefont
  {Haegeman}, \citenamefont {Lubich}, \citenamefont {Oseledets}, \citenamefont
  {Vandereycken},\ and\ \citenamefont {Verstraete}}]{haegeman_unifying_2016}%
  \BibitemOpen
  \bibfield  {author} {\bibinfo {author} {\bibfnamefont {J.}~\bibnamefont
  {Haegeman}}, \bibinfo {author} {\bibfnamefont {C.}~\bibnamefont {Lubich}},
  \bibinfo {author} {\bibfnamefont {I.}~\bibnamefont {Oseledets}}, \bibinfo
  {author} {\bibfnamefont {B.}~\bibnamefont {Vandereycken}},\ and\ \bibinfo
  {author} {\bibfnamefont {F.}~\bibnamefont {Verstraete}},\ }\bibfield  {title}
  {\bibinfo {title} {Unifying time evolution and optimization with matrix
  product states},\ }\href {https://doi.org/10.1103/physrevb.94.165116}
  {\bibfield  {journal} {\bibinfo  {journal} {Phys. Rev. B}\ }\textbf {\bibinfo
  {volume} {94}},\ \bibinfo {pages} {165116} (\bibinfo {year} {2016})},\
  \Eprint {https://arxiv.org/abs/1408.5056} {arXiv:1408.5056 [quant-ph]}
  \BibitemShut {NoStop}%
\bibitem [{\citenamefont {Milsted}(2020)}]{milsted_evomps}%
  \BibitemOpen
  \bibfield  {author} {\bibinfo {author} {\bibfnamefont {A.}~\bibnamefont
  {Milsted}},\ }\href {https://github.com/amilsted/evoMPS} {\bibinfo {title}
  {{\emph{evoMPS}}}} (\bibinfo {year} {2020})\BibitemShut {NoStop}%
\bibitem [{\citenamefont {Li}\ \emph {et~al.}(2020)\citenamefont {Li},
  \citenamefont {Ren},\ and\ \citenamefont {Shuai}}]{li_tddmrg_2020}%
  \BibitemOpen
  \bibfield  {author} {\bibinfo {author} {\bibfnamefont {W.}~\bibnamefont
  {Li}}, \bibinfo {author} {\bibfnamefont {J.}~\bibnamefont {Ren}},\ and\
  \bibinfo {author} {\bibfnamefont {Z.}~\bibnamefont {Shuai}},\ }\bibfield
  {title} {\bibinfo {title} {Numerical assessment for accuracy and gpu
  acceleration of td-dmrg time evolution schemes},\ }\href
  {https://doi.org/10.1063/1.5135363} {\bibfield  {journal} {\bibinfo
  {journal} {J. Chem. Phys.}\ }\textbf {\bibinfo {volume} {152}},\ \bibinfo
  {pages} {024127} (\bibinfo {year} {2020})},\ \Eprint
  {https://arxiv.org/abs/1907.12044} {arXiv:1907.12044} \BibitemShut {NoStop}%
\bibitem [{\citenamefont {Verstraete}\ \emph {et~al.}(2006)\citenamefont
  {Verstraete}, \citenamefont {Wolf}, \citenamefont {P{\'e}rez-Garc{\'\i}a},\
  and\ \citenamefont {Cirac}}]{verstraete_peps_2006}%
  \BibitemOpen
  \bibfield  {author} {\bibinfo {author} {\bibfnamefont {F.}~\bibnamefont
  {Verstraete}}, \bibinfo {author} {\bibfnamefont {M.}~\bibnamefont {Wolf}},
  \bibinfo {author} {\bibfnamefont {D.}~\bibnamefont {P{\'e}rez-Garc{\'\i}a}},\
  and\ \bibinfo {author} {\bibfnamefont {J.~I.}\ \bibnamefont {Cirac}},\
  }\bibfield  {title} {\bibinfo {title} {Projected entangled states: Properties
  and applications},\ }\href {https://doi.org/10.1142/S021797920603620X}
  {\bibfield  {journal} {\bibinfo  {journal} {Int. J. Mod. Phys. B}\ }\textbf
  {\bibinfo {volume} {20}},\ \bibinfo {pages} {5142} (\bibinfo {year}
  {2006})}\BibitemShut {NoStop}%
\bibitem [{\citenamefont {Vanderstraeten}\ \emph
  {et~al.}(2019{\natexlab{a}})\citenamefont {Vanderstraeten}, \citenamefont
  {Haegeman},\ and\ \citenamefont {Verstraete}}]{vanderstraeten_2019}%
  \BibitemOpen
  \bibfield  {author} {\bibinfo {author} {\bibfnamefont {L.}~\bibnamefont
  {Vanderstraeten}}, \bibinfo {author} {\bibfnamefont {J.}~\bibnamefont
  {Haegeman}},\ and\ \bibinfo {author} {\bibfnamefont {F.}~\bibnamefont
  {Verstraete}},\ }\bibfield  {title} {\bibinfo {title} {Simulating excitation
  spectra with projected entangled-pair states},\ }\href
  {https://doi.org/10.1103/PhysRevB.99.165121} {\bibfield  {journal} {\bibinfo
  {journal} {Phys. Rev. B}\ }\textbf {\bibinfo {volume} {99}},\ \bibinfo
  {pages} {165121} (\bibinfo {year} {2019}{\natexlab{a}})},\ \Eprint
  {https://arxiv.org/abs/1809.06747} {arXiv:1809.06747} \BibitemShut {NoStop}%
\bibitem [{\citenamefont {Motruk}\ \emph {et~al.}(2016)\citenamefont {Motruk},
  \citenamefont {Zaletel}, \citenamefont {Mong},\ and\ \citenamefont
  {Pollmann}}]{motruk_2016}%
  \BibitemOpen
  \bibfield  {author} {\bibinfo {author} {\bibfnamefont {J.}~\bibnamefont
  {Motruk}}, \bibinfo {author} {\bibfnamefont {M.~P.}\ \bibnamefont {Zaletel}},
  \bibinfo {author} {\bibfnamefont {R.~S.~K.}\ \bibnamefont {Mong}},\ and\
  \bibinfo {author} {\bibfnamefont {F.}~\bibnamefont {Pollmann}},\ }\bibfield
  {title} {\bibinfo {title} {Density matrix renormalization group on a cylinder
  in mixed real and momentum space},\ }\href
  {https://doi.org/10.1103/PhysRevB.93.155139} {\bibfield  {journal} {\bibinfo
  {journal} {Phys. Rev. B}\ }\textbf {\bibinfo {volume} {93}},\ \bibinfo
  {pages} {155139} (\bibinfo {year} {2016})},\ \Eprint
  {https://arxiv.org/abs/1512.03318} {arXiv:1512.03318} \BibitemShut {NoStop}%
\bibitem [{\citenamefont {Vanderstraeten}\ \emph
  {et~al.}(2019{\natexlab{b}})\citenamefont {Vanderstraeten}, \citenamefont
  {Haegeman},\ and\ \citenamefont {Verstraete}}]{vanderstraeten_tangent_2019}%
  \BibitemOpen
  \bibfield  {author} {\bibinfo {author} {\bibfnamefont {L.}~\bibnamefont
  {Vanderstraeten}}, \bibinfo {author} {\bibfnamefont {J.}~\bibnamefont
  {Haegeman}},\ and\ \bibinfo {author} {\bibfnamefont {F.}~\bibnamefont
  {Verstraete}},\ }\bibfield  {title} {\bibinfo {title} {Tangent-space methods
  for uniform matrix product states},\ }\href
  {https://doi.org/10.21468/SciPostPhysLectNotes.7} {\bibfield  {journal}
  {\bibinfo  {journal} {SciPost Phys. Lect. Notes}\ ,\ \bibinfo {pages} {7}}
  (\bibinfo {year} {2019}{\natexlab{b}})},\ \Eprint
  {https://arxiv.org/abs/1810.07006} {arXiv:1810.07006 [cond-mat.str-el]}
  \BibitemShut {NoStop}%
\bibitem [{\citenamefont {Tang}(2020)}]{tang_private}%
  \BibitemOpen
  \bibfield  {author} {\bibinfo {author} {\bibfnamefont {E.}~\bibnamefont
  {Tang}},\ }\href@noop {} {\bibfield  {journal} {\bibinfo  {journal} {private
  communication}\ } (\bibinfo {year} {2020})}\BibitemShut {NoStop}%
\end{thebibliography}%

\end{document}